\documentclass[10pt,prl,twocolumn,showpacs,floatfix,superscriptaddress]{revtex4-1}
\usepackage{amsmath,amsbsy,graphics,epsfig,mathrsfs}
\usepackage{bm}
\usepackage[]{color}

\newcommand{\tx}{\text}

\newcommand{\nn}{\nonumber}

\newcommand{\alp}{\alpha}

\begin{document}
\title{Spin Hall and spin swapping torques in diffusive ferromagnets}
\author{Christian Ortiz Pauyac}
\affiliation{King Abdullah University of Science and Technology (KAUST),
Physical Science and Engineering Division, Thuwal 23955-6900, Saudi Arabia}
\affiliation{Univ. Grenoble Alpes, CNRS, CEA, Grenoble-INP, INAC-Spintec, 38000 Grenoble, France}
\author{Mairbek Chshiev}
\affiliation{Univ. Grenoble Alpes, CNRS, CEA, Grenoble-INP, INAC-Spintec, 38000 Grenoble, France}
\author{Aurelien Manchon}
\affiliation{King Abdullah University of Science and Technology (KAUST),
Physical Science and Engineering Division, Thuwal 23955-6900, Saudi Arabia}
\affiliation{King Abdullah University of Science and Technology (KAUST),
Computer, Electrical and Mathematical Science and Engineering (CEMSE), Thuwal 23955-6900, Saudi Arabia}
\author{Sergey A. Nikolaev}
\email[]{saishi@inbox.ru}
\affiliation{Ural Federal University, 10 Mira str., 620002 Ekaterinburg, Russia}
\affiliation{Univ. Grenoble Alpes, CNRS, CEA, Grenoble-INP, INAC-Spintec, 38000 Grenoble, France}
\date{\today}

\begin{abstract}
A complete set of the generalized drift-diffusion equations for a coupled charge and spin dynamics in ferromagnets in the presence of extrinsic spin-orbit coupling is derived from the quantum kinetic approach, covering major transport phenomena, such as the spin and anomalous Hall effects, spin swapping, spin precession and relaxation processes. We argue that the spin swapping effect in ferromagnets is enhanced due to spin polarization, while the overall spin texture induced by the interplay of spin-orbital and spin precessional effects displays a complex spatial dependence that can be exploited to generate torques and nucleate/propagate domain walls in centrosymmetric geometries without use of external polarizers, as opposed to the conventional understanding of spin-orbit mediated torques. 
\end{abstract}
\pacs{85.75.-d, 73.40.Gk, 72.25.-b, 72.10.-d}
\maketitle

\par \emph{Introduction}. The exploitation of spin-orbit coupling (SOC) effects to probe and control the magnetization in nanodevices has been extensively studied, uncovering many physical phenomena, such as the anomalous Hall effect \cite{ahe}, spin Hall effect \cite{she}, tunneling anisotropic magnetoresistance~\cite{tamr}, electrically controlled perpendicular magnetic anisotropy \cite{ecma}, and relativistic spin torques \cite{mihai,miron,liu}. The latter observed in multilayers comprising ferromagnets and normal metals display both spin-orbit torques induced by the interfacial inverse spin-galvanic effect \cite{miron} and spin-transfer torques associated with the spin Hall effect in an adjacent non-magnetic layer~\cite{liu}. Spin-orbit torques generated in a single ferromagnetic layer are of great importance to enable electrical control of the magnetization without use of an external polarizer, and they offer many promising advantages compared to spin-transfer torques, such as high scalability and stability. Thus, finding novel routes to excite magnetization dynamics by means of spin-orbit torques is essential for realizing high-performance spintronic devices. 

\par Meanwhile, new ways to generate spin accumulation are also of strong interest. More recently, a new mechanism referred to as spin swapping, which converts a primary spin current into a secondary spin current with interchanged spin and flow directions, was proposed to exist in normal metals and semiconductors in the presence of spin-orbit coupled impurities~\cite{dyakonov}. However, whether spin swapping in ferromagnets can produce a measurable effect remains an open question that has not yet been addressed. On the one hand, the exchange magnetic field present in ferromagnets tends to destroy the induced spin accumulation. On the other hand, not only does SOC act constructively in generating spin accumulation, it also leads to the spin-memory loss \cite{fabian}. Overall, the possibility to employ these effects in ferromagnets strongly depends on the transport regime as a function of many parameters describing a given system. 

\par In this letter, we explore the nature of the extrinsic spin Hall and spin swapping effects in diffusive ferromagnets and demonstrate that these effects can offer potential advantages in contrast to non-centrosymmetric magnetic multilayers involving heavy metals. To this end, we develop a set of coupled spin-charge diffusive equations by using the non-equilibrium Green's function formalism and taking into account scattering off the impurity induced SOC potential. Based on these equations, we proceed to study the interplay between spin-orbital and spin precessional effects that can be used to demonstrate current-driven manipulation of the magnetization in centrosymmetric magnets.

\par  \emph{Derivation of the coupled spin-charge drift-diffusion equation}. We consider a single ferromagnetic layer in the standard $s$-$d$ model~\cite{shubin} defined as $\hat{\mathcal{H}}=\frac{\hat{\boldsymbol{p}}^{2}}{2m}\hat{\sigma}_{0}+J\hat{\boldsymbol{\sigma}}\cdot\boldsymbol{m}+\hat{\mathcal{H}}_{\mathrm{imp}}$, where $m$ is the effective electron's mass, $\hat{\boldsymbol{p}}$ is the momentum operator, $J$ is the exchange coupling, $\hat{\sigma}_{0} $ is the identity matrix, $\hat{\boldsymbol{\sigma}}$ is the Pauli matrix vector, and $\boldsymbol{m}$ is the unit vector of the spatial magnetization profile. Here, the third term stands for the impurity potential given by randomly distributed $N$ impurities $\boldsymbol{R}_{j}$, $ \hat{\mathcal{H}}_{\mathrm{imp}} = \sum_j^N[V(\boldsymbol{r} - \boldsymbol{R}_{j})\hat{\sigma}_{0}+\frac{\xi_{SO}}{\hbar k_F^2}\hat{\boldsymbol{\sigma}} \cdot (\nabla V(\boldsymbol{r} - \boldsymbol{R}_{j})\times \hat{\boldsymbol{p}})]$, where $V(\boldsymbol{r} - \boldsymbol{R}_{j})=v_{i}\delta(\boldsymbol{r} - \boldsymbol{R}_{j})$ is the on-site impurity potential, $\xi_{SO}$ is the SOC parameter (defined as a dimensionless quantity), and $k_F$ is the Fermi wave vector. In the Keldysh formalism for an interacting system driven out of equilibrium, the Dyson equation for the non-equilibrium Green's function $\hat{G}^{K}$ is written as:

\begin{equation}
\label{eq:dyson}
[\hat{G}^{R}]^{-1}\ast\hat{G}^{K} - \hat{G}^{K}\ast[\hat{G}^A]^{-1} = \hat{\Sigma}^{K}\ast\hat{G}^{A} - \hat{G}^{R}\ast\hat{\Sigma}^{K},
\end{equation}

\noindent where $\hat{G}^{i}\equiv \hat{G}^{i}(\boldsymbol{r},t;\boldsymbol{r'},t')$ and $\hat{\Sigma}^{i}\equiv\hat{\Sigma}^{i}(\boldsymbol{r},t;\boldsymbol{r'},t')$ with $i=K,R,A$ are the real space real time Keldysh, retarded and advanced Green's functions and self-energies, respectively, and $\hat{G}^{R(A)}=\hat{G}_{0}^{-1}-\hat{\Sigma}^{R(A)}_{}$, where $\hat{G}_{0}^{-1}=i\hbar\partial_{t}-\hat{\mathcal{H}}$ is a non-interacting Green's function for the system without impurities \cite{rammer}. Having applied the Wigner transformation $\hat{G}^{K}(\boldsymbol{r},t;\boldsymbol{r'},t')=\int \frac{dE}{2\pi} \frac{d\boldsymbol{k}}{(2\pi)^{3}} e^{i\boldsymbol{k}\cdot(\boldsymbol{r}-\boldsymbol{r'})-i\frac{E}{\hbar}(t-t')}\hat{g}^{K}_{\boldsymbol{k},E}(\boldsymbol{R},T)$ with $\bm{R}=(\bm{r}+\bm{r'})/2$ and $T=(t+t')/2$, we employ the so-called gradient approximation to linearize convolutions $(\ast)$ in the Dyson equation~(\ref{eq:dyson}) and obtain the following quantum kinetic equation for the non-equilibrium distribution function $\hat{g}_{\boldsymbol{k}}=i\int\frac{dE}{2\pi}\hat{g}^{K}_{\boldsymbol{k},E}(\boldsymbol{R},T)$:

\begin{equation}
\label{qke2}
\hbar\partial_{T}\hat{g}_{\boldsymbol{k}}-i[\hat{g}_{\boldsymbol{k}},J \hat{\boldsymbol{\sigma}} \cdot\boldsymbol{m}]+\frac{\hbar^{2}}{m}(\boldsymbol{k} \cdot \nabla_{\boldsymbol{R}})\,\hat{g}_{\boldsymbol{k}} = \int\frac{dE}{2\pi}\mathscr{C},\vspace{0.5cm}
\end{equation}

\noindent where $\mathscr{C} =  (\hat{\Sigma}^{K} \hat{G}^{A} - \hat{G}^{R}\hat{\Sigma}^{K}) + (\hat{\Sigma}^{R}\hat{g}^{K} - \hat{g}^{K}\hat{\Sigma}^{A})$ is the collision integral that accounts for the scattering and relaxation events, respectively. Scattering off the impurity potential in the right-hand side of Eq.~(\ref{qke2}) is considered up to third order by impurity averaging over disorder with concentration $n_{i}$, so that the spin-dependent momentum and spin-flip relaxations, side-jump \cite{berger}, spin swapping and skew-scattering \cite{maekawa} processes are properly taken into account. In the diffusive limit, where the mean-free path is much less compared to the size of the system, one can partition the distribution function $\hat{g}_{\boldsymbol{k}}=\hat{\sigma}_{0}-2\hat{h}_{\boldsymbol{k}}$ into the isotropic charge $\mu_c$, spin $\boldsymbol{\mu}$ and anisotropic $\hat{\boldsymbol{j}}$ components, $\hat{h}_{\boldsymbol{k}}=\mu_c\hat{\sigma}_{0}+\boldsymbol{\mu}\cdot \hat{\boldsymbol{\sigma}}+\hat{\boldsymbol{j}}\cdot\check{\boldsymbol{k}}$, where $\check{\boldsymbol{k}}=\boldsymbol{k}/|\boldsymbol{k}|$ \cite{brataas}. First, integrating Eq.~(\ref{qke2}) multiplied by $\check{\boldsymbol{k}}$ over the Brillouin zone gives us the corresponding expression for $\hat{\boldsymbol{j}}\equiv\hat{\boldsymbol{j}}(\mu_{c},\boldsymbol{\mu})$. Secondly, integrating Eq.~(\ref{qke2}) itself leads to the generalized continuity equation for the charge $\mu_{c}$ and spin $\boldsymbol{\mu}$ densities, so that their time dependence is given as a divergence of the charge $\boldsymbol{j}^{C}$ and spin $\boldsymbol{J}_{j}^{S}$ (its $j$th spin component) currents, respectively. We refer the reader to Ref.~\cite{suppl} for more detail concerning the derivation.  Finally, in the weak exchange coupling limit ($J\ll\varepsilon_{F}$, where $\varepsilon_{F}$ is the Fermi energy), the resulting drift-diffusion equations up to leading orders in the exchange interaction and SOC have the following form:

\begin{figure*}[ht]
\centering
\includegraphics[trim = 0mm 0mm 0mm 0mm, clip, scale=0.31]{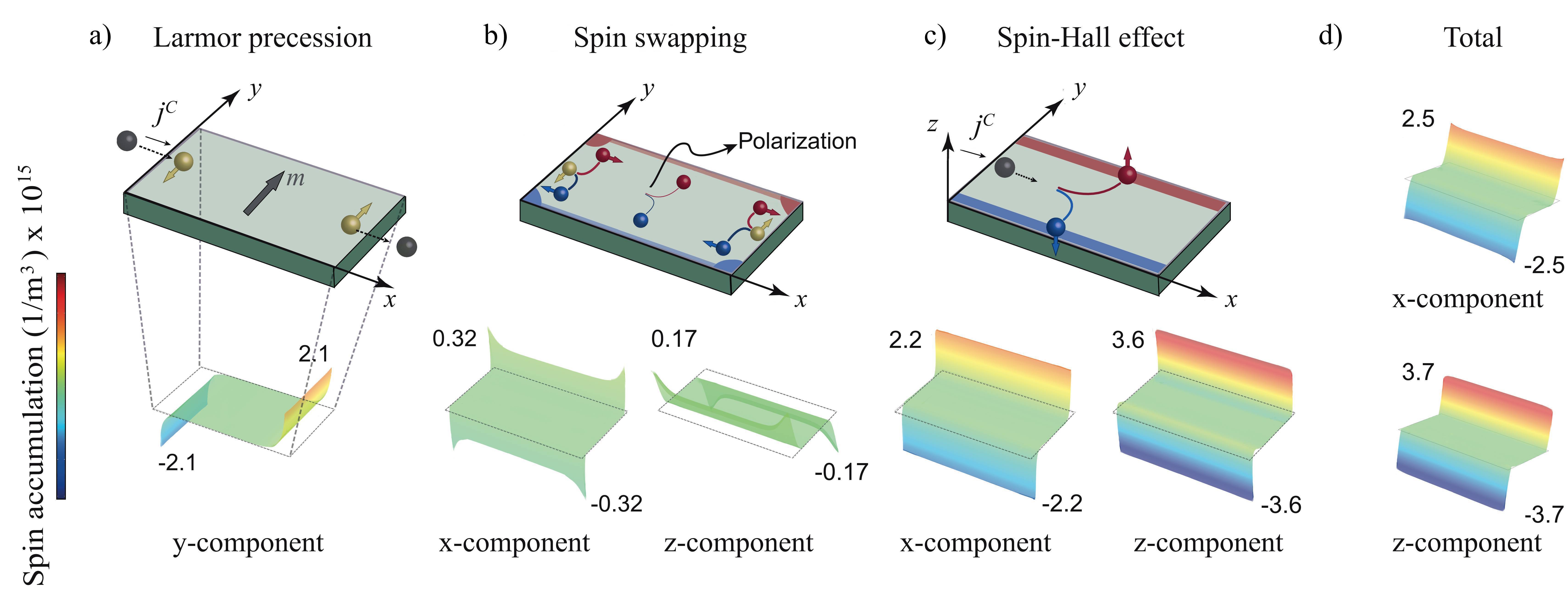}
\caption{Spin accumulation profiles $\mu_x$, $\mu_y$, and $\mu_z$, as calculated for the rectangular geometry of $100\times 50$ nm$^{2}$: a)~$\mu_{y}$ in the case of Larmor precession only, b) $\mu_{x}$ and $\mu_{z}$ when only the spin swapping and Larmor precession terms are considered, c) $\mu_{x}$ and $\mu_{z}$ when only the spin Hall effect and Larmor precession are considered, d) refers to the full drift-diffusion equations. Here, the spin current polarized along the $y$ axis is flowing along the $x$ axis, $\varepsilon_{F}=0.7$~eV, $J=0.02$~eV, and $\xi_{SO}= 0.3$. The spin diffusion length is  $l_{sf} = \sqrt{D\tau_{sf}}=5$~nm, the Larmor precession length is $l_{L}=\sqrt{D\tau_{L}} = 2.6$ nm, the spin dephasing length is $l_{\phi}=\sqrt{D\tau_{\phi}}=4.8$~nm, the mean free path is $l_{F}= \tau_{0}v_{F}=2.5$~nm, the Fermi velocity is $v_{F}=5\times10^5$~m/s, and the Fermi wave-vector is $k_{F}=4.3$~nm$^{-1}$. The grey and yellow spheres denote the non- and spin-polarized charge currents, while the red and blue spheres denote the spin-up and spin-down components, respectively.} \label{fig2}
\end{figure*}

\begin{widetext}
\begin{align}
\label{muc}
\partial_{T}\mu_{c}&=D\nabla^2 [ \mu_c + \beta \bm{\mu}\cdot\bm{m}]=-\nabla\cdot\boldsymbol{j}^{C}, \\
\label{mu}
\partial_{T}\boldsymbol{\mu}&=-\nabla\cdot\boldsymbol{J}^{S}  + \frac{1}{\tau_L}\boldsymbol{m}\times\boldsymbol{\mu} + \frac{1}{\tau_\phi}\boldsymbol{m}\times(\boldsymbol{m}\times\boldsymbol{\mu}) - \frac{1}{\tau_{sf}}\boldsymbol{\mu},\\
\label{jeo}
\boldsymbol{j}^{C}/D&= \tilde{\boldsymbol{j}}^{C}/D + \nabla \times \Big[ \alp_{sj}(2\bm{\mu}-\beta\mu_c \bm m) +\alp_{sk}(\boldsymbol{\mu} - \beta\mu_c \boldsymbol{m}) +(\alp_{sj} \frac{\tau_0}{\tau_L} + \alp_{sk}\frac{\tau_0}{\tau_L} -\alp_{sw}\beta)\boldsymbol{m}\times\boldsymbol{\mu} \Big], \\
\label{qs}
\boldsymbol{J}_{j}^{S}/D& = \tilde{\boldsymbol{J}}_{j}^{S}/D + \nabla\times\Big[\alp_{sj}\boldsymbol{e}_j(2\mu_c-\beta\boldsymbol{\mu}\cdot\boldsymbol{m}) + \alp_{sk}\boldsymbol{e}_j(\mu_c-\beta\bm \mu \cdot \bm m) - \alp_{sw}\boldsymbol{e}_{j}\times(\boldsymbol{\mu} - \beta \mu_c\boldsymbol{m}) \\
& + \frac{\tau_0}{\tau_L}(\alp_{sj}+\alp_{sk})(\boldsymbol{e}_j\times\boldsymbol{m})\mu_{c} + (\alp_{sj}\beta+\alp_{sk}\beta-\alp_{sw}\frac{\tau_0}{\tau_L}) (\boldsymbol{e}_j\times\boldsymbol{m})\times\boldsymbol{\mu} - (\alp_{sw}\frac{\tau_0}{\tau_L}+\alp_{sj}\beta+\alp_{sk}\beta)\boldsymbol{e}_j\times(\boldsymbol{m}\times\boldsymbol{\mu}) \Big],\nn \\\nn
\vspace{1.0cm}
\end{align}
\end{widetext}

\noindent where $\tilde{\boldsymbol{j}}^{C} = -D\nabla (\mu_c+\beta\boldsymbol{m}\cdot\boldsymbol{\mu})$ and $\tilde{\boldsymbol{J}}_{j}^{S}/D=-\nabla(\mu_j+\beta \mu_c m_j) +\frac{\tau_{0}}{\tau_{L}}\nabla(\boldsymbol{m}\times\boldsymbol{\mu})_{j} + \frac{\tau_0}{\tau_\phi}\nabla(\boldsymbol{m}\times(\boldsymbol{m}\times\boldsymbol{\mu}))_{j}$ are the charge and spin currents in the absence of SOC, respectively; $\beta=\frac{J}{2\varepsilon_{F}}$ is the polarization factor; $\alp_{sw}=\frac{2\xi_{SO}}{3}$, $\alp_{sj}=\frac{\xi_{SO}}{l_F k_F}$, and $\alp_{sk}=\frac{v_{i} m k_F\xi_{SO}}{3\pi\hbar^{2}}$ are the dimensionless spin swapping, side-jump, and skew scattering coefficients, respectively; $D=\frac{\tau_{0}v_{F}^{2}}{3}$ is the diffusion coefficient, $v_{F}$ is the Fermi velocity, and $l_{F}=\tau_{0}v_{F}$ is the mean-free path. Here, $\frac{1}{\tau_{0}}=\frac{2\pi v_{i}^{2}n_{i}D_{0}}{\hbar}$ is the spin independent relaxation time, where $D_{0}=\frac{mk_{F}}{2\pi^{2}\hbar^{2}}$ is the spin independent density of states, $\frac{1}{\tau_{sf}}=\frac{8}{9}\frac{\xi_{SO}^2}{\tau_0}$ is the spin-flip relaxation time, $\frac{1}{\tau_{L}}=\frac{2J}{\hbar}$ is the spin precession time around the magnetization, known as the Larmor precession time, and $\frac{1}{\tau_{\phi}}=\frac{4J^2\tau_0}{\hbar^{2}}$ refers to the spin dephasing term. The set of the drift-diffusion equations (\ref{muc})--(\ref{qs}) is the central result of this paper. On the one hand, in the absence of SOC, our approach is in-line with the generalized drift-diffusion theory \cite{waintal}, which captures main features of the transverse spin transport in ferromagnets, such as the Larmor precession and spin dephasing terms. On the other hand, in the case of normal metals ($\beta\rightarrow0$, $\tau_{L}\rightarrow\infty$, $\tau_{\phi}\rightarrow\infty$) our equations are in agreement with Shen {\it et al.} \cite{vignale}, while some other works fail to include the correct symmetry of the spin swapping term \cite{soc2,hamed,hamed2,suppl}. In the presence of both, the exchange interaction and extrinsic SOC, the anomalous Hall effect is present in Eq.~\eqref{jeo} (second and third terms) from both the side-jump and skew scattering processes \cite{ahe1}, while spin polarization and spin precession give rise to additional terms to the anomalous charge and spin currents in Eqs.~(\ref{jeo}) and~(\ref{qs}). Overall, the resulting charge and spin accumulation profiles appear to be much more complex, as opposed to normal metals. In magnetic systems with SOC, the competition between these effects is governed by the ratio of the corresponding characteristic lengths: spin precession, spin dephasing, and spin diffusion lengths. In ferromagnets with a strong exchange coupling, where the spin dephasing length is shorter than the spin-flip relaxation, the spin Hall and spin swapping effects are expected to vanish far from the interface, as the strong exchange field tends to destroy induced spin currents. Consequently, any transverse spin component will eventually align or anti-align with the magnetization \cite{stiles}. In contrast, in the weak exchange coupling limit ($J\ll\varepsilon_{F}$) the spin dephasing length is larger or comparable with the spin-flip relaxation, and these coupled effects can become prominent.

\par \emph{Spin accumulation profile}. To study the interplay of the effects in question, the drift-diffusion equations (\ref{muc})-(\ref{qs}) can be solved numerically by putting $\boldsymbol{m}$ along the $y$ axis and imposing adequate boundary conditions \cite{bound}. The results calculated for the rectangular geometry are shown in Fig.~\ref{fig2}. In the absence of SOC, the spin current density $J^{S}_{xy}\sim \nabla_{x}\mu_{y}$ is induced in the magnetic layer and due to Larmor precession the spin accumulation $\mu_{y}$ is localized at the normal metal/ferromagnet interfaces along the transport direction (Fig.~\ref{fig2}a), in agreement with the Valet-Fert theory \cite{vfert}. When the impurity induced SOC is present, the transverse spin accumulation is expected to build up at the lateral edges. First off,  let us analyze additional contributions to the spin swapping term. As seen from Eq.~(\ref{qs}), the ``spin swapping'' spin current has the following form:

\begin{equation}
\label{eq:cursw}
\begin{aligned}
J_{ij}^{sw}/D&=\alpha_{sw}(\nabla_{j}\mu_{i}-\delta_{ij}\nabla_{k}\mu_{k}) \\
&+ \alpha_{sw}\beta(\delta_{ij}m_{k}\nabla_{k}-m_{i}\nabla_{j})\mu_{c} \\
&+(\alpha_{sj}\frac{\tau_{0}}{\tau_{L}}+\alpha_{sk}\frac{\tau_{0}}{\tau_{L}})(\delta_{ij}m_{k}\nabla_{k}-m_{i}\nabla_{j})\mu_{c},
\end{aligned}
\end{equation}

\noindent where summation over repeated indexes is implied. In normal metals, due the spin swapping mechanism the primary spin current $\tilde{J}^{S}_{ji}\sim \nabla_{j}\mu_{i}$ gives rise to the secondary spin current $J_{ij}^{sw}\sim \alpha_{sw}(\nabla_{j}\mu_{i}-\delta_{ij}\nabla_{k}\mu_{k})$ and the generated spin accumulation $\mu_{i}$ decays over a length scale given by the spin-flip relaxation length and survives only close to the interface \cite{hamed}. However, in ferromagnets an additional non-vanishing spin accumulation $\sim\alpha_{sw}\beta$ builds up due to spin polarization and develops smoothly at the lateral edges (Fig.~\ref{fig2}b, $x$-component). Interestingly, there is an extra term coming from the spin Hall effect and precessional motion (third term in Eq.~(\ref{eq:cursw})), which also exhibits the symmetry of the spin-swapping current, even though it is not actually attributed to the spin swapping effect itself. The rest of the effects are related to Larmor precession. For example, let us consider a toy model, where $\boldsymbol{m}$ points along the $y$ axis and the spin Hall effect, spin polarization, and higher order terms to the precessional motion are discarded, so the spin current density in Eq.~(\ref{qs}) is reduced to:
\begin{align}
\label{swq1}
\frac{J_{yx}^{S}}{D} &\sim -\nabla_y \mu_x - \frac{\tau_0}{\tau_L} \nabla_y \mu_z + \alp_{sw}\nabla_x \mu_y,\\
\frac{J_{yz}^{S}}{D} &\sim -\nabla_y \mu_z +\frac{\tau_0}{\tau_L} \nabla_y \mu_x.
\label{swq2}
\end{align} 
\noindent  As one can see, there is an additional term in Eq.~(\ref{swq2}) that couples $J^{S}_{yx}$ to $J^{S}_{yz}$ and builds up $\mu_{z}$, which eventually manifests itself due to the spin swapping effect (Fig.~\ref{fig2}b, $z$-component). Similar results are obtained if we neglect spin swapping and focus on the spin Hall effect instead, where the spin accumulations do not vanish far from the interfaces (Fig.~\ref{fig2}c). If we solve our toy models separately for the side-jump and spin swapping terms, the following relation holds for the maxima of the corresponding spin accumulations:
\begin{equation}
\frac{\mu_x}{\mu_z} \propto \frac{\alp_{sw}\beta}{\alp_{sj}} = \frac{1}{3}l_F k_F \beta,
\label{17}
\end{equation}
\noindent so that spin swapping is about one order of magnitude smaller than the spin Hall effect (as also seen in Fig.~\ref{fig2}) in diffusive ferromagnets ($J\ll\varepsilon_{F}$) as a result of $\mu_x$ depending strongly on the polarization factor $\beta$. 

\begin{figure}[t!]
\centering
\includegraphics[trim = 0mm 0mm 0mm -10mm, clip, scale=0.26]{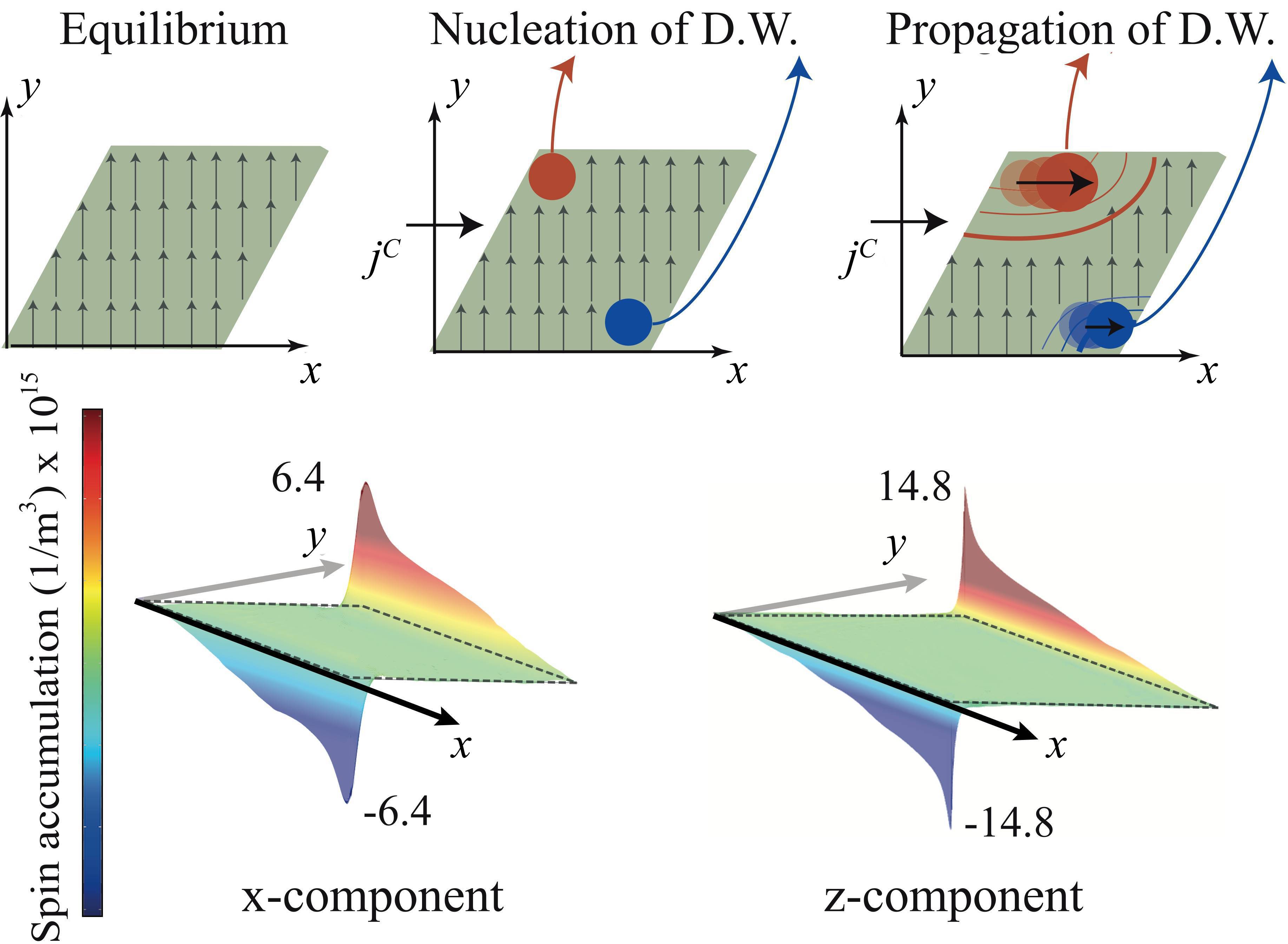}
\caption{Spin accumulation profiles $\mu_x$ (left) and $\mu_z$ (right), as calculated for the diamond-shaped geometry of $100\times50$~nm$^{2}$. The top panel describes the idea of nucleation and propagation of reversed magnetic domains, where the red and blue circles correspond to the spin-up and spin-down components of the spin accumulation, respectively. Here, the black arrows show the magnetization direction in the $xy$ plane, and its orientation with respect to the $z$ axis is given by $\theta=60^\circ$. The rest of the calculation parameters are given in Fig.~1.} \label{fig4}
\end{figure}

\par \emph{Current driven magnetization switching}. We propose a possible way to exploit the spin Hall and spin swapping effects to reversibly control the magnetization in centrosymmetric ferromagnets that can be realized even in the absence of adjacent non-magnetic layers.~Normally, spin-orbit torques are observed in ferromagnetic films lacking inversion symmetry through the Rashba effect, which is essentially inherent to non-centrosymmetric structures. However, geometry itself can play an important role building up distorted spin accumulation profiles and giving rise to non-zero local spin-orbit torques. For example, let us consider a centrosymmetric diamond-shaped geometry. The resulting spin accumulation presented in Fig.~\ref{fig4} turns out to be highly asymmetric (while the net spin accumulation is zero) and peaks at the opposite edges that can be used to nucleate reversed magnetic domains. Once nucleated at the corners, the flowing current can either expand or shrink the reversed magnetic domain by current-driven domain wall motion, as shown in the top panel of Fig.~\ref{fig4}. A similar scenario in controlling the magnetization (albeit without considering the spin Hall or spin swapping mechanisms) has been studied in Ref.~\cite{safeer}. 

\par \emph{Conclusion}. We derived a complete set of the drift-diffusion equations for the coupled charge and spin transport in diffusive ferromagnets in the presence of extrinsic SOC.
While combining major effects, such as the spin and inverse spin Hall effects, anomalous Hall effect and spin swapping, these equations reveal some new intriguing features. In particular, we showed that in ferromagnets the resulting spin accumulation exhibits a complex spatial profile, where the spin swapping effect is enhanced due to spin polarization, while spin precession gives rise to additional contributions to the anomalous charge and spin currents. These effects can be employed to generate spin-orbit mediated torques and reversibly control the magnetization in centrosymmetric structures. Our results call for experimental approbation in current-driven magnetization dynamics, where suitable materials may include magnetic alloys with heavy impurities, such as Co-Pt, Fe-Au \cite{fert-fm} or CuMn-Pt \cite{prejean-fm}.

\par \emph{Acknowledgements.}~A.M and C.O.P were supported by the King Abdullah University and Technology (KAUST). M.C. and S.A.N. were supported by the programme ``Emergence et partenariat strat\'egique''  at Univ. Grenoble Alpes and the PRESTIGE Postdoctoral Fellowship Programme co-funded by EU. S.A.N. acknowledges support from the Act~211 Government of the Russian Federation, contract No~02.A03.21.0006
\label{sec:conclusion}



\begin{thebibliography}{999}
\bibitem{bound}
In our calculations the following boundary conditions are imposed: $\boldsymbol{J}^{S}_{j}=0$ at the boundary, while the charge densities at the edges in the propagation direction are $\mu_{c\tx{L}}=k_F^2/2\pi$ and $\mu_{c\tx{R}} =0$, respectively.
\end{thebibliography}
\end{document}


\title{Supplementary Materials:\\Spin Hall and spin swapping torques in diffusive ferromagnets}
\author{Christian Ortiz Pauyac}
\affiliation{King Abdullah University of Science and Technology (KAUST),
Physical Science and Engineering Division, Thuwal 23955-6900, Saudi Arabia}
\affiliation{Univ. Grenoble Alpes, CNRS, CEA, Grenoble-INP, INAC-Spintec, 38000 Grenoble, France}
\author{Mairbek Chshiev}
\affiliation{Univ. Grenoble Alpes, CNRS, CEA, Grenoble-INP, INAC-Spintec, 38000 Grenoble, France}
\author{Aurelien Manchon}
\affiliation{King Abdullah University of Science and Technology (KAUST),
Physical Science and Engineering Division, Thuwal 23955-6900, Saudi Arabia}
\affiliation{King Abdullah University of Science and Technology (KAUST),
Computer, Electrical and Mathematical Science and Engineering (CEMSE), Thuwal 23955-6900, Saudi Arabia}
\author{Sergey A. Nikolaev}
\email[]{saishi@inbox.ru}
\affiliation{Department of Theoretical Physics and Applied Mathematics, Ural Federal University, 19 Mira str., 620002 Ekaterinburg, Russia}
\affiliation{Univ. Grenoble Alpes, CNRS, CEA, Grenoble-INP, INAC-Spintec, 38000 Grenoble, France}
\date{\today}
\maketitle

\tableofcontents

\newpage
\section*{List of model and transport parameters}
\vspace{-6.0cm}
\begin{table*}[h!]
\begin{center}
\begin{tabular}{ll}
$\begin{aligned} k_{F} \end{aligned}$ & $\qquad$ Fermi wavevector \\
& \\ 
$\begin{aligned} m \end{aligned}$ & $\qquad$ Electron's effective mass \\
& \\ 
$\begin{aligned} J \end{aligned}$ & $\qquad$ Exchange constant in the $s$-$d$ model \\
& \\
$\begin{aligned} v_{i} \end{aligned}$ & $\qquad$ Impurity potential \\
& \\
$\begin{aligned} n_{i} \end{aligned}$ & $\qquad$ Impurity concentration \\
& \\
$\begin{aligned} \xi_{SO} \end{aligned}$ & $\qquad$ Dimensionless spin-orbit coupling constant \\
& \\
$\begin{aligned} \varepsilon_{F}=\frac{\hbar^{2}k_{F}^{2}}{2m} \end{aligned}$ & $\qquad$ Fermi energy \\
& \\
$\begin{aligned} v_{F}=\frac{\hbar k_{F}}{m} \end{aligned}$ & $\qquad$ Fermi velocity \\
& \\
$\begin{aligned} \beta=\frac{J}{2\varepsilon_{F}} \end{aligned}$ & $\qquad$ Spin polarization factor  \\
& \\
$\begin{aligned} D_{0}=\frac{mk_{F}}{2\pi^{2}\hbar^{2}} \end{aligned}$ & $\qquad$ Spin independent density of states per spin at the Fermi level \\
& \\
$\begin{aligned} \frac{1}{\tau_{0}}=\frac{2\pi v_{i}^{2}n_{i}D_{0}}{\hbar} \end{aligned}$ & $\qquad$ Spin independent relaxation time \\
& \\
$\begin{aligned} \frac{1}{\tau_{L}}=\frac{2J}{\hbar} \end{aligned}$ & $\qquad$ Larmor precession time \\
& \\
$\begin{aligned} \frac{1}{\tau_{\phi}}=\frac{4J^{2}\tau_{0}}{\hbar^{2}} \end{aligned}$ & $\qquad$ Spin dephasing relaxation time \\
& \\
$\begin{aligned} \frac{1}{\tau_{sf}}=\frac{8}{9}\frac{\xi_{SO}^{2}}{\tau_{0}} \end{aligned}$ & $\qquad$ Spin-flip relaxation time \\
& \\
$\begin{aligned} l_{F}=\tau_{0}v_{F} \end{aligned}$ & $\qquad$ Mean-free path \\
& \\
$\begin{aligned} D=\frac{\tau_{0}v_{F}^{2}}{3} \end{aligned}$ & $\qquad$ Diffusion coefficient \\
& \\
$\begin{aligned} \alpha_{sw}=\frac{2}{3}\xi_{SO} \end{aligned}$ & $\qquad$ Dimensionless spin swapping constant \\
& \\
$\begin{aligned} \alpha_{sj}=\frac{\xi_{SO}}{l_{F}k_{F}} \end{aligned}$ & $\qquad$ Dimensionless side-jump constant \\
& \\
$\begin{aligned} \alpha_{sk}=\frac{v_{i}mk_{F}}{3\pi\hbar^{2}}\xi_{SO} \end{aligned}$ & $\qquad$ Dimensionless skew scattering constant \\
& \\
\end{tabular}
\end{center}
\end{table*}

\pagebreak
\section{General formalism}
\par We start with a free-electron Hamiltonian $\hat{\mathcal{H}}_{0}$ and its Fourier transform $\hat{\mathcal{H}}_{\boldsymbol{k}}$: 
\begin{equation}
\hat{\mathcal{H}}_{0}=-\frac{\hbar^{2}}{2m}\nabla^{2}\hat{\sigma}_{0}+J\hat{\boldsymbol{\sigma}}\cdot\boldsymbol{m}\qquad\xrightarrow{\,\,\mathcal{F}\,\,}\qquad\hat{\mathcal{H}}_{\boldsymbol{k}}=\frac{\hbar^{2}_{}\boldsymbol{k}^{2}_{}}{2m}\hat{\sigma}_{0}+J\hat{\boldsymbol{\sigma}}\cdot\boldsymbol{m}.
\end{equation}
\noindent Here, the first term stands for the kinetic energy, where $m$ and $\boldsymbol{k}$ are the electron's effective mass and wave vector, respectively; the second term refers to the exchange interaction in the so-called $s$-$d$ model, where $\boldsymbol{m}=(\cos{\phi}\sin{\theta},\sin{\phi}\sin{\theta},\cos{\theta})$ is the magnetization unit vector parametrized in spherical coordinates, $J$ is the exchange coupling parameter, $\hat{\boldsymbol{\sigma}}$ is the Pauli matrix vector, and $\hat{\sigma}_{0}$ is the identity matrix. The unperturbed Green's function is defined by $\hat{\mathcal{H}}_{\boldsymbol{k}}$:
\begin{equation}
\hat{G}_{0,\boldsymbol{k}E}^{R(A)}=\left[E-\hat{\mathcal{H}}_{\boldsymbol{k}}\pm i\eta\right]^{-1}=\sum\limits_{s=\pm}\frac{|s\rangle\langle s|}{E-E_{\boldsymbol{k}s}\pm i\eta}=\frac{1}{2}\sum\limits_{s=\pm}\frac{\hat{\sigma}_{0}+s\hat{\boldsymbol{\sigma}}\cdot\boldsymbol{m}}{E-E_{\boldsymbol{k}s}\pm i\eta},
\label{eq:gzero}
\end{equation}
\noindent where $s$ refers to the spin index, $|s\rangle$ is the corresponding eigenstate:
\begin{equation}
|s\rangle=\left( \begin{array}{c}
se^{-i\phi}\sqrt{\frac{1+s\cos{\theta}}{2}}\\
\sqrt{\frac{1-s\cos{\theta}}{2}}\end{array} \right),
\end{equation}
\noindent $E_{\boldsymbol{k}s}=E_{\boldsymbol{k}}+sJ$, $E_{\boldsymbol{k}}=\hbar^{2}\boldsymbol{k}^{2}/2m$, and $\eta$ is a positive infinitesimal. 
\par Next, we consider the impurity Hamiltonian with spin-orbit coupling:
\begin{equation}
\label{eq:imppot}
\hat{\mathcal{H}}_{\mathrm{imp}}=\sum\limits_{\boldsymbol{R}_{i}}V(\boldsymbol{r}-\boldsymbol{R}_{i})\hat{\sigma}_{0}+\frac{\xi_{SO}}{\hbar k_{F}^{2}}\sum_{\boldsymbol{R}_{i}}\left(\nabla V(\boldsymbol{r}-\boldsymbol{R}_{i})\times\hat{\boldsymbol{p}}\right)\cdot\hat{\boldsymbol{\sigma}},
\end{equation}
\noindent where $\hat{\boldsymbol{p}}=-i\hbar\partial_{\boldsymbol{r}}$ is the momentum operator, $V(\boldsymbol{r}-\boldsymbol{R}_{i})=v_{i}\delta(\boldsymbol{r}-\boldsymbol{R}_{i})$ is the on-site potential at the impurity site $\boldsymbol{R}_{i}$, $\xi_{SO}$ is the spin-orbit coupling parameter (defined as a dimensionless quantity), and $k_{F}$ is the Fermi wavevector. Here, we neglect the localization effects and electron-electron correlations, and assume a short-range impurity potential. In the reciprocal space, it can be written as:\cite{rammer1}
\begin{equation}
\hat{\mathcal{H}}_{\boldsymbol{k}\boldsymbol{k'}}=\Omega\langle\boldsymbol{k}|\hat{\mathcal{H}}_{\mathrm{imp}}|\boldsymbol{k'}\rangle,
\end{equation}
\noindent where the momentum eigenstates are defined as $\langle\boldsymbol{r}|\boldsymbol{k}\rangle=\Omega^{-1/2}e^{i\boldsymbol{k}\cdot\boldsymbol{r}}$, and $\Omega$ is the volume of the system. Then, by using the following identities:
\begin{equation}
\int\limits_{\Omega}d\boldsymbol{r}\,f(\boldsymbol{r})\delta(\boldsymbol{r}-\boldsymbol{r}_{i})=f(\boldsymbol{r}_{i}),\qquad\int\limits_{\Omega}d\boldsymbol{r}\,f(\boldsymbol{r})\nabla\delta(\boldsymbol{r}-\boldsymbol{r}_{i})=-\int\limits_{\Omega}d\boldsymbol{r}\,\nabla f(\boldsymbol{r})\delta(\boldsymbol{r}-\boldsymbol{r}_{i}),
\end{equation}
\noindent we obtain:
\begin{equation}
\begin{aligned}
\hat{\mathcal{H}}_{\boldsymbol{k}\boldsymbol{k'}}&=\sum\limits_{\boldsymbol{R}_{i}}\int\limits_{\Omega}d\boldsymbol{r}\,\left[v_{i}\delta(\boldsymbol{r}-\boldsymbol{R}_{i})\hat{\sigma}_{0}e^{-i(\boldsymbol{k}-\boldsymbol{k'})\cdot\boldsymbol{r}}-iv_{i}\frac{\xi_{SO}}{k_{F}^{2}}e^{-i\boldsymbol{k}\cdot\boldsymbol{r}}\Big(\nabla\delta(\boldsymbol{r}-\boldsymbol{R}_{i})\times\partial_{\boldsymbol{r}}\Big)\cdot\hat{\boldsymbol{\sigma}}e^{i\boldsymbol{k'}\cdot\boldsymbol{r}} \right]\\
&=V(\boldsymbol{k}-\boldsymbol{k'})\left[\hat{\sigma}_{0}+i\frac{\xi_{SO}}{k_{F}^{2}}\hat{\boldsymbol{\sigma}}\cdot(\boldsymbol{k}\times\boldsymbol{k'})\right],
\end{aligned}
\label{eq:potsoi}
\end{equation}
\noindent where $V(\boldsymbol{k}-\boldsymbol{k'})$ is the Fourier transform of the impurity on-site potential:
\begin{equation}
V(\boldsymbol{k}-\boldsymbol{k'})=v_{i}\sum_{\boldsymbol{R}_{i}}e^{-i(\boldsymbol{k}-\boldsymbol{k'})\cdot\boldsymbol{R}_{i}}.
\end{equation}
\par We proceed to write a kinetic equation by means of the Keldysh formalism:
\begin{equation}
\underline{\hat{G}}^{-1}=\hat{G}_{0}^{-1}-\underline{\Sigma},\qquad 
\underline{\hat{G}}=\left( \begin{array}{cc}
\hat{G}^{R} &\hat{G}^{K}\\
0&\hat{G}^{A}\end{array} \right),\qquad
\underline{\hat{\Sigma}}=\left( \begin{array}{cc}
\hat{\Sigma}^{R} &\hat{\Sigma}^{K}\\
0&\hat{\Sigma}^{A}\end{array} \right),
\end{equation}
\noindent where $\underline{\hat{G}}$ and $\underline{\hat{\Sigma}}$ are the Green's function and self-energy in the Keldysh space; the indexes $R$, $A$ and $K$ stand for the retarded, advanced and Keldysh components, respectively, and $\hat{G}_{0}^{-1}=i\hbar\partial_{t}-\hat{\mathcal{H}}_{0}$. In the semiclassical approximation, a set of diffusive equations for the non-equilibrium charge and spin densities can be derived through the distribution function $\hat{g}_{\boldsymbol{k}}\equiv \hat{g}_{\boldsymbol{k}}(\boldsymbol{R},T)$ defined as the Wigner representation of the Keldysh Green's function $\hat{G}^{K}$:\cite{rammer2}
\begin{equation}
\begin{aligned}
\hat{G}^{K}(\boldsymbol{r}_{1},t_{1};\boldsymbol{r}_{2},t_{2})&\,\,\xrightarrow{\mathcal{W}}\hat{G}^{K}(\boldsymbol{R}+\frac{\boldsymbol{r}}{2},T+\frac{t}{2};\boldsymbol{R}-\frac{\boldsymbol{r}}{2},T-\frac{t}{2})\equiv\hat{G}^{K}(\boldsymbol{r},t;\boldsymbol{R},T)\\
&\,\,\xrightarrow{\mathcal{F}}\hat{G}^{K}(\boldsymbol{r},t;\boldsymbol{R},T)=\int\frac{dE}{2\pi}\frac{d\boldsymbol{k}}{(2\pi)^{3}}\,\,\hat{g}^{K}_{\boldsymbol{k}E}(\boldsymbol{R},T)e^{-\frac{i}{\hbar}Et}e^{i\boldsymbol{r}\cdot\boldsymbol{k}}\\
&\,\,\xrightarrow{\,\,}\hat{g}_{\boldsymbol{k}}=i\int\frac{dE}{2\pi}\,\hat{g}^{K}_{\boldsymbol{k}E}(\boldsymbol{R},T),
\end{aligned}
\end{equation}
\noindent where the relative $\boldsymbol{r}=\boldsymbol{r}_{1}-\boldsymbol{r}_{2}$, $t=t_{1}-t_{2}$ and center-of-mass $\boldsymbol{R}=(\boldsymbol{r}_{1}+\boldsymbol{r}_{2})/2$, $T=(t_{1}+t_{2})/2$ coordinates are introduced. In the dilute limit, we can employ the Kadanoff-Baym anzats:
\begin{equation}
\underline{\hat{G}}_{\boldsymbol{k}E}(\boldsymbol{R},T)=\left( \begin{array}{cc}
\hat{G}^{R}_{\boldsymbol{k}E} &\hat{g}^{K}_{\boldsymbol{k}E}(\boldsymbol{R},T)\\
0&\hat{G}^{A}_{\boldsymbol{k}E}\end{array} \right)
\end{equation}
\noindent and
\begin{equation}
\hat{g}^{K}_{\boldsymbol{k}E}(\boldsymbol{R},T)=\hat{G}^{R}_{\boldsymbol{k}E}\,\hat{g}_{\boldsymbol{k}}(\boldsymbol{R},T)-\hat{g}_{\boldsymbol{k}}(\boldsymbol{R},T)\hat{G}^{A}_{\boldsymbol{k}E}.
\label{eq:anzats}
\end{equation}
\noindent The Keldysh Green's function $\hat{G}^{K}$ satisfies the Kadanoff-Baym equation:\cite{rammer2}
\begin{equation}
[\hat{G}^{R}]^{-1}\ast\hat{G}^{K}-\hat{G}^{K}\ast[\hat{G}^{A}]^{-1}=\hat{\Sigma}^{K}\ast\hat{G}^{A}-\hat{G}^{R}\ast\hat{\Sigma}^{K},
\end{equation}
\noindent Having applied the Wigner transformation, we use the so-called gradient approximation, where the convolution $\mathcal{A}\ast\mathcal{B}$ of two functions is expressed as:
\begin{equation}
\begin{aligned}
\left(\mathcal{A}\ast\mathcal{B}\right)_{\boldsymbol{k}E}(\boldsymbol{R},T)&\simeq\mathcal{A}\mathcal{B}-\frac{i\hbar}{2}\left(\partial_{T}\mathcal{A}\partial_{E}\mathcal{B}-\partial_{E}\mathcal{A}\partial_{T}\mathcal{B}\right)\\
&-\frac{i}{2}\left(\nabla_{\boldsymbol{k}}\mathcal{A}\cdot\nabla_{\boldsymbol{R}}\mathcal{B}-\nabla_{\boldsymbol{R}}\mathcal{A}\cdot\nabla_{\boldsymbol{k}}\mathcal{B}\right).
\end{aligned}
\end{equation}
\noindent Taking into account that $\hat{G}^{R(A)}$ and $\hat{\Sigma}^{R(A)}$ do not depend on the center-of-mass coordinates, we obtain:
\begin{equation}
\begin{aligned}
i\hbar\partial_{T}\hat{g}^{K}+[\hat{g}^{K},J\hat{\boldsymbol{\sigma}}\cdot\boldsymbol{m}]&+\frac{i}{2}\left\{\nabla_{\boldsymbol{k}}\hat{\mathcal{H}}_{\boldsymbol{k}},\nabla_{\boldsymbol{R}}\hat{g}^{K}\right\}=\hat{\Sigma}^{K}\hat{G}^{A}-\hat{G}^{R}\hat{\Sigma}^{K}+\hat{\Sigma}^{R}\hat{g}^{K}-\hat{g}^{K}\hat{\Sigma}^{A}\\
&-\frac{i\hbar}{2}\left(\partial_{T}\hat{\Sigma}^{K}\partial_{E}\hat{G}^{A}+\partial_{E}\hat{G}^{R}\partial_{T}\hat{\Sigma}^{K} \right)
+\frac{i\hbar}{2}\left(\partial_{E}\hat{\Sigma}^{R}\partial_{T}\hat{g}^{K}+\partial_{T}\hat{g}^{K}\partial_{E}\hat{\Sigma}^{A} \right)\\
&-\frac{i}{2}\left(\nabla_{\boldsymbol{k}}\hat{\Sigma}^{R}\cdot\nabla_{\boldsymbol{R}}\hat{g}^{K}+\nabla_{\boldsymbol{R}}\hat{g}^{K}\cdot\nabla_{\boldsymbol{k}}\hat{\Sigma}^{A}\right)
+\frac{i}{2}\left(\nabla_{\boldsymbol{R}}\hat{\Sigma}^{K}\cdot\nabla_{\boldsymbol{k}}\hat{G}^{A}+\nabla_{\boldsymbol{k}}\hat{G}^{R}\cdot\nabla_{\boldsymbol{R}}\hat{\Sigma}^{K}\right),
\label{eq:full}
\end{aligned}
\end{equation}
\noindent where $[\cdot\,,\cdot]$ and $\{\cdot\,,\cdot\}$ stand for a commutator and anticommutator, respectively. In steady state, we have:
\begin{equation}
\begin{aligned}
\,[\hat{g}^{K},J\hat{\boldsymbol{\sigma}}\cdot\boldsymbol{m}]&+\frac{i}{2}\left\{\nabla_{\boldsymbol{k}}\hat{\mathcal{H}}_{\boldsymbol{k}},\nabla_{\boldsymbol{R}}\hat{g}^{K}\right\}=\hat{\Sigma}^{K}\hat{G}^{A}-\hat{G}^{R}\hat{\Sigma}^{K}+\hat{\Sigma}^{R}\hat{g}^{K}-\hat{g}^{K}\hat{\Sigma}^{A}\\
&-\frac{i}{2}\left(\nabla_{\boldsymbol{k}}\hat{\Sigma}^{R}\cdot\nabla_{\boldsymbol{R}}\hat{g}^{K}+\nabla_{\boldsymbol{R}}\hat{g}^{K}\cdot\nabla_{\boldsymbol{k}}\hat{\Sigma}^{A}\right)
+\frac{i}{2}\left(\nabla_{\boldsymbol{R}}\hat{\Sigma}^{K}\cdot\nabla_{\boldsymbol{k}}\hat{G}^{A}+\nabla_{\boldsymbol{k}}\hat{G}^{R}\cdot\nabla_{\boldsymbol{R}}\hat{\Sigma}^{K}\right).
\label{eq:kelfull}
\end{aligned}
\end{equation}
\noindent Finally, in the dilute limit, we can assume that the self-energy is almost constant and neglect its derivatives on the right-hand side:
\begin{equation}
[\hat{g}^{K},J\hat{\boldsymbol{\sigma}}\cdot\boldsymbol{m}]+\frac{i}{2}\left\{\nabla_{\boldsymbol{k}}\hat{\mathcal{H}}_{\boldsymbol{k}},\nabla_{\boldsymbol{R}}\hat{g}^{K}\right\}=\hat{\Sigma}^{K}\hat{G}^{A}-\hat{G}^{R}\hat{\Sigma}^{K}+\hat{\Sigma}^{R}\hat{g}^{K}-\hat{g}^{K}\hat{\Sigma}^{A},
\end{equation}
\noindent or
\begin{equation}
[\hat{g}^{K},J\hat{\boldsymbol{\sigma}}\cdot\boldsymbol{m}]+i\frac{\hbar^{2}}{m}(\boldsymbol{k}\cdot\nabla_{\boldsymbol{R}})\,\hat{g}^{K}=\hat{\Sigma}^{K}\hat{G}^{A}-\hat{G}^{R}\hat{\Sigma}^{K}+\hat{\Sigma}^{R}\hat{g}^{K}-\hat{g}^{K}\hat{\Sigma}^{A}.
\label{eq:keldysh}
\end{equation}

\begin{figure}
\begin{center}
\includegraphics[scale=0.55]{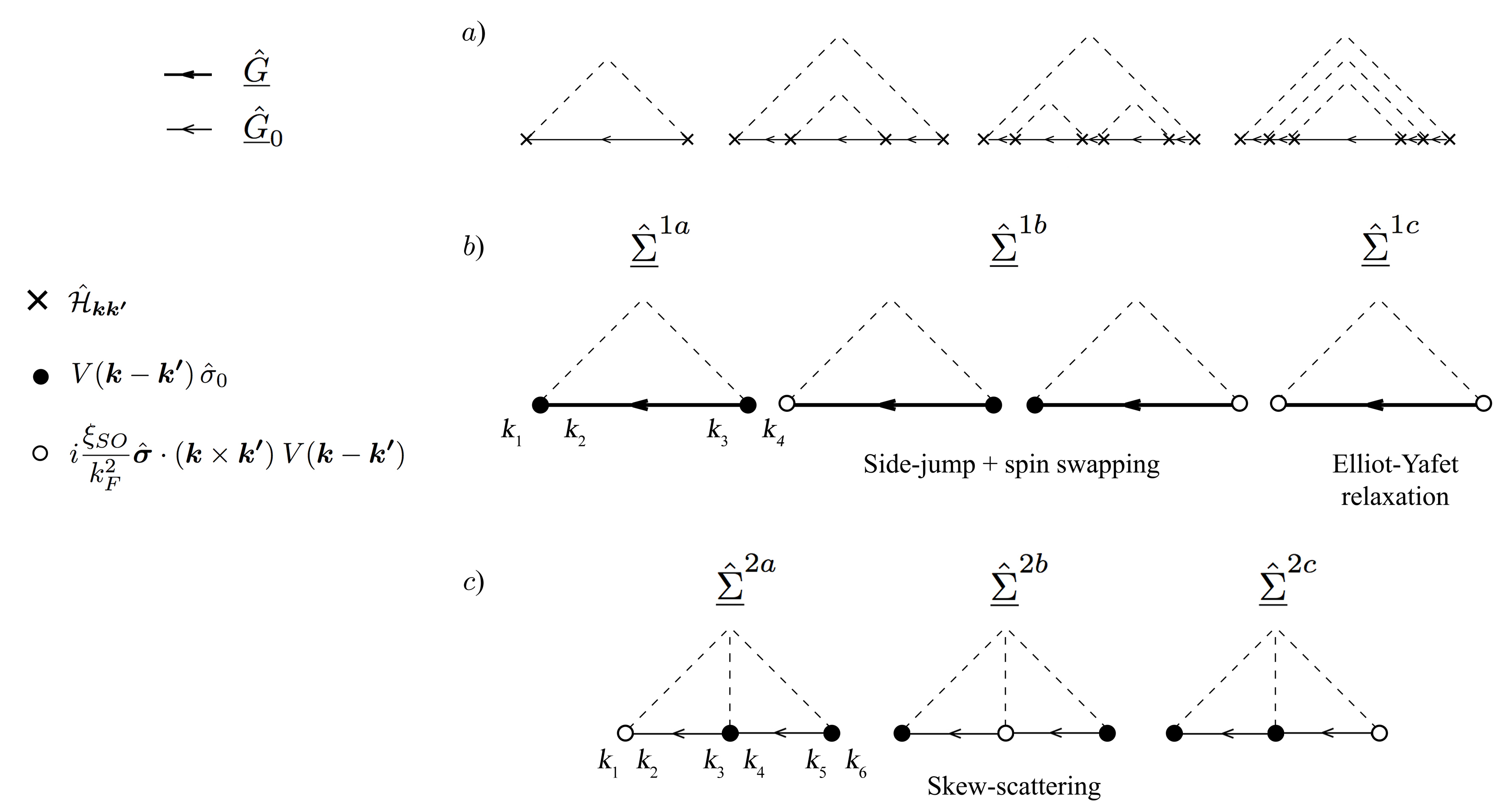}
\end{center}
\caption{$a)$ Diagrammatic expansion for scatterings off the static impurity potential. $b)$ Self-consistent Born approximation. $c)$ Skew-scattering diagrams.}
\end{figure}

\section{Self-energy}
\subsection{Self-consistent Born approximation}
\par Let us consider first and second orders of the diagrammatic expansion for the scattering off the static impurity potential (Fig. 1$a$). The Green's function $\underline{\hat{G}}$ and the corresponding self-energy $\underline{\hat{\Sigma}}$ are defined in the wave vector representation as follows:
\begin{equation}
\begin{aligned}
\underline{\hat{G}}(\boldsymbol{k},t_{1};\boldsymbol{k'},t_{2})&=\underline{\hat{G}}_{0}(\boldsymbol{k},t_{1};\boldsymbol{k'},t_{2})+\sum\limits_{\{\boldsymbol{k}_{i}\}} \underline{\hat{G}}_{0}(\boldsymbol{k},t_{1};\boldsymbol{k}_{1},t_{2}) \langle\boldsymbol{k}_{1}|\hat{\mathcal{H}}_{\mathrm{imp}}|\boldsymbol{k}_{2}\rangle\underline{\hat{G}}_{0}(\boldsymbol{k}_{2},t_{1};\boldsymbol{k'},t_{2})\\
&+\sum\limits_{\{\boldsymbol{k}_{i}\}} \underline{\hat{G}}_{0}(\boldsymbol{k},t_{1};\boldsymbol{k}_{1},t_{2}) \langle\boldsymbol{k}_{1}|\hat{\mathcal{H}}_{\mathrm{imp}}|\boldsymbol{k}_{2}\rangle\underline{\hat{G}}_{0}(\boldsymbol{k}_{2},t_{1};\boldsymbol{k}_{3},t_{2})\langle\boldsymbol{k}_{3}|\hat{\mathcal{H}}_{\mathrm{imp}}|\boldsymbol{k}_{4}\rangle\underline{\hat{G}}_{0}(\boldsymbol{k}_{4},t_{1};\boldsymbol{k'},t_{2})+...\\
&=\underline{\hat{G}}_{0}(\boldsymbol{k},t_{1};\boldsymbol{k'},t_{2})+\sum\limits_{\{\boldsymbol{k}_{i}\}} \underline{\hat{G}}_{0}(\boldsymbol{k},t_{1};\boldsymbol{k}_{1},t_{2})\underline{\hat{\Sigma}}(\boldsymbol{k}_{1},t_{1};\boldsymbol{k}_{2},t_{2})\underline{\hat{G}}(\boldsymbol{k}_{2},t_{1};\boldsymbol{k'},t_{2}),
\end{aligned}
\label{eq:selfscat}
\end{equation}
\noindent where $\underline{\hat{G}}_{0}$ is a free propagator. To consider a particle moving in a random potential we take the average over different spatial configurations of the ensemble of $N$ impurities. Upon impurity-averaging the first order term in Eq.~(\ref{eq:selfscat}) is a constant and can be renormalized away. For the second order term, we take into account all two-line irreducible diagrams corresponding to the double scattering off the same impurity and neglect the so-called crossing diagrams, where impurity lines cross and give a small contribution to the region of interest, $E\simeq E_{F}$ and $k\simeq k_{F}$.\cite{rammer1} This is nothing else but the self-consistent Born approximation (Fig. 1$b$). Using Eq.~(\ref{eq:potsoi}) for the impurity potential, the self-energy is given by three terms:
\begin{equation}
\begin{aligned}
\underline{\hat{\Sigma}}^{1a}(\boldsymbol{k}_{1},t_{1};\boldsymbol{k}_{4},t_{2})=\frac{1}{\Omega^{2}}\sum\limits_{\boldsymbol{k}_{2}\boldsymbol{k}_{3}}\underline{\hat{G}}(\boldsymbol{k}_{2},t_{1};\boldsymbol{k}_{3},t_{2})\left\langle V(\boldsymbol{k}_{1}-\boldsymbol{k}_{2})V(\boldsymbol{k}_{3}-\boldsymbol{k}_{4})\right\rangle,
\end{aligned}
\end{equation}
\begin{equation}
\begin{aligned}
\underline{\hat{\Sigma}}^{1b}(\boldsymbol{k}_{1},t_{1};\boldsymbol{k}_{4},t_{2})&=\frac{i}{\Omega^{2}}\frac{\xi_{SO}}{k_{F}^{2}}\sum\limits_{\boldsymbol{k}_{2}\boldsymbol{k}_{3}}\left[[(\boldsymbol{k}_{1}\times\boldsymbol{k}_{2})\cdot\hat{\boldsymbol{\sigma}}]\underline{\hat{G}}(\boldsymbol{k}_{2},t_{1};\boldsymbol{k}_{3},t_{2})\right.\\
&\left.+\,\underline{\hat{G}}(\boldsymbol{k}_{2},t_{1};\boldsymbol{k}_{3},t_{2})[(\boldsymbol{k}_{3}\times\boldsymbol{k}_{4})\cdot\hat{\boldsymbol{\sigma}}]\right]\left\langle V(\boldsymbol{k}_{1}-\boldsymbol{k}_{2})V(\boldsymbol{k}_{3}-\boldsymbol{k}_{4})\right\rangle,
\end{aligned}
\end{equation}
\begin{equation}
\begin{aligned}
\underline{\hat{\Sigma}}^{1c}(\boldsymbol{k}_{1},t_{1};\boldsymbol{k}_{4},t_{2})&=-\frac{1}{\Omega^{2}}\frac{\xi_{SO}^{2}}{k^{4}_{F}}\sum\limits_{\boldsymbol{k}_{2}\boldsymbol{k}_{3}}[(\boldsymbol{k}_{1}\times\boldsymbol{k}_{2})\cdot\hat{\boldsymbol{\sigma}}]\underline{\hat{G}}(\boldsymbol{k}_{2},t_{1};\boldsymbol{k}_{3},t_{2})[(\boldsymbol{k}_{3}\times\boldsymbol{k}_{4})\cdot\hat{\boldsymbol{\sigma}}]\left\langle V(\boldsymbol{k}_{1}-\boldsymbol{k}_{2})V(\boldsymbol{k}_{3}-\boldsymbol{k}_{4})\right\rangle,\\
\end{aligned}
\end{equation}
\noindent where impurity averaging leads to:
\begin{equation}
\begin{aligned}
 \left\langle V(\boldsymbol{k}_{1}-\boldsymbol{k}_{2})V(\boldsymbol{k}_{3}-\boldsymbol{k}_{4})\right\rangle&= v_{i}^{2} \left\langle\sum_{\boldsymbol{R}_{i}}e^{-i(\boldsymbol{k}_{1}-\boldsymbol{k}_{2})\cdot\boldsymbol{R}_{i}}e^{-i(\boldsymbol{k}_{3}-\boldsymbol{k}_{4})\cdot\boldsymbol{R}_{i}}\right\rangle=v_{i}^{2}N\delta_{\boldsymbol{k}_{1}+\boldsymbol{k}_{3},\boldsymbol{k}_{2}+\boldsymbol{k}_{4}}.
 \end{aligned}
\end{equation}
\noindent To proceed with the Wigner transformation, we change the variables:
$$
\boldsymbol{k}_{1}=\boldsymbol{k}+\frac{\boldsymbol{q}}{2}\qquad\boldsymbol{k}_{2}=\boldsymbol{k'}+\frac{\boldsymbol{q'}}{2}\qquad\boldsymbol{k}_{3}=\boldsymbol{k'}-\frac{\boldsymbol{q'}}{2} \qquad\boldsymbol{k}_{4}=\boldsymbol{k}-\frac{\boldsymbol{q}}{2}
$$
\noindent and
$$
T=\frac{t_{1}+t_{2}}{2}\qquad t=t_{1}-t_{2},
$$
\noindent that gives:
\begin{equation}
\begin{aligned}
\underline{\hat{\Sigma}}^{1a}(\boldsymbol{k},t;\boldsymbol{q},T)=\frac{1}{\Omega^{2}}\sum\limits_{\boldsymbol{k}_{2}\boldsymbol{k}_{3}}\underline{\hat{G}}(\boldsymbol{k'},t;\boldsymbol{q'},T)\delta_{\boldsymbol{q},\boldsymbol{q'}},
\end{aligned}
\end{equation}
\begin{equation}
\begin{aligned}
\underline{\hat{\Sigma}}^{1b}(\boldsymbol{k},t;\boldsymbol{q},T)&=\frac{i}{\Omega^{2}}\frac{\xi_{SO}}{k_{F}^{2}}\sum\limits_{\boldsymbol{k}_{2}\boldsymbol{k}_{3}}\left[\big[((\boldsymbol{k}+\frac{\boldsymbol{q}}{2})\times(\boldsymbol{k'}+\frac{\boldsymbol{q'}}{2}))\cdot\hat{\boldsymbol{\sigma}}\big]\underline{\hat{G}}(\boldsymbol{k'},t;\boldsymbol{q'},T)\right.\\
&\left.+\,\underline{\hat{G}}(\boldsymbol{k'},t;\boldsymbol{q'},T)\big[((\boldsymbol{k'}-\frac{\boldsymbol{q'}}{2})\times(\boldsymbol{k}-\frac{\boldsymbol{q'}}{2})))\cdot\hat{\boldsymbol{\sigma}}\big]\right]\delta_{\boldsymbol{q},\boldsymbol{q'}},
\end{aligned}
\end{equation}
\begin{equation}
\begin{aligned}
\underline{\hat{\Sigma}}^{1c}(\boldsymbol{k},t;\boldsymbol{q},T)&=-\frac{1}{\Omega^{2}}\frac{\xi_{SO}^{2}}{k^{4}_{F}}\sum\limits_{\boldsymbol{k}_{2}\boldsymbol{k}_{3}}\big[((\boldsymbol{k}+\frac{\boldsymbol{q}}{2})\times(\boldsymbol{k'}+\frac{\boldsymbol{q'}}{2}))\cdot\hat{\boldsymbol{\sigma}}\big]\underline{\hat{G}}(\boldsymbol{k'},t;\boldsymbol{q'},T)\big[((\boldsymbol{k'}-\frac{\boldsymbol{q'}}{2})\times(\boldsymbol{k}-\frac{\boldsymbol{q}}{2}))\cdot\hat{\boldsymbol{\sigma}}\big]\delta_{\boldsymbol{q},\boldsymbol{q'}}.
\end{aligned}
\end{equation}
\noindent The Kronecker function reflects that translation invariance is recovered, and we have in the continuum limit:
\begin{equation}
\begin{aligned}
\underline{\hat{\Sigma}}^{1a}(\boldsymbol{k},t;\boldsymbol{q},T)=v_{i}^{2}n_{i}\int\frac{d\boldsymbol{k'}}{(2\pi)^{3}}\,\underline{\hat{G}}(\boldsymbol{k'},t;\boldsymbol{q},T),
\end{aligned}
\end{equation}
\begin{equation}
\begin{aligned}
\underline{\hat{\Sigma}}^{1b}(\boldsymbol{k},t;\boldsymbol{q},T)&=\underline{\hat{\Sigma}}^{sw}(\boldsymbol{k},t;\boldsymbol{q},T)+\underline{\hat{\Sigma}}^{sj}(\boldsymbol{k},t;\boldsymbol{q},T)\\
&= iv_{i}^{2}n_{i}\frac{\xi_{SO}}{k_{F}^{2}}\int\frac{d\boldsymbol{k'}}{(2\pi)^{3}}\left[(\boldsymbol{k}\times\boldsymbol{k'})\cdot\hat{\boldsymbol{\sigma}},\underline{\hat{G}}(\boldsymbol{k'},t;\boldsymbol{q},T)\right]\\
&+iv_{i}^{2}n_{i}\frac{\xi_{SO}}{2k_{F}^{2}}\int\frac{d\boldsymbol{k'}}{(2\pi)^{3}}\left\{[\boldsymbol{q}\times(\boldsymbol{k}'-\boldsymbol{k})]\cdot\hat{\boldsymbol{\sigma}},\underline{\hat{G}}(\boldsymbol{k'},t;\boldsymbol{q},T)\right\},
\end{aligned}
\end{equation}
\begin{equation}
\begin{aligned}
\underline{\hat{\Sigma}}^{1c}(\boldsymbol{k},t;\boldsymbol{q},T)&=-v_{i}^{2}n_{i}\frac{\xi^{2}_{SO}}{k_{F}^{4}}\int\frac{d\boldsymbol{k'}}{(2\pi)^{3}}[(\boldsymbol{k}\times\boldsymbol{k'}+\frac{1}{2}\boldsymbol{q}\times\boldsymbol{k'}+\frac{1}{2}\boldsymbol{k}\times\boldsymbol{q})\cdot\hat{\boldsymbol{\sigma}}]\\
&\cdot\underline{\hat{G}}(\boldsymbol{k'},t;\boldsymbol{q},T)[(\boldsymbol{k'}\times\boldsymbol{k}-\frac{1}{2}\boldsymbol{k'}\times\boldsymbol{q}-\frac{1}{2}\boldsymbol{q}\times\boldsymbol{k})\cdot\hat{\boldsymbol{\sigma}}].
\end{aligned}
\end{equation}
\noindent Having Fourier transformed with respect to $\boldsymbol{q}$, that gives the Wigner coordinate $\boldsymbol{R}$:
\begin{equation}
\begin{aligned}
\hat{G}(\boldsymbol{r}_{1};\boldsymbol{r}_{4})&=\int\frac{d\boldsymbol{k}_{1}}{(2\pi)^{3}}\int\frac{d\boldsymbol{k}_{4}}{(2\pi)^{3}}\hat{G}(\boldsymbol{k}_{1};\boldsymbol{k}_{4})e^{i(\boldsymbol{k}_{1}\boldsymbol{r}_{1}-\boldsymbol{k}_{4}\boldsymbol{r}_{4})}\\
&=\int\frac{d\boldsymbol{k}}{(2\pi)^{3}}\int\frac{d\boldsymbol{q}}{(2\pi)^{3}}\hat{G}(\boldsymbol{k}+\frac{1}{2}\boldsymbol{q};\boldsymbol{k}-\frac{1}{2}\boldsymbol{q})e^{i(\boldsymbol{k}+\frac{1}{2}\boldsymbol{q})\cdot(\boldsymbol{R}+\frac{1}{2}\boldsymbol{r})}e^{-i(\boldsymbol{k}-\frac{1}{2}\boldsymbol{q})\cdot(\boldsymbol{R}-\frac{1}{2}\boldsymbol{r})}\\
&=\int\frac{d\boldsymbol{k}}{(2\pi)^{3}}\int\frac{d\boldsymbol{q}}{(2\pi)^{3}}\hat{G}(\boldsymbol{k};\boldsymbol{q})e^{i\boldsymbol{q}\cdot\boldsymbol{R}}e^{i\boldsymbol{k}\cdot\boldsymbol{r}}=\hat{G}(\boldsymbol{r};\boldsymbol{R}),
\end{aligned}
\end{equation}
\noindent we get the final form for the self-energy in the mixed representation:
\begin{equation}
\begin{aligned}
\underline{\hat{\Sigma}}^{1a}_{\boldsymbol{k}E}(\boldsymbol{R},T)&=v_{i}^{2}n_{i}\int\frac{d\boldsymbol{k'}}{(2\pi)^{3}}\,\underline{\hat{G}}_{\boldsymbol{k'}E}(\boldsymbol{R},T),\\
\underline{\hat{\Sigma}}^{1b}_{\boldsymbol{k}E}(\boldsymbol{R},T)&=\underline{\hat{\Sigma}}^{sw}_{\boldsymbol{k}E}(\boldsymbol{R},T)+\underline{\hat{\Sigma}}^{sj}_{\boldsymbol{k}E}(\boldsymbol{R},T)\\
&= iv_{i}^{2}n_{i}\frac{\xi_{SO}}{k_{F}^{2}}\int\frac{d\boldsymbol{k'}}{(2\pi)^{3}}\left[(\boldsymbol{k}\times\boldsymbol{k'})\cdot\hat{\boldsymbol{\sigma}},\underline{\hat{G}}_{\boldsymbol{k'}E}(\boldsymbol{R},T)\right]\\
&+v_{i}^{2}n_{i}\frac{\xi_{SO}}{2k_{F}^{2}}\int\frac{d\boldsymbol{k'}}{(2\pi)^{3}}\nabla_{\boldsymbol{R}}\left\{(\boldsymbol{k}'-\boldsymbol{k})\times\hat{\boldsymbol{\sigma}},\underline{\hat{G}}_{\boldsymbol{k'}E}(\boldsymbol{R};T)\right\},\\
\underline{\hat{\Sigma}}^{1c}_{\boldsymbol{k}E}(\boldsymbol{R},T)&=-v_{i}^{2}n_{i}\frac{\xi^{2}_{SO}}{k_{F}^{4}}\int\frac{d\boldsymbol{k'}}{(2\pi)^{3}}[(\boldsymbol{k}\times\boldsymbol{k'})\cdot\hat{\boldsymbol{\sigma}}]\underline{\hat{G}}_{\boldsymbol{k'}E}(\boldsymbol{R},T)[(\boldsymbol{k'}\times\boldsymbol{k})\cdot\hat{\boldsymbol{\sigma}}].
\end{aligned}
\end{equation}
\noindent At the level of the self-consistent Born approximation, the self-energy is given by the following contributions. The first term $\underline{\hat{\Sigma}}^{1a}$ stands for the standard elastic scattering off the on-site impurity potential. To first order of $\xi_{SO}$, there are two terms, $\underline{\hat{\Sigma}}^{sw}$ and $\underline{\hat{\Sigma}}^{sj}$, related to the side-jump and spin swapping contributions, respectively. Finally, the second order of $\xi_{SO}$ yields the Elliot-Yafet spin relaxation mechanism (all gradient terms $\sim\xi_{SO}^{2}\nabla_{\boldsymbol{R}}$ are neglected on account of their smallness). 
\par Let us rewrite these contributions for the retarded, advanced and Keldysh Green's functions (taking into account that $\hat{G}^{R(A)}$ does not depend on the center-of-mass coordinates):
\begin{equation}
\underline{\hat{\Sigma}}_{\boldsymbol{k}E}(\boldsymbol{R},T)=\underline{\hat{\Sigma}}^{1a}_{\boldsymbol{k}E}(\boldsymbol{R},T)+\underline{\hat{\Sigma}}^{1b}_{\boldsymbol{k}E}(\boldsymbol{R},T)+\underline{\hat{\Sigma}}^{1c}_{\boldsymbol{k}E}(\boldsymbol{R},T),
\end{equation}
\noindent which is equivalent to:
\begin{equation}
\hat{\Sigma}^{R}_{\boldsymbol{k}E}=v_{i}^{2}n_{i}\int\frac{d\boldsymbol{k'}}{(2\pi)^{3}}\,\left[\hat{\sigma}_{0}+i\frac{\xi_{SO}}{k_{F}^{2}}(\boldsymbol{k}\times\boldsymbol{k'})\cdot\hat{\boldsymbol{\sigma}}\right]\hat{G}^{R}_{\boldsymbol{k'}E}\left[\hat{\sigma}_{0}-i\frac{\xi_{SO}}{k_{F}^{2}}(\boldsymbol{k}\times\boldsymbol{k'})\cdot\hat{\boldsymbol{\sigma}}\right],
\label{eq:sret}
\end{equation}
\begin{equation}
\hat{\Sigma}^{A}_{\boldsymbol{k}E}=v_{i}^{2}n_{i}\int\frac{d\boldsymbol{k'}}{(2\pi)^{3}}\,\left[\hat{\sigma}_{0}+i\frac{\xi_{SO}}{k_{F}^{2}}(\boldsymbol{k}\times\boldsymbol{k'})\cdot\hat{\boldsymbol{\sigma}}\right]\hat{G}^{A}_{\boldsymbol{k'}E}\left[\hat{\sigma}_{0}-i\frac{\xi_{SO}}{k_{F}^{2}}(\boldsymbol{k}\times\boldsymbol{k'})\cdot\hat{\boldsymbol{\sigma}}\right],
\label{eq:sadv}
\end{equation}
\noindent and
\begin{equation}
\begin{aligned}
\hat{\Sigma}^{K}_{\boldsymbol{k}E}(\boldsymbol{R},T)&=v_{i}^{2}n_{i}\int\frac{d\boldsymbol{k'}}{(2\pi)^{3}}\,\left[\hat{\sigma}_{0}+i\frac{\xi_{SO}}{k_{F}^{2}}(\boldsymbol{k}\times\boldsymbol{k'})\cdot\hat{\boldsymbol{\sigma}}\right]\hat{g}^{K}_{\boldsymbol{k'}E}(\boldsymbol{R},T)\left[\hat{\sigma}_{0}-i\frac{\xi_{SO}}{k_{F}^{2}}(\boldsymbol{k}\times\boldsymbol{k'})\cdot\hat{\boldsymbol{\sigma}}\right]\\
&+v_{i}^{2}n_{i}\frac{\xi_{SO}}{2k_{F}^{2}}\int\frac{d\boldsymbol{k'}}{(2\pi)^{3}}\nabla_{\boldsymbol{R}}\left\{(\boldsymbol{k}'-\boldsymbol{k})\times\hat{\boldsymbol{\sigma}},\hat{g}^{K}_{\boldsymbol{k'}E}(\boldsymbol{R};T)\right\}.
\label{eq:skel}
\end{aligned}
\end{equation}

\subsection{Skew-scattering}
\par To take into account skew-scattering, one has to go beyond the Born approximation. Starting from third order diagrams (Fig.~1$c$), we obtain the following expressions for the self-energy to first order in $\xi_{SO}$:
\begin{equation}
\begin{aligned}
\underline{\hat{\Sigma}}^{2a}(\boldsymbol{k}_{1},t_{1};\boldsymbol{k}_{6},t_{2})=\frac{i}{\Omega^{3}}\frac{\xi_{SO}}{k_{F}^{2}}\sum\limits_{\boldsymbol{k}_{2},\boldsymbol{k}_{3},\boldsymbol{k}_{4},\boldsymbol{k}_{5}}&[(\boldsymbol{k}_{1}\times\boldsymbol{k}_{2})\cdot\hat{\boldsymbol{\sigma}}]\underline{\hat{G}}_{0}(\boldsymbol{k}_{2},t_{1};\boldsymbol{k}_{3},t_{2})\underline{\hat{G}}_{0}(\boldsymbol{k}_{4},t_{1};\boldsymbol{k}_{5},t_{2})\\
&\cdot\left\langle V(\boldsymbol{k}_{1}-\boldsymbol{k}_{2})V(\boldsymbol{k}_{3}-\boldsymbol{k}_{4})V(\boldsymbol{k}_{5}-\boldsymbol{k}_{6})\right\rangle,
\end{aligned}
\end{equation}
\begin{equation}
\begin{aligned}
\underline{\hat{\Sigma}}^{2b}(\boldsymbol{k}_{1},t_{1};\boldsymbol{k}_{6},t_{2})=\frac{i}{\Omega^{3}}\frac{\xi_{SO}}{k_{F}^{2}}\sum\limits_{\boldsymbol{k}_{2},\boldsymbol{k}_{3},\boldsymbol{k}_{4},\boldsymbol{k}_{5}}&\underline{\hat{G}}_{0}(\boldsymbol{k}_{2},t_{1};\boldsymbol{k}_{3},t_{2})[(\boldsymbol{k}_{3}\times\boldsymbol{k}_{4})\cdot\hat{\boldsymbol{\sigma}}]\underline{\hat{G}}_{0}(\boldsymbol{k}_{4},t_{1};\boldsymbol{k}_{5},t_{2})\\
&\cdot\left\langle V(\boldsymbol{k}_{1}-\boldsymbol{k}_{2})V(\boldsymbol{k}_{3}-\boldsymbol{k}_{4})V(\boldsymbol{k}_{5}-\boldsymbol{k}_{6})\right\rangle,
\end{aligned}
\end{equation}
\begin{equation}
\begin{aligned}
\underline{\hat{\Sigma}}^{2c}(\boldsymbol{k}_{1},t_{1};\boldsymbol{k}_{6},t_{2})=\frac{i}{\Omega^{3}}\frac{\xi_{SO}}{k_{F}^{2}}\sum\limits_{\boldsymbol{k}_{2},\boldsymbol{k}_{3},\boldsymbol{k}_{4},\boldsymbol{k}_{5}}&\underline{\hat{G}}_{0}(\boldsymbol{k}_{2},t_{1};\boldsymbol{k}_{3},t_{2})\underline{\hat{G}}_{0}(\boldsymbol{k}_{4},t_{1};\boldsymbol{k}_{5},t_{2})[(\boldsymbol{k}_{5}\times\boldsymbol{k}_{6})\cdot\hat{\boldsymbol{\sigma}}]\\
&\cdot\left\langle V(\boldsymbol{k}_{1}-\boldsymbol{k}_{2})V(\boldsymbol{k}_{3}-\boldsymbol{k}_{4})V(\boldsymbol{k}_{5}-\boldsymbol{k}_{6})\right\rangle,
\end{aligned}
\end{equation}
\noindent where impurity averaging in the triple scattering off the same impurity potential is implied:
\begin{equation}
\begin{aligned}
\left\langle V(\boldsymbol{k}_{1}-\boldsymbol{k}_{2})V(\boldsymbol{k}_{3}-\boldsymbol{k}_{4})V(\boldsymbol{k}_{5}-\boldsymbol{k}_{6})\right\rangle&= v_{i}^{3}\left\langle \sum_{\boldsymbol{R}_{i}}e^{-i(\boldsymbol{k}_{1}-\boldsymbol{k}_{2})\cdot\boldsymbol{R}_{i}}e^{-i(\boldsymbol{k}_{3}-\boldsymbol{k}_{4})\cdot\boldsymbol{R}_{i}}e^{-i(\boldsymbol{k}_{5}-\boldsymbol{k}_{6})\cdot\boldsymbol{R}_{i}}\right\rangle\\
 &=v_{i}^{3}N\delta_{\boldsymbol{k}_{1}+\boldsymbol{k}_{3}+\boldsymbol{k}_{5},\boldsymbol{k}_{2}+\boldsymbol{k}_{4}+\boldsymbol{k}_{6}}.
\end{aligned}
\end{equation}
\noindent Here, we do not consider triple scatterings off the on-site impurity potential without spin-orbit coupling, which gives a negligible correction to the elastic relaxation time ($\sim\frac{\beta v_{i}}{\tau_{0}}$). Changing the variables:
\begin{equation}
\begin{aligned}
\boldsymbol{k}_{1}&=\boldsymbol{k}+\frac{\boldsymbol{q}}{2}\qquad \boldsymbol{k}_{2}=\boldsymbol{k'}+\frac{\boldsymbol{q'}}{2}\qquad \boldsymbol{k}_{3}=\boldsymbol{k'}-\frac{\boldsymbol{q'}}{2},\\
\boldsymbol{k}_{4}&=\boldsymbol{k''}+\frac{\boldsymbol{q''}}{2}\qquad \boldsymbol{k}_{5}=\boldsymbol{k''}-\frac{\boldsymbol{q''}}{2}\qquad \boldsymbol{k}_{6}=\boldsymbol{k}-\frac{\boldsymbol{q}}{2}
\end{aligned}
\end{equation}
\noindent and
\begin{equation}
T=\frac{t_{1}+t_{2}}{2}\qquad t=t_{1}-t_{2}
\end{equation}
\noindent leads in the continuum limit to:
\begin{equation}
\begin{aligned}
\underline{\hat{\Sigma}}^{2a}(\boldsymbol{k},t;\boldsymbol{q},T)&=iv_{i}^{3}n_{i}\frac{\xi_{SO}}{k_{F}^{2}}\int\frac{d\boldsymbol{k'}}{(2\pi)^{3}}\int\frac{d\boldsymbol{k''}}{(2\pi)^{3}}\int\frac{d\boldsymbol{q'}}{(2\pi)^{3}}\\
&\left[\left(\boldsymbol{k}+\frac{\boldsymbol{q}}{2}\right)\times\left(\boldsymbol{k'}+\frac{\boldsymbol{q'}}{2}\right)\cdot\hat{\boldsymbol{\sigma}}\right]\underline{\hat{G}}(\boldsymbol{k'},t;\boldsymbol{q'},T)\underline{\hat{G}}(\boldsymbol{k''},t;\boldsymbol{q}-\boldsymbol{q'},T),
\end{aligned}
\end{equation}
\begin{equation}
\begin{aligned}
\underline{\hat{\Sigma}}^{2b}(\boldsymbol{k},t;\boldsymbol{q},T)&=iv_{i}^{3}n_{i}\frac{\xi_{SO}}{k_{F}^{2}}\int\frac{d\boldsymbol{k'}}{(2\pi)^{3}}\int\frac{d\boldsymbol{k''}}{(2\pi)^{3}}\int\frac{d\boldsymbol{q'}}{(2\pi)^{3}}\\
&\underline{\hat{G}}(\boldsymbol{k'},t;\boldsymbol{q'},T)\left[\left(\boldsymbol{k'}-\frac{\boldsymbol{q'}}{2}\right)\times\left(\boldsymbol{k''}+\frac{\boldsymbol{q}}{2}-\frac{\boldsymbol{q'}}{2}\right)\cdot\hat{\boldsymbol{\sigma}}\right]\underline{\hat{G}}(\boldsymbol{k''},t;\boldsymbol{q}-\boldsymbol{q'},T),
\end{aligned}
\end{equation}
\begin{equation}
\begin{aligned}
\underline{\hat{\Sigma}}^{2c}(\boldsymbol{k},t;\boldsymbol{q},T)&=iv_{i}^{3}n_{i}\frac{\xi_{SO}}{k_{F}^{2}}\int\frac{d\boldsymbol{k'}}{(2\pi)^{3}}\int\frac{d\boldsymbol{k''}}{(2\pi)^{3}}\int\frac{d\boldsymbol{q'}}{(2\pi)^{3}}\\
&\underline{\hat{G}}(\boldsymbol{k'},t;\boldsymbol{q'},T)\underline{\hat{G}}(\boldsymbol{k''},t;\boldsymbol{q}-\boldsymbol{q'},T)\left[\left(\boldsymbol{k''}-\frac{\boldsymbol{q}}{2}+\frac{\boldsymbol{q'}}{2}\right)\times\left(\boldsymbol{k}-\frac{\boldsymbol{q}}{2}\right)\cdot\hat{\boldsymbol{\sigma}}\right].
\end{aligned}
\end{equation}
\noindent Let us assume that any inhomogeneity in a system is smooth, so we can neglect all gradient terms $\sim\boldsymbol{q}$. Having Fourier transformed with respect to $\boldsymbol{q}$ and $t$, we obtain:
\begin{equation}
\underline{\hat{\Sigma'}}_{\boldsymbol{k}E}(\boldsymbol{R},T)=\underline{\hat{\Sigma}}^{2a}_{\boldsymbol{k}E}(\boldsymbol{R},T)+\underline{\hat{\Sigma}}^{2b}_{\boldsymbol{k}E}(\boldsymbol{R},T)+\underline{\hat{\Sigma}}^{2c}_{\boldsymbol{k}E}(\boldsymbol{R},T),
\end{equation}
\noindent where 
\begin{equation}
\underline{\hat{\Sigma}}^{2a}_{\boldsymbol{k}E}(\boldsymbol{R},T)=iv_{i}^{3}n_{i}\frac{\xi_{SO}}{k_{F}^{2}}\int\frac{d\boldsymbol{k'}}{(2\pi)^{3}}\int\frac{d\boldsymbol{k''}}{(2\pi)^{3}}[(\boldsymbol{k}\times\boldsymbol{k'})\cdot\hat{\boldsymbol{\sigma}}]\underline{\hat{G}}_{\boldsymbol{k'}E}(\boldsymbol{R},T)\underline{\hat{G}}_{\boldsymbol{k''}E}(\boldsymbol{R},T),
\end{equation}
\begin{equation}
\underline{\hat{\Sigma}}^{2b}_{\boldsymbol{k}E}(\boldsymbol{R},T)=iv_{i}^{3}n_{i}\frac{\xi_{SO}}{k_{F}^{2}}\int\frac{d\boldsymbol{k'}}{(2\pi)^{3}}\int\frac{d\boldsymbol{k''}}{(2\pi)^{3}}\underline{\hat{G}}_{\boldsymbol{k'}E}(\boldsymbol{R},T)[(\boldsymbol{k'}\times\boldsymbol{k''})\cdot\hat{\boldsymbol{\sigma}}]\underline{\hat{G}}_{\boldsymbol{k''}E}(\boldsymbol{R},T),
\end{equation}
\begin{equation}
\underline{\hat{\Sigma}}^{2c}_{\boldsymbol{k}E}(\boldsymbol{R},T)=iv_{i}^{3}n_{i}\frac{\xi_{SO}}{k_{F}^{2}}\int\frac{d\boldsymbol{k'}}{(2\pi)^{3}}\int\frac{d\boldsymbol{k''}}{(2\pi)^{3}}\underline{\hat{G}}_{\boldsymbol{k'}E}(\boldsymbol{R},T)\underline{\hat{G}}_{\boldsymbol{k''}E}(\boldsymbol{R},T)[(\boldsymbol{k''}\times\boldsymbol{k})\cdot\hat{\boldsymbol{\sigma}}].
\end{equation}
\par To first order in $\xi_{SO}$, we can express the retarded and advanced Green's functions by using the Sokhotski formula:
\begin{equation}
\begin{aligned}
\hat{G}^{R(A)}_{\boldsymbol{k}E}&=\left(\hat{E}-\hat{\mathcal{H}}_{\boldsymbol{k}}-\hat{\Sigma}^{R(A)}_{\boldsymbol{k}E}\right)^{-1}\approx\frac{1}{2}\sum\limits_{s=\pm}\frac{\hat{\sigma}_{0}+s\hat{\boldsymbol{\sigma}}\cdot\boldsymbol{m}}{E-E_{\boldsymbol{k}s}\pm i\frac{\hbar}{2\tau_{s}}}\\
&=\frac{1}{2}\sum\limits_{s=\pm}\left(\hat{\sigma}_{0}+s\hat{\boldsymbol{\sigma}}\cdot\boldsymbol{m}\right)\left[\mp i\pi\delta(E-E_{\boldsymbol{k}s})\right],
\end{aligned}
\end{equation}
\noindent where $\tau_{s}$ is the spin dependent relaxation time. As a result, the retarded and advanced components of the skew-scattering self-energy vanish, and we deal with its Keldysh part, which survives for $\hat{\Sigma}^{2a}$ and $\hat{\Sigma}^{2c}$ only:
\begin{equation}
\begin{aligned}
\hat{\Sigma'}^{K}_{\boldsymbol{k}E}(\boldsymbol{R},T)&=i\frac{\xi_{SO}}{k_{F}^{2}}v_{i}^{3}n_{i}\int\frac{d\boldsymbol{k'}}{(2\pi)^{3}}\int\frac{d\boldsymbol{k''}}{(2\pi)^{3}}(\boldsymbol{k}\times\boldsymbol{k'})\cdot\hat{\boldsymbol{\sigma}}\left(\hat{G}^{R}_{\boldsymbol{k'}E}\hat{g}^{K}_{\boldsymbol{k''}E}(\boldsymbol{R},T)+\hat{g}^{K}_{\boldsymbol{k'}E}(\boldsymbol{R},T)\hat{G}^{A}_{\boldsymbol{k''}E} \right)\\
&+i\frac{\xi_{SO}}{k_{F}^{2}}v_{i}^{3}n_{i}\int\frac{d\boldsymbol{k'}}{(2\pi)^{3}}\int\frac{d\boldsymbol{k''}}{(2\pi)^{3}}\left(\hat{G}^{R}_{\boldsymbol{k'}E}\hat{g}^{K}_{\boldsymbol{k''}E}(\boldsymbol{R},T)+\hat{g}^{K}_{\boldsymbol{k'}E}(\boldsymbol{R},T)\hat{G}^{A}_{\boldsymbol{k''}E} \right)(\boldsymbol{k''}\times\boldsymbol{k})\cdot\hat{\boldsymbol{\sigma}}
\end{aligned}
\end{equation}
\noindent  or
\begin{equation}
\begin{aligned}
\hat{\Sigma'}^{K}_{\boldsymbol{k}E}(\boldsymbol{R},T)&=i\frac{\xi_{SO}}{k_{F}^{2}}v_{i}^{3}n_{i}\int\frac{d\boldsymbol{k'}}{(2\pi)^{3}}\int\frac{d\boldsymbol{k''}}{(2\pi)^{3}}(\boldsymbol{k}\times\boldsymbol{k'})\cdot\hat{\boldsymbol{\sigma}}\hat{g}^{K}_{\boldsymbol{k'}E}(\boldsymbol{R},T)\hat{G}^{A}_{\boldsymbol{k''}E}\\
&+i\frac{\xi_{SO}}{k_{F}^{2}}v_{i}^{3}n_{i}\int\frac{d\boldsymbol{k'}}{(2\pi)^{3}}\int\frac{d\boldsymbol{k''}}{(2\pi)^{3}}\hat{G}^{R}_{\boldsymbol{k'}E}\hat{g}^{K}_{\boldsymbol{k''}E}(\boldsymbol{R},T)(\boldsymbol{k''}\times\boldsymbol{k})\cdot\hat{\boldsymbol{\sigma}}.
\end{aligned}
\end{equation}
\noindent This expression can be further simplified, as we integrate over $\boldsymbol{k}$:
\begin{equation}
\begin{aligned}
\mp\frac{i\pi}{2}\int\frac{d\boldsymbol{k'}}{(2\pi)^{3}}\sum\limits_{s=\pm}\left(\hat{\sigma}_{0}+s\hat{\boldsymbol{\sigma}}\cdot\boldsymbol{m}\right)\delta(E-E_{\boldsymbol{k}s})=\mp\frac{i\pi}{2}(D^{\uparrow}+D^{\downarrow})\left(\hat{\sigma}_{0}+\delta\hat{\boldsymbol{\sigma}}\cdot\boldsymbol{m}\right),
\end{aligned}
\end{equation}
\noindent where $D^{\uparrow(\downarrow)}$ is the spin-dependent density of states and $\delta=(D^{\uparrow}-D^{\downarrow})/(D^{\uparrow}+D^{\downarrow})$, so the final form of $\hat{\Sigma'}^{K}_{\boldsymbol{k}E}(\boldsymbol{R},T)$ is given by:
\begin{equation}
\begin{aligned}
\hat{\Sigma'}^{K}_{\boldsymbol{k}E}(\boldsymbol{R},T)&=-\frac{\pi}{2}\frac{\xi_{SO}}{k_{F}^{2}}v_{i}^{3}n_{i}(D^{\uparrow}+D^{\downarrow})\int\frac{d\boldsymbol{k'}}{(2\pi)^{3}}(\boldsymbol{k}\times\boldsymbol{k'})\cdot\hat{\boldsymbol{\sigma}}\hat{g}^{K}_{\boldsymbol{k'}E}(\boldsymbol{R},T)\left[\hat{\sigma}_{0}+\delta\hat{\boldsymbol{\sigma}}\cdot\boldsymbol{m} \right]\\
&\,\,\,\,\,-\frac{\pi}{2}\frac{\xi_{SO}}{k_{F}^{2}}v_{i}^{3}n_{i}(D^{\uparrow}+D^{\downarrow})\int\frac{d\boldsymbol{k'}}{(2\pi)^{3}}\left[\hat{\sigma}_{0}+\delta\hat{\boldsymbol{\sigma}}\cdot\boldsymbol{m} \right]\hat{g}^{K}_{\boldsymbol{k'}E}(\boldsymbol{R},T)(\boldsymbol{k}\times\boldsymbol{k'})\cdot\hat{\boldsymbol{\sigma}}.
\end{aligned}
\end{equation}
\noindent In the weak exchange coupling limit $(J\ll\varepsilon_{F})$, we can express $D^{\uparrow(\downarrow)}\approx D_{0}(1\mp\beta)$, where $\beta=\frac{J}{2\varepsilon_{F}}$ is the spin polarization factor and $D_{0}=\frac{mk_{F}}{2\pi^{2}\hbar^{2}}$ is the spin independent density of states per spin at the Fermi level. Then, we obtain:
\begin{equation}
\begin{aligned}
\hat{\Sigma'}^{K}_{\boldsymbol{k}E}(\boldsymbol{R},T)&=-\pi v_{i}^{3}n_{i}D_{0}\frac{\xi_{SO}}{k_{F}^{2}}\int\frac{d\boldsymbol{k'}}{(2\pi)^{3}}(\boldsymbol{k}\times\boldsymbol{k'})\cdot\hat{\boldsymbol{\sigma}}\hat{g}^{K}_{\boldsymbol{k'}E}(\boldsymbol{R},T)\left[\hat{\sigma}_{0}-\beta\hat{\boldsymbol{\sigma}}\cdot\boldsymbol{m} \right]\\
&\,\,\,\,\,-\pi v_{i}^{3}n_{i}D_{0}\frac{\xi_{SO}}{k_{F}^{2}}\int\frac{d\boldsymbol{k'}}{(2\pi)^{3}}\left[\hat{\sigma}_{0}-\beta\hat{\boldsymbol{\sigma}}\cdot\boldsymbol{m} \right]\hat{g}^{K}_{\boldsymbol{k'}E}(\boldsymbol{R},T)(\boldsymbol{k}\times\boldsymbol{k'})\cdot\hat{\boldsymbol{\sigma}}
\end{aligned}
\end{equation}
\noindent or 
\begin{equation}
\begin{aligned}
\hat{\Sigma'}^{K}_{\boldsymbol{k}E}(\boldsymbol{R},T)&=-\frac{\hbar}{2}\frac{v_{i}}{\tau_{0}}\frac{\xi_{SO}}{k_{F}^{2}}\int\frac{d\boldsymbol{k'}}{(2\pi)^{3}}(\boldsymbol{k}\times\boldsymbol{k'})\cdot\hat{\boldsymbol{\sigma}}\hat{g}^{K}_{\boldsymbol{k'}E}(\boldsymbol{R},T)\left[\hat{\sigma}_{0}-\beta\hat{\boldsymbol{\sigma}}\cdot\boldsymbol{m} \right]\\
&\,\,\,\,\,-\frac{\hbar}{2}\frac{v_{i}}{\tau_{0}}\frac{\xi_{SO}}{k_{F}^{2}}\int\frac{d\boldsymbol{k'}}{(2\pi)^{3}}\left[\hat{\sigma}_{0}-\beta\hat{\boldsymbol{\sigma}}\cdot\boldsymbol{m} \right]\hat{g}^{K}_{\boldsymbol{k'}E}(\boldsymbol{R},T)(\boldsymbol{k}\times\boldsymbol{k'})\cdot\hat{\boldsymbol{\sigma}},
\end{aligned}
\label{eq:sskew}
\end{equation}
\noindent where $\frac{1}{\tau_{0}}=2\pi v_{i}^{2}n_{i}D_{0}/\hbar$ is the spin independent relaxation time.

\section{Relaxation time}
\par The imaginary part of the retarded and advanced self-energies is related to the momentum relaxation time, which is given by the elastic scattering off the on-site impurity potential and Elliot-Yafet mechanism:
\begin{equation}
\hat{\Sigma}^{R(A)}_{\boldsymbol{k}E}=\mp i\frac{\hbar}{2\hat{\tau}_{\boldsymbol{k}}}.
\end{equation}
\noindent Taking into account Eqs. (\ref{eq:sret}) and (\ref{eq:gzero}), we get:
\begin{equation}
\frac{1}{\hat{\tau}_{\boldsymbol{k}}}=\frac{\pi v_{i}^{2}n_{i}}{\hbar}\sum_{s=\pm}\int\frac{d\boldsymbol{k'}}{(2\pi)^{3}}\,\left[\hat{\sigma}_{0}+i\frac{\xi_{SO}}{k_{F}^{2}}\boldsymbol{n}\cdot\hat{\boldsymbol{\sigma}}\right]\left(\hat{\sigma}_{0}+s\hat{\boldsymbol{\sigma}}\cdot\boldsymbol{m} \right)\left[\hat{\sigma}_{0}-i\frac{\xi_{SO}}{k_{F}^{2}}\boldsymbol{n}\cdot\hat{\boldsymbol{\sigma}}\right]\delta(E-E_{\boldsymbol{k'}s}),
\end{equation}
\noindent where $\boldsymbol{n}=\boldsymbol{k}\times\boldsymbol{k'}$. In terms of spherical coordinates, $d\boldsymbol{k}=k^{2}dk\,\sin{\theta}d\theta\,d\phi$ with $k\in[0,k_{F}]$, $\theta\in[0,\pi]$ and $\phi\in[0,2\pi]$, it is easy to show:
\begin{equation}
\int k_{i}d\boldsymbol{k}=0,
\end{equation}
\noindent where $k_{i}$ is the $i$th cartesian coordinate of $\boldsymbol{k}$, and all terms linear in $\boldsymbol{n}$ vanish. Thus, we get:
\begin{equation}
\frac{1}{\hat{\tau}_{\boldsymbol{k}}}=\frac{\pi v_{i}^{2}n_{i}}{\hbar}\sum_{s=\pm}\int\frac{d\boldsymbol{k'}}{(2\pi)^{3}}\,\left[\hat{\sigma}_{0}+s\hat{\boldsymbol{\sigma}}\cdot\boldsymbol{m}+\frac{\xi_{SO}^{2}}{k_{F}^{4}}(\boldsymbol{n}\cdot\hat{\boldsymbol{\sigma}})(\boldsymbol{n}\cdot\hat{\boldsymbol{\sigma}})+s\,\frac{\xi_{SO}^{2}}{k_{F}^{4}}(\boldsymbol{n}\cdot\hat{\boldsymbol{\sigma}})(\hat{\boldsymbol{\sigma}}\cdot\boldsymbol{m})(\boldsymbol{n}\cdot\hat{\boldsymbol{\sigma}}) \right]\delta(E-E_{\boldsymbol{k'}s}),
\end{equation}
\noindent or having used $(\boldsymbol{a}\cdot\hat{\boldsymbol{\sigma}})(\boldsymbol{b}\cdot\hat{\boldsymbol{\sigma}})=(\boldsymbol{a}\cdot\boldsymbol{b})\hat{\sigma}_{0}+i(\boldsymbol{a}\times\boldsymbol{b})\cdot\hat{\boldsymbol{\sigma}}$:
\begin{equation}
\begin{aligned}
\frac{1}{\hat{\tau}_{\boldsymbol{k}}}&=\frac{\pi v_{i}^{2}n_{i}}{\hbar}\int\frac{d\boldsymbol{k'}}{(2\pi)^{3}}\,\left[(1+\frac{\xi_{SO}^{2}}{k_{F}^{4}}\boldsymbol{n}^{2})\hat{\sigma}_{0}(\delta(E-E_{\boldsymbol{k'}+})+\delta(E-E_{\boldsymbol{k'}-}))\right.\\
&\left.+\left(\hat{\boldsymbol{\sigma}}\cdot\boldsymbol{m}+\frac{\xi_{SO}^{2}}{k_{F}^{4}}\hat{\boldsymbol{\sigma}}\cdot(2\boldsymbol{n}(\boldsymbol{n}\cdot\boldsymbol{m})-\boldsymbol{m}\boldsymbol{n}^{2})\right)(\delta(E-E_{\boldsymbol{k'}+})-\delta(E-E_{\boldsymbol{k'}-}))\right].
\label{eq:tau2}
\end{aligned}
\end{equation}
\noindent Let us also consider the following expressions:
\begin{equation}
\begin{aligned}
\int\frac{d\boldsymbol{k'}}{(2\pi)^{3}}k'_{i}k'_{j}\,(\delta(E-E_{\boldsymbol{k'}+})\pm\delta(E-E_{\boldsymbol{k'}-}))=0\qquad\mathrm{for}\,\,\, i\ne j,\\
\int\frac{d\boldsymbol{k'}}{(2\pi)^{3}}{k'}_{i}^{2}\,(\delta(E-E_{\boldsymbol{k'}+})\pm\delta(E-E_{\boldsymbol{k'}-}))=\frac{1}{6\pi^{2}}\int d{k'}{k'}^{4}\,(\delta(E-E_{\boldsymbol{k'}+})\pm\delta(E-E_{\boldsymbol{k'}-})),\\
\int\frac{d\boldsymbol{k'}}{(2\pi)^{3}}\boldsymbol{n}^{2}\,(\delta(E-E_{\boldsymbol{k'}+})\pm\delta(E-E_{\boldsymbol{k'}-}))=\frac{1}{3\pi^{2}}\,k^{2}\int d{k'}{k'}^{4}\,(\delta(E-E_{\boldsymbol{k'}+})\pm\delta(E-E_{\boldsymbol{k'}-})).
\end{aligned}
\end{equation}
\noindent We can rewrite Eq. (\ref{eq:tau2}) as:
\begin{equation}
\begin{aligned}
\frac{1}{\hat{\tau}_{\boldsymbol{k}}}&=\frac{v_{i}^{2}n_{i}}{2\pi\hbar}\int dk'\,\left[({k'}^{2}+\frac{2}{3}\frac{\xi_{SO}^{2}}{k_{F}^{4}}k^{2}{k'}^{4})\hat{\sigma}_{0}(\delta(E-E_{\boldsymbol{k'}+})+\delta(E-E_{\boldsymbol{k'}-}))\right.\\
&\left.+\left(\hat{\boldsymbol{\sigma}}\cdot\boldsymbol{m}{k'}^{2}-\frac{2}{3}\frac{\xi_{SO}^{2}}{k_{F}^{4}}{k'}^{4}(\hat{\boldsymbol{\sigma}}\cdot\boldsymbol{k})(\boldsymbol{k}\cdot\boldsymbol{m})\right)(\delta(E-E_{\boldsymbol{k'}+})-\delta(E-E_{\boldsymbol{k'}-}))\right].
\end{aligned}
\end{equation}
\noindent Next, we can employ the following relation for the delta-function:
\begin{equation}
\delta(E-E_{\boldsymbol{k}s})=\frac{\delta(k-k_{s})}{\frac{\hbar^{2}k_{s}}{m}},
\end{equation}
\noindent where $k_{\pm}=\sqrt{2m(\varepsilon_{F}\mp J)}/\hbar$, and $\varepsilon_{F}=\frac{\hbar^{2}k_{F}^{2}}{2m}$ is the Fermi energy. Then, integrating over $k'$ gives:
\begin{equation}
\begin{aligned}
\frac{1}{\hat{\tau}_{\boldsymbol{k}}}=\frac{v_{i}^{2}n_{i}}{2\pi\hbar}\Big(\frac{m}{\hbar^{2}}(k_{+}+k_{-})\hat{\sigma}_{0}&+\frac{2m}{3\hbar^{2}}\frac{\xi_{SO}^{2}}{k_{F}^{4}}k^{2}(k^{3}_{+}+k^{3}_{-})\hat{\sigma}_{0} +\frac{m}{\hbar^{2}}(k_{+}-k_{-}) \hat{\boldsymbol{\sigma}}\cdot\boldsymbol{m}\\
&-\frac{2m}{3\hbar^{2}}\frac{\xi_{SO}^{2}}{k_{F}^{4}}(k^{3}_{+}-k^{3}_{-})(\hat{\boldsymbol{\sigma}}\cdot\boldsymbol{k})(\boldsymbol{k}\cdot\boldsymbol{m}) \Big).
\end{aligned}
\end{equation}
\noindent Finally, in the weak exchange coupling limit $(J\ll\varepsilon_{F})$, we can perform a Taylor expansion, $k_{\pm}\approx k_{F}(1\mp\beta)$. Neglecting higher order terms $\sim\beta\xi_{SO}^{2}$, we obtain:
\begin{equation}
\frac{1}{\hat{\tau}_{\boldsymbol{k}}}=\frac{1}{\tau_{0}}\left(\hat{\sigma_{0}}-\beta\hat{\boldsymbol{\sigma}}\cdot\boldsymbol{m}+\frac{2}{3}\frac{\xi_{SO}^{2}}{k_{F}^{2}}k^{2}\hat{\sigma}_{0} \right),
\end{equation}
\noindent where $\frac{1}{\tau_{0}}=2\pi v^{2}_{i}n_{i}D_{0}/\hbar$ is the spin-independent relaxation time due to scattering off the impurity potential.

\section{Averaged velocity operator}
\par For educational purposes, let us derive the averaged velocity operator in diffusive ferromagnets with extrinsic spin-orbit coupling.\cite{velo} Within the Lippmann-Schwinger equation, the scattered state $\parallel\!\boldsymbol{k},s\rangle$ can be written in the first order of $\hat{\mathcal{H}}_{\mathrm{imp}}$:
\begin{equation}
\begin{aligned}
\parallel\!\boldsymbol{k},s\rangle&=|\boldsymbol{k},s\rangle+\sum_{\boldsymbol{k'}}\hat{G}^{R}_{0,\boldsymbol{k'}}\langle\boldsymbol{k'}|\hat{\mathcal{H}}_{\mathrm{imp}}|\boldsymbol{k}\rangle|\boldsymbol{k'},s\rangle\\
&=|\boldsymbol{k},s\rangle-\frac{1}{\Omega}\frac{i\pi}{2}\sum_{\boldsymbol{k'}s'}(\hat{\sigma}_{0}+s'\hat{\boldsymbol{\sigma}}\cdot\boldsymbol{m})(\hat{\sigma
_{0}}-i\frac{\xi_{SO}}{k_{F}^{2}}\hat{\boldsymbol{\sigma}}\cdot(\boldsymbol{k}\times\boldsymbol{k'}))V(\boldsymbol{k'}-\boldsymbol{k})\delta(E-E_{\boldsymbol{k'}s'})|\boldsymbol{k'},s\rangle,
\end{aligned}
\end{equation}
\noindent and 
\begin{equation}
\begin{aligned}
\langle\boldsymbol{k},s\!\parallel&=\langle\boldsymbol{k},s|+\sum_{\boldsymbol{k'}}\langle\boldsymbol{k'},s|\langle\boldsymbol{k}|\hat{\mathcal{H}}_{\mathrm{imp}}|\boldsymbol{k'}\rangle\hat{G}^{R}_{0,\boldsymbol{k'}}\\
&=\langle\boldsymbol{k},s|+\frac{1}{\Omega}\frac{i\pi}{2}\sum_{\boldsymbol{k'}s'}\langle\boldsymbol{k'},s| (\hat{\sigma
_{0}}+i\frac{\xi_{SO}}{k_{F}^{2}}\hat{\boldsymbol{\sigma}}\cdot(\boldsymbol{k}\times\boldsymbol{k'}))(\hat{\sigma}_{0}+s'\hat{\boldsymbol{\sigma}}\cdot\boldsymbol{m})V(\boldsymbol{k}-\boldsymbol{k'})\delta(E-E_{\boldsymbol{k'}s'}).
\end{aligned}
\end{equation}
\noindent The corresponding matrix elements of the velocity operator  can be found as:
\begin{equation}
\begin{aligned}
\boldsymbol{v}_{\boldsymbol{k}\boldsymbol{k'}}^{ss'}=-\frac{i}{\hbar}\langle\boldsymbol{k},s\!\parallel[\hat{\boldsymbol{r}},\hat{\mathcal{H}}]\parallel\!\boldsymbol{k'},s'\rangle=-\frac{i}{\hbar}\langle\boldsymbol{k},s\!\parallel[\hat{\boldsymbol{r}},\hat{\mathcal{H}}_{0}+\hat{\mathcal{H}}_{\mathrm{imp}}]\parallel\!\boldsymbol{k'},s'\rangle,
\end{aligned}
\end{equation}
\noindent where
\begin{equation}
-\frac{i}{\hbar}[\hat{\boldsymbol{r}},\hat{\mathcal{H}}]=-\frac{i}{\hbar}[\hat{\boldsymbol{r}},\hat{\mathcal{H}}_{0}+\hat{\mathcal{H}}_{\mathrm{imp}}]=-\frac{i\hbar}{m}\nabla\hat{\sigma}_{0}+\frac{\xi_{SO}}{\hbar k_{F}^{2}}\sum_{\boldsymbol{R}_{i}}\hat{\boldsymbol{\sigma}}\times\nabla V(\boldsymbol{r}-\boldsymbol{R}_{i}).
\end{equation}
\noindent Let us express these terms separately neglecting higher order terms $\sim\xi_{SO}^{2}$:
\begin{equation}
\begin{aligned}
&-\frac{i}{\hbar}\langle\boldsymbol{k},s\!\parallel[\hat{\boldsymbol{r}},\hat{\mathcal{H}}_{0}]\parallel\!\boldsymbol{k'},s'\rangle=-\frac{i\hbar}{m}\langle\boldsymbol{k},s\!\parallel\nabla\parallel\!\boldsymbol{k'},s'\rangle\\
&=-\frac{i\hbar}{m}\langle\boldsymbol{k}|\nabla|\boldsymbol{k'}\rangle\delta_{ss'} - \frac{i\pi}{2}\sum_{s''}\int\frac{d\boldsymbol{k''}}{(2\pi)^{3}}V(\boldsymbol{k''}-\boldsymbol{k'})\delta(E-E_{\boldsymbol{k''}s''})\langle\boldsymbol{k}|\nabla|\boldsymbol{k''}\rangle\langle s|(\hat{\sigma}_{0}+s''\hat{\boldsymbol{\sigma}}\cdot\boldsymbol{m})(\hat{\sigma
_{0}}-i\frac{\xi_{SO}}{k_{F}^{2}}\hat{\boldsymbol{\sigma}}\cdot(\boldsymbol{k'}\times\boldsymbol{k''}))|s'\rangle\\
&\qquad\qquad\qquad\qquad\,\,\,\,\,\, + \frac{i\pi}{2}\sum_{s''}\int\frac{d\boldsymbol{k''}}{(2\pi)^{3}}V(\boldsymbol{k}-\boldsymbol{k''})\delta(E-E_{\boldsymbol{k''}s''})\langle s|(\hat{\sigma
_{0}}+i\frac{\xi_{SO}}{k_{F}^{2}}\hat{\boldsymbol{\sigma}}\cdot(\boldsymbol{k}\times\boldsymbol{k''}))(\hat{\sigma}_{0}+s''\hat{\boldsymbol{\sigma}}\cdot\boldsymbol{m})|s'\rangle\langle\boldsymbol{k''}|\nabla|\boldsymbol{k'}\rangle,
\end{aligned}
\end{equation}
\noindent and
\begin{equation}
\begin{aligned}
&-\frac{i}{\hbar}\langle\boldsymbol{k},s\!\parallel[\hat{\boldsymbol{r}},\hat{\mathcal{H}}_{\mathrm{imp}}]\parallel\!\boldsymbol{k'},s'\rangle=\frac{\xi_{SO}}{\hbar k_{F}^{2}}\sum_{\boldsymbol{R}_{i}}\langle\boldsymbol{k},s\!\parallel\hat{
\boldsymbol{\sigma}}\times\nabla V(\boldsymbol{r}-\boldsymbol{R}_{i})\parallel\!\boldsymbol{k'},s'\rangle\\
&=\frac{i}{\Omega}\frac{\xi_{SO}}{\hbar k_{F}^{2}}\langle s|\hat{\boldsymbol{\sigma}}|s'\rangle\times(\boldsymbol{k}-\boldsymbol{k'})V(\boldsymbol{k}-\boldsymbol{k'})\\
& +\frac{1}{\Omega} \frac{\pi}{2}\frac{\xi_{SO}}{\hbar k_{F}^{2}}\sum_{s''}\int\frac{d\boldsymbol{k''}}{(2\pi)^{3}}V(\boldsymbol{k''}-\boldsymbol{k'})V(\boldsymbol{k}-\boldsymbol{k''})\delta(E-E_{\boldsymbol{k''}s''})\langle s|\hat{
\boldsymbol{\sigma}}\times(\boldsymbol{k}-\boldsymbol{k''})(\hat{\sigma}_{0}+s''\hat{\boldsymbol{\sigma}}\cdot\boldsymbol{m})|s'\rangle\\
& -\frac{1}{\Omega} \frac{\pi}{2}\frac{\xi_{SO}}{\hbar k_{F}^{2}}\sum_{s''}\int\frac{d\boldsymbol{k''}}{(2\pi)^{3}}V(\boldsymbol{k}-\boldsymbol{k''})V(\boldsymbol{k''}-\boldsymbol{k'})\delta(E-E_{\boldsymbol{k''}s''})\langle s|(\hat{\sigma}_{0}+s''\hat{\boldsymbol{\sigma}}\cdot\boldsymbol{m})\hat{
\boldsymbol{\sigma}}\times(\boldsymbol{k''}-\boldsymbol{k'})| s'\rangle.
\end{aligned}
\end{equation}
\noindent Upon impurity averaging we obtain:
\begin{equation}
\begin{aligned}
 \boldsymbol{v}_{\boldsymbol{k}}^{ss'}&=\frac{\hbar}{m}\boldsymbol{k}\,\delta_{ss'}+v_{i}^{2}n_{i}\frac{\pi}{2}\frac{\xi_{SO}}{\hbar k_{F}^{2}}\sum_{s''}\int\frac{d\boldsymbol{k''}}{(2\pi)^{3}}\delta(E-E_{\boldsymbol{k''}s''})\langle s|\left\{\hat{
\boldsymbol{\sigma}}\times(\boldsymbol{k}-\boldsymbol{k''}),(\hat{\sigma}_{0}+s''\hat{\boldsymbol{\sigma}}\cdot\boldsymbol{m})\right\}|s'\rangle\\
&=\frac{\hbar}{m}\boldsymbol{k}\,\delta_{ss'}+v_{i}^{2}n_{i}\pi\frac{\xi_{SO}}{\hbar k_{F}^{2}}\sum_{s''}\int\frac{d\boldsymbol{k''}}{(2\pi)^{3}}\delta(E-E_{\boldsymbol{k''}s''})\langle s|\hat{\boldsymbol{\sigma}}\times\boldsymbol{k}+s''\boldsymbol{m}\times\boldsymbol{k}|s'\rangle,
\end{aligned}
\end{equation}
\noindent or in the limit $J\ll\varepsilon_{F}$:
\begin{equation}
\hat{\boldsymbol{v}}_{\boldsymbol{k}}=\frac{\hbar}{m}\boldsymbol{k}\,\hat{\sigma}_{0}+\frac{1}{\tau_{0}}\frac{\xi_{SO}}{k_{F}^{2}}\hat{\boldsymbol{\sigma}}\times\boldsymbol{k}-\frac{\beta}{\tau_{0}}\frac{\xi_{SO}}{k_{F}^{2}}\boldsymbol{m}\times\boldsymbol{k}\,\hat{\sigma}_{0}.
\label{eq:av}
\end{equation}

\section{Quantum transport equations}
\par Having integrated Eq.~(\ref{eq:keldysh}) over energy, we arrive at the kinetic equation written for the distribution function $\hat{g}_{\boldsymbol{k}}$:
\begin{equation}
-i[\hat{g}_{\boldsymbol{k}},J\hat{\boldsymbol{\sigma}}\cdot\boldsymbol{m}]+\frac{\hbar^{2}}{m}(\boldsymbol{k}\cdot\nabla_{\boldsymbol{R}})\,\hat{g}_{\boldsymbol{k}}=coll,
\label{eq:keld2}
\end{equation}
\noindent where the collision integral is defined as:
\begin{equation}
coll=\mathcal{J}_{\boldsymbol{k}}+\mathcal{I}_{\boldsymbol{k}},
\end{equation}
\noindent and
\begin{equation}
\mathcal{J}_{\boldsymbol{k}}=\int\frac{dE}{2\pi}\left(\hat{\Sigma}^{K}_{\boldsymbol{k}E}\hat{G}^{A}_{\boldsymbol{k}E}-\hat{G}^{R}_{\boldsymbol{k}E}\hat{\Sigma}^{K}_{\boldsymbol{k}E} \right),
\end{equation} 
\begin{equation}
\mathcal{I}_{\boldsymbol{k}}=\int\frac{dE}{2\pi}\left(\hat{\Sigma}^{R}_{\boldsymbol{k}E}\hat{g}^{K}_{\boldsymbol{k}E}-\hat{g}^{K}_{\boldsymbol{k}E}\hat{\Sigma}^{A}_{\boldsymbol{k}E} \right).
\end{equation}
\noindent Let us proceed with its detailed derivation. Taking into account the Kadanoff-Baym anzats~(\ref{eq:anzats}) for $\hat{g}^{K}_{\boldsymbol{k}E}$, the integration over energy (up to a given Fermi level $\varepsilon_{F}$) can be performed by using the residue theorem:
\begin{equation}
\begin{aligned}
\int\frac{dE}{2\pi}\hat{G}^{R}_{\boldsymbol{k}E}\hat{g}_{\boldsymbol{k'}}\hat{G}^{A}_{\boldsymbol{k'}E}&=\int\frac{dE}{8\pi}\sum\limits_{s,s'}\frac{\hat{\sigma}_{0}+s\hat{\boldsymbol{\sigma}}\cdot\boldsymbol{m}}{E-E_{\boldsymbol{k}s}+i\frac{\hbar}{2\tau_{s}}}\hat{g}_{\boldsymbol{k'}}\frac{\hat{\sigma}_{0}+s'\hat{\boldsymbol{\sigma}}\cdot\boldsymbol{m}}{E-E_{\boldsymbol{k'}s'}-i\frac{\hbar}{2\tau_{s'}}}\\
&=-\frac{i}{4}\sum_{s,s'}\frac{(\hat{\sigma}_{0}+s\hat{\boldsymbol{\sigma}}\cdot\boldsymbol{m})\hat{g}_{\boldsymbol{k'}}(\hat{\sigma}_{0}+s'\hat{\boldsymbol{\sigma}}\cdot\boldsymbol{m})}{\varepsilon_{F}-E_{\boldsymbol{k'}s'}-i\frac{\hbar}{2}\left(\frac{1}{\tau_{s'}}+\frac{1}{\tau_{s}}\right)},
\end{aligned}
\end{equation}
\noindent and
\begin{equation}
\begin{aligned}
\int\frac{dE}{2\pi}\hat{G}^{R}_{\boldsymbol{k'}E}\hat{g}_{\boldsymbol{k'}}\hat{G}^{A}_{\boldsymbol{k}E}&=\int\frac{dE}{8\pi}\sum\limits_{s,s'}\frac{\hat{\sigma}_{0}+s'\hat{\boldsymbol{\sigma}}\cdot\boldsymbol{m}}{E-E_{\boldsymbol{k'}s'}+i\frac{\hbar}{2\tau_{s'}}}\hat{g}_{\boldsymbol{k'}}\frac{\hat{\sigma}_{0}+s\hat{\boldsymbol{\sigma}}\cdot\boldsymbol{m}}{E-E_{\boldsymbol{k}s}-i\frac{\hbar}{2\tau_{s}}}\\
&=\frac{i}{4}\sum_{s,s'}\frac{(\hat{\sigma}_{0}+s'\hat{\boldsymbol{\sigma}}\cdot\boldsymbol{m})\hat{g}_{\boldsymbol{k'}}(\hat{\sigma}_{0}+s\hat{\boldsymbol{\sigma}}\cdot\boldsymbol{m})}{\varepsilon_{F}-E_{\boldsymbol{k'}s'}+i\frac{\hbar}{2}\left(\frac{1}{\tau_{s}}+\frac{1}{\tau_{s'}}\right)},
\end{aligned}
\end{equation}
\noindent while 
\begin{equation}
\int\frac{dE}{2\pi}\hat{G}^{R(A)}_{\boldsymbol{k}E}\hat{g}_{\boldsymbol{k'}}\hat{G}^{R(A)}_{\boldsymbol{k'}E}=0,
\end{equation}
\noindent where the retarded and advanced Green's functions are defined as:
\begin{equation}
\hat{G}_{\boldsymbol{k}E}^{R(A)}=\frac{1}{2}\sum\limits_{s=\pm}\frac{\hat{\sigma}_{0}+s\hat{\boldsymbol{\sigma}}\cdot\boldsymbol{m}}{E-E_{\boldsymbol{k}s}\pm i\frac{\hbar}{2\tau_{s}}}.
\end{equation}
\noindent Assuming the scattering term in the denominator to be small and transport properties to be described solely by the electrons close to the Fermi level, we can rewrite these expressions with the Sokhotski formula:
\begin{equation}
\begin{aligned}
\int\frac{dE}{2\pi}\hat{G}^{R}_{\boldsymbol{k}E}\hat{g}_{\boldsymbol{k'}}\hat{G}^{A}_{\boldsymbol{k'}E}&=\frac{\pi}{2}\sum_{s'}\hat{g}_{\boldsymbol{k'}}(\hat{\sigma}_{0}+s'\hat{\boldsymbol{\sigma}}\cdot\boldsymbol{m})\delta(\varepsilon_{F}-E_{\boldsymbol{k'}s'}),
\end{aligned}
\end{equation}
\noindent and
\begin{equation}
\begin{aligned}
\int\frac{dE}{2\pi}\hat{G}^{R}_{\boldsymbol{k'}E}\hat{g}_{\boldsymbol{k'}}\hat{G}^{A}_{\boldsymbol{k}E}=\frac{\pi}{2}\sum_{s'}(\hat{\sigma}_{0}+s'\hat{\boldsymbol{\sigma}}\cdot\boldsymbol{m})\hat{g}_{\boldsymbol{k'}}\delta(\varepsilon_{F}-E_{\boldsymbol{k'}s'}).
\end{aligned}
\end{equation}
\par Starting from the Born approximation~(\ref{eq:skel}), we have:
\begin{equation}
\begin{aligned}
\hat{\Sigma}^{K}_{\boldsymbol{k}E}\hat{G}^{A}_{\boldsymbol{k}E}&=v_{i}^{2}n_{i}\int\frac{d\boldsymbol{k'}}{(2\pi)^{3}}\,\left[\hat{G}^{R}_{\boldsymbol{k'}E}\hat{g}_{\boldsymbol{k'}}\hat{G}^{A}_{\boldsymbol{k}E}+i\frac{\xi_{SO}}{k^{2}_{F}}\boldsymbol{n}\cdot\hat{\boldsymbol{\sigma}}\hat{G}^{R}_{\boldsymbol{k'}E}\hat{g}_{\boldsymbol{k'}}\hat{G}^{A}_{\boldsymbol{k}E}  \right.\\
&\left.-i\frac{\xi_{SO}}{k_{F}^{2}}\hat{G}^{R}_{\boldsymbol{k'}E}\hat{g}_{\boldsymbol{k'}}\boldsymbol{n}\cdot\hat{\boldsymbol{\sigma}}\hat{G}^{A}_{\boldsymbol{k}E}+\frac{\xi^{2}_{SO}}{k_{F}^{4}}\boldsymbol{n}\cdot\hat{\boldsymbol{\sigma}}\hat{G}^{R}_{\boldsymbol{k'}E}\hat{g}_{\boldsymbol{k'}}\boldsymbol{n}\cdot\hat{\boldsymbol{\sigma}}\hat{G}^{A}_{\boldsymbol{k}E}\right]\\
&+\frac{v_{i}^{2}n_{i}}{2}\frac{\xi_{SO}}{k_{F}^{2}}\int\frac{d\boldsymbol{k'}}{(2\pi)^{3}}\left\{(\boldsymbol{k'}-\boldsymbol{k}\}\times\hat{\boldsymbol{\sigma}},\hat{G}^{R}_{\boldsymbol{k'}E}\nabla_{\boldsymbol{R}}\hat{g}_{\boldsymbol{k'}}\right\}\hat{G}^{A}_{\boldsymbol{k}E},
\end{aligned}
\end{equation}
\noindent and 
\begin{equation}
\begin{aligned}
\hat{G}^{R}_{\boldsymbol{k}E}\hat{\Sigma}^{K}_{\boldsymbol{k}E}&=-v_{i}^{2}n_{i}\int\frac{d\boldsymbol{k'}}{(2\pi)^{3}}\,\left[\hat{G}^{R}_{\boldsymbol{k}E}\hat{g}_{\boldsymbol{k'}}\hat{G}^{A}_{\boldsymbol{k'}E}+i\frac{\xi_{SO}}{k_{F}^{2}}\hat{G}^{R}_{\boldsymbol{k}E}\boldsymbol{n}\cdot\hat{\boldsymbol{\sigma}}\hat{g}_{\boldsymbol{k'}}\hat{G}^{A}_{\boldsymbol{k'}E}  \right.\\
&\left.-i\frac{\xi_{SO}}{k_{F}^{2}}\hat{G}^{R}_{\boldsymbol{k}E}\hat{g}_{\boldsymbol{k'}}\hat{G}^{A}_{\boldsymbol{k'}E}\boldsymbol{n}\cdot\hat{\boldsymbol{\sigma}}+\frac{\xi^{2}_{SO}}{k_{F}^{4}}\hat{G}^{R}_{\boldsymbol{k}E}\boldsymbol{n}\cdot\hat{\boldsymbol{\sigma}}\hat{g}_{\boldsymbol{k'}}\hat{G}^{A}_{\boldsymbol{k'}E}\boldsymbol{n}\cdot\hat{\boldsymbol{\sigma}}\right]\\
&-\frac{v_{i}^{2}n_{i}}{2}\frac{\xi_{SO}}{k_{F}^{2}}\int\frac{d\boldsymbol{k'}}{(2\pi)^{3}}\hat{G}^{R}_{\boldsymbol{k}E}\left\{(\boldsymbol{k'}-\boldsymbol{k})\times\hat{\boldsymbol{\sigma}},\nabla_{\boldsymbol{R}}\hat{g}_{\boldsymbol{k'}}\hat{G}^{A}_{\boldsymbol{k'}E}\right\}.
\end{aligned}
\end{equation}
\noindent By using the following relations: 
\begin{equation}
\begin{aligned}
\hat{\sigma}_{a}\hat{\sigma}_{b}&=i\varepsilon_{abc}\,\hat{\sigma}_{c}+\delta_{ab}\hat{\sigma}_{0},\\
(\boldsymbol{a}\cdot\hat{\boldsymbol{\sigma}})(\boldsymbol{b}\cdot\hat{\boldsymbol{\sigma}})&=(\boldsymbol{a}\cdot\boldsymbol{b})\hat{\sigma}_{0}+i(\boldsymbol{a}\times\boldsymbol{b})\cdot\hat{\boldsymbol{\sigma}},
\end{aligned}
\end{equation}
\noindent these terms give:
\begin{equation}
\begin{aligned}
\int\frac{dE}{2\pi}\left[\hat{G}^{R}_{\boldsymbol{k'}E}\hat{g}_{\boldsymbol{k'}}\hat{G}^{A}_{\boldsymbol{k}E}+\hat{G}^{R}_{\boldsymbol{k}E}\hat{g}_{\boldsymbol{k'}}\hat{G}^{A}_{\boldsymbol{k'}E}\right]&=\pi\hat{g}_{\boldsymbol{k'}}\delta_{T}+\pi\{\hat{g}_{\boldsymbol{k'}},\hat{\boldsymbol{\sigma}}\cdot\boldsymbol{m}\}\delta_{J},
\end{aligned}
\end{equation}
\begin{equation}
\begin{aligned}
&\int\frac{dE}{2\pi} i\frac{\xi_{SO}}{k_{F}^{2}}\left[[\boldsymbol{n}\cdot\hat{\boldsymbol{\sigma}},\hat{G}^{R}_{\boldsymbol{k'}E}\hat{g}_{\boldsymbol{k'}}]\hat{G}^{A}_{\boldsymbol{k}E}+\hat{G}^{R}_{\boldsymbol{k}E}[\boldsymbol{n}\cdot\hat{\boldsymbol{\sigma}},\hat{g}_{\boldsymbol{k'}}\hat{G}^{A}_{\boldsymbol{k'}E}]\right]=\\
&=i\pi\frac{\xi_{SO}}{k_{F}^{2}}[\boldsymbol{n}\cdot\hat{\boldsymbol{\sigma}},\hat{g}_{\boldsymbol{k'}}]\delta_{T}+i\pi\frac{\xi_{SO}}{k_{F}^{2}}\left(\boldsymbol{n}\cdot\hat{\boldsymbol{\sigma}}\hat{g}_{\boldsymbol{k'}}\boldsymbol{m}\cdot\hat{\boldsymbol{\sigma}}-\boldsymbol{m}\cdot\hat{\boldsymbol{\sigma}}\hat{g}_{\boldsymbol{k'}}\boldsymbol{n}\cdot\hat{\boldsymbol{\sigma}}  \right)\delta_{J}\\
&-\pi\frac{\xi_{SO}}{k_{F}^{2}}\{(\boldsymbol{n}\times\boldsymbol{m})\cdot\hat{\boldsymbol{\sigma}},\hat{g}_{\boldsymbol{k'}}\}\delta_{J},
\end{aligned}
\end{equation}
\begin{equation}
\begin{aligned}
&\int\frac{dE}{2\pi} \frac{\xi^{2}_{SO}}{k_{F}^{4}}\left[\boldsymbol{n}\cdot\hat{\boldsymbol{\sigma}}\hat{G}^{R}_{\boldsymbol{k'}E}\hat{g}_{\boldsymbol{k'}}\boldsymbol{n}\cdot\hat{\boldsymbol{\sigma}}\hat{G}^{A}_{\boldsymbol{k}E}+\hat{G}^{R}_{\boldsymbol{k}E}\boldsymbol{n}\cdot\hat{\boldsymbol{\sigma}}\hat{g}_{\boldsymbol{k'}}\hat{G}^{A}_{\boldsymbol{k'}E}\boldsymbol{n}\cdot\hat{\boldsymbol{\sigma}}\right]=\\
&=\pi\frac{\xi^{2}_{SO}}{k_{F}^{4}}\boldsymbol{n}\cdot\hat{\boldsymbol{\sigma}}\hat{g}_{\boldsymbol{k'}}\boldsymbol{n}\cdot\hat{\boldsymbol{\sigma}}\delta_{T}
+\pi\frac{\xi^{2}_{SO}}{k_{F}^{4}}\boldsymbol{n}\cdot\boldsymbol{m}\{\hat{g}_{\boldsymbol{k'}},\boldsymbol{n}\cdot\hat{\boldsymbol{\sigma}} \}\delta_{J}\\
&+i\pi\frac{\xi^{2}_{SO}}{k_{F}^{4}}(\boldsymbol{n}\times\boldsymbol{m})\cdot\hat{\boldsymbol{\sigma}}\hat{g}_{\boldsymbol{k'}}\boldsymbol{n}\cdot\hat{\boldsymbol{\sigma}}\delta_{J}
-i\pi\frac{\xi^{2}_{SO}}{k_{F}^{4}}\boldsymbol{n}\cdot\hat{\boldsymbol{\sigma}}\hat{g}_{\boldsymbol{k'}}(\boldsymbol{n}\times\boldsymbol{m})\cdot\hat{\boldsymbol{\sigma}}\delta_{J},
\end{aligned}
\end{equation}
\begin{equation}
\begin{aligned}
&\int\frac{dE}{2\pi} \left[\left\{(\boldsymbol{k'}-\boldsymbol{k}\}\times\hat{\boldsymbol{\sigma}},\hat{G}^{R}_{\boldsymbol{k'}E}\nabla_{\boldsymbol{R}}\hat{g}_{\boldsymbol{k'}}\right\}\hat{G}^{A}_{\boldsymbol{k}E}+\hat{G}^{R}_{\boldsymbol{k}E}\left\{(\boldsymbol{k'}-\boldsymbol{k})\times\hat{\boldsymbol{\sigma}},\nabla_{\boldsymbol{R}}\hat{g}_{\boldsymbol{k'}}\hat{G}^{A}_{\boldsymbol{k'}E}\right\} \right]=\\
&=\pi\left\{(\boldsymbol{k'}-\boldsymbol{k})\times\hat{\boldsymbol{\sigma}},\nabla_{\boldsymbol{R}}\hat{g}_{\boldsymbol{k'}}\right\}\delta_{T}
+\pi\left\{(\boldsymbol{k'}-\boldsymbol{k})\times\hat{\boldsymbol{\sigma}},\{\nabla_{\boldsymbol{R}}\hat{g}_{\boldsymbol{k'}},\boldsymbol{m}\cdot\hat{\boldsymbol{\sigma}}\}\right\}\delta_{J},
\end{aligned}
\end{equation}
\noindent where the following notations are used:
\begin{equation}
\begin{aligned}
\delta_{T}&=\delta(\varepsilon_{F}-E_{\boldsymbol{k'}+})+\delta(\varepsilon_{F}-E_{\boldsymbol{k'}-}),\\
\delta_{J}&=\frac{1}{2}\left[\delta(\varepsilon_{F}-E_{\boldsymbol{k'}+})-\delta(\varepsilon_{F}-E_{\boldsymbol{k'}-})\right].
\end{aligned}
\end{equation}
\noindent For the skew-scattering self-energy Eq.~(\ref{eq:sskew}), we have:
\begin{equation}
\begin{aligned}
\hat{\Sigma'}^{K}_{\boldsymbol{k}E}\hat{G}^{A}_{\boldsymbol{k}E}&=-\frac{\hbar}{2}\frac{v_{i}}{\tau_{0}}\frac{\xi_{SO}}{k_{F}^{2}}\int\frac{d\boldsymbol{k'}}{(2\pi)^{3}}\boldsymbol{n}\cdot\hat{\boldsymbol{\sigma}}\hat{G}^{R}_{\boldsymbol{k'}E}\hat{g}_{\boldsymbol{k'}}\left[\hat{\sigma}_{0}-\beta\hat{\boldsymbol{\sigma}}\cdot\boldsymbol{m} \right]\hat{G}^{A}_{\boldsymbol{k}E}\\
&-\frac{\hbar}{2}\frac{v_{i}}{\tau_{0}}\frac{\xi_{SO}}{k_{F}^{2}}\int\frac{d\boldsymbol{k'}}{(2\pi)^{3}}\left[\hat{\sigma}_{0}-\beta\hat{\boldsymbol{\sigma}}\cdot\boldsymbol{m} \right]\hat{G}^{R}_{\boldsymbol{k'}E}\hat{g}_{\boldsymbol{k'}}\boldsymbol{n}\cdot\hat{\boldsymbol{\sigma}}\hat{G}^{A}_{\boldsymbol{k}E},
\end{aligned}
\end{equation}
\noindent and
\begin{equation}
\begin{aligned}
\hat{G}^{R}_{\boldsymbol{k}E}\hat{\Sigma'}^{K}_{\boldsymbol{k}E}&=\frac{\hbar}{2}\frac{v_{i}}{\tau_{0}}\frac{\xi_{SO}}{k_{F}^{2}}\int\frac{d\boldsymbol{k'}}{(2\pi)^{3}}\hat{G}^{R}_{\boldsymbol{k}E}\boldsymbol{n}\cdot\hat{\boldsymbol{\sigma}}\hat{g}_{\boldsymbol{k'}}\hat{G}^{A}_{\boldsymbol{k'}E}\left[\hat{\sigma}_{0}-\beta\hat{\boldsymbol{\sigma}}\cdot\boldsymbol{m} \right]\\
&+\frac{\hbar}{2}\frac{v_{i}}{\tau_{0}}\frac{\xi_{SO}}{k_{F}^{2}}\int\frac{d\boldsymbol{k'}}{(2\pi)^{3}}\hat{G}^{R}_{\boldsymbol{k}E}\left[\hat{\sigma}_{0}-\beta\hat{\boldsymbol{\sigma}}\cdot\boldsymbol{m}\right]\hat{g}_{\boldsymbol{k'}}\hat{G}^{A}_{\boldsymbol{k'}E}\boldsymbol{n}\cdot\hat{\boldsymbol{\sigma}},
\end{aligned}
\end{equation}
\noindent that gives:
\begin{equation}
\begin{aligned}
&\int\frac{dE}{2\pi}\,\left[\left\{\boldsymbol{n}\cdot\hat{\boldsymbol{\sigma}},\hat{G}^{R}_{\boldsymbol{k'}E}\hat{g}_{\boldsymbol{k'}}\right\}\hat{G}^{A}_{\boldsymbol{k}E}+\hat{G}^{R}_{\boldsymbol{k}E}\left\{\boldsymbol{n}\cdot\hat{\boldsymbol{\sigma}},\hat{g}_{\boldsymbol{k'}}\hat{G}^{A}_{\boldsymbol{k'}E}\right\}\right]=\\
&=\pi\{\hat{g}_{\boldsymbol{k'}},\boldsymbol{n}\cdot\hat{\boldsymbol{\sigma}}\}\delta_{T}+\pi\left\{\boldsymbol{n}\cdot\hat{\boldsymbol{\sigma}},\{\hat{g}_{\boldsymbol{k'}},\hat{\boldsymbol{\sigma}}\cdot\boldsymbol{m}\}\right\}\delta_{J},
\end{aligned}
\end{equation}
\begin{equation}
\begin{aligned}
&\int\frac{dE}{2\pi}\,\left[\boldsymbol{n}\cdot\hat{\boldsymbol{\sigma}}\hat{G}^{R}_{\boldsymbol{k'}E}\hat{g}_{\boldsymbol{k'}}\hat{\boldsymbol{\sigma}}\cdot\boldsymbol{m}\hat{G}^{A}_{\boldsymbol{k}E}+\hat{\boldsymbol{\sigma}}\cdot\boldsymbol{m}\hat{G}^{R}_{\boldsymbol{k'}E}\hat{g}_{\boldsymbol{k'}}\boldsymbol{n}\cdot\hat{\boldsymbol{\sigma}}\hat{G}^{A}_{\boldsymbol{k}E}\right.\\
&\qquad\qquad\left.+\hat{G}^{R}_{\boldsymbol{k}E}\boldsymbol{n}\cdot\hat{\boldsymbol{\sigma}}\hat{g}_{\boldsymbol{k'}}\hat{G}^{A}_{\boldsymbol{k'}E}\hat{\boldsymbol{\sigma}}\cdot\boldsymbol{m}+\hat{G}^{R}_{\boldsymbol{k}E}\hat{\boldsymbol{\sigma}}\cdot\boldsymbol{m}\hat{g}_{\boldsymbol{k'}}\hat{G}^{A}_{\boldsymbol{k'}E}\boldsymbol{n}\cdot\hat{\boldsymbol{\sigma}}\right]=\\
&=\pi\left(\hat{\boldsymbol{\sigma}}\cdot\boldsymbol{m}\hat{g}_{\boldsymbol{k'}}\boldsymbol{n}\cdot\hat{\boldsymbol{\sigma}}+\boldsymbol{n}\cdot\hat{\boldsymbol{\sigma}}\hat{g}_{\boldsymbol{k'}}\hat{\boldsymbol{\sigma}}\cdot\boldsymbol{m} \right) \delta_{T}+\pi\boldsymbol{n}\cdot\boldsymbol{m}\{\hat{g}_{\boldsymbol{k'}},\hat{\boldsymbol{\sigma}}\cdot\boldsymbol{m} \}\delta_{J}+\pi\boldsymbol{m}^{2}\{\hat{g}_{\boldsymbol{k'}},\boldsymbol{n}\cdot\hat{\boldsymbol{\sigma}} \}\delta_{J}\\
&\qquad\qquad+i\pi\left((\boldsymbol{n}\times\boldsymbol{m})\cdot\hat{\boldsymbol{\sigma}}\hat{g}_{\boldsymbol{k'}}\hat{\boldsymbol{\sigma}}\cdot\boldsymbol{m}-\hat{\boldsymbol{\sigma}}\cdot\boldsymbol{m} \hat{g}_{\boldsymbol{k'}} (\boldsymbol{n}\times\boldsymbol{m})\cdot\hat{\boldsymbol{\sigma}} \right)\delta_{J}.
\end{aligned}
\end{equation}
\par Finally, we obtain:
\begin{equation}
\begin{aligned}
\mathcal{J}_{\boldsymbol{k}}&=\pi a_{1}\int\frac{d\boldsymbol{k'}}{(2\pi)^{3}}\Big[\hat{g}_{\boldsymbol{k'}}\delta_{T}+\{\hat{g}_{\boldsymbol{k'}},\hat{\boldsymbol{\sigma}}\cdot\boldsymbol{m}\}\delta_{J}\\
&\qquad\qquad+i\frac{\xi_{SO}}{k_{F}^{2}}[\boldsymbol{n}\cdot\hat{\boldsymbol{\sigma}},\hat{g}_{\boldsymbol{k'}}]\delta_{T}+i\frac{\xi_{SO}}{k_{F}^{2}}\left(\boldsymbol{n}\cdot\hat{\boldsymbol{\sigma}}\hat{g}_{\boldsymbol{k'}}\boldsymbol{m}\cdot\hat{\boldsymbol{\sigma}}-\boldsymbol{m}\cdot\hat{\boldsymbol{\sigma}}\hat{g}_{\boldsymbol{k'}}\boldsymbol{n}\cdot\hat{\boldsymbol{\sigma}}  \right)\delta_{J}-\frac{\xi_{SO}}{k_{F}^{2}}\{(\boldsymbol{n}\times\boldsymbol{m})\cdot\hat{\boldsymbol{\sigma}},\hat{g}_{\boldsymbol{k'}}\}\delta_{J}\\
&\qquad\qquad+\frac{\xi^{2}_{SO}}{k_{F}^{4}}\boldsymbol{n}\cdot\hat{\boldsymbol{\sigma}}\hat{g}_{\boldsymbol{k'}}\boldsymbol{n}\cdot\hat{\boldsymbol{\sigma}}\delta_{T}
+\frac{\xi^{2}_{SO}}{k_{F}^{4}}\boldsymbol{n}\cdot\boldsymbol{m}\{\hat{g}_{\boldsymbol{k'}},\boldsymbol{n}\cdot\hat{\boldsymbol{\sigma}} \}\delta_{J}\\
&\qquad\qquad+i\frac{\xi^{2}_{SO}}{k_{F}^{4}}(\boldsymbol{n}\times\boldsymbol{m})\cdot\hat{\boldsymbol{\sigma}}\hat{g}_{\boldsymbol{k'}}\boldsymbol{n}\cdot\hat{\boldsymbol{\sigma}}\delta_{J}
-i\frac{\xi^{2}_{SO}}{k_{F}^{4}}\boldsymbol{n}\cdot\hat{\boldsymbol{\sigma}}\hat{g}_{\boldsymbol{k'}}(\boldsymbol{n}\times\boldsymbol{m})\cdot\hat{\boldsymbol{\sigma}}\delta_{J}\\
&\qquad\qquad+\frac{1}{2}\frac{\xi_{SO}}{k_{F}^{2}}\left\{(\boldsymbol{k'}-\boldsymbol{k})\times\hat{\boldsymbol{\sigma}},\nabla_{\boldsymbol{R}}\hat{g}_{\boldsymbol{k'}}\right\}\delta_{T}
+\frac{1}{2}\frac{\xi_{SO}}{k_{F}^{2}}\left\{(\boldsymbol{k'}-\boldsymbol{k})\times\hat{\boldsymbol{\sigma}},\{\nabla_{\boldsymbol{R}}\hat{g}_{\boldsymbol{k'}},\boldsymbol{m}\cdot\hat{\boldsymbol{\sigma}}\}\right\}\delta_{J}\\
&-\pi a_{2}\int\frac{d\boldsymbol{k'}}{(2\pi)^{3}}\Big[\{\hat{g}_{\boldsymbol{k'}},\boldsymbol{n}\cdot\hat{\boldsymbol{\sigma}}\}\delta_{T}+\left\{\boldsymbol{n}\cdot\hat{\boldsymbol{\sigma}},\{\hat{g}_{\boldsymbol{k'}},\hat{\boldsymbol{\sigma}}\cdot\boldsymbol{m}\}\right\}\delta_{J}\\
&\qquad\qquad-\beta\left(\hat{\boldsymbol{\sigma}}\cdot\boldsymbol{m}\hat{g}_{\boldsymbol{k'}}\boldsymbol{n}\cdot\hat{\boldsymbol{\sigma}}+\boldsymbol{n}\cdot\hat{\boldsymbol{\sigma}}\hat{g}_{\boldsymbol{k'}}\hat{\boldsymbol{\sigma}}\cdot\boldsymbol{m} \right) \delta_{T}-\beta\boldsymbol{n}\cdot\boldsymbol{m}\{\hat{g}_{\boldsymbol{k'}},\hat{\boldsymbol{\sigma}}\cdot\boldsymbol{m} \}\delta_{J}-\beta\boldsymbol{m}^{2}\{\hat{g}_{\boldsymbol{k'}},\boldsymbol{n}\cdot\hat{\boldsymbol{\sigma}} \}\delta_{J}\\
&\qquad\qquad-i\beta\left((\boldsymbol{n}\times\boldsymbol{m})\cdot\hat{\boldsymbol{\sigma}}\hat{g}_{\boldsymbol{k'}}\hat{\boldsymbol{\sigma}}\cdot\boldsymbol{m}-\hat{\boldsymbol{\sigma}}\cdot\boldsymbol{m} \hat{g}_{\boldsymbol{k'}} (\boldsymbol{n}\times\boldsymbol{m})\cdot\hat{\boldsymbol{\sigma}} \right)\delta_{J}\Big],
\end{aligned}
\end{equation}
\noindent where $a_{1}=n_{i}v_{i}^{2}$ and $a_{2}=\frac{\hbar}{2}\frac{v_{i}}{\tau_{0}}\frac{\xi_{SO}}{k_{F}^{2}}$. 
\par In a similar manner, by using Eqs. (\ref{eq:sret}) and (\ref{eq:sadv}) we proceed with the second part of the collision integral $\mathcal{I}_{\boldsymbol{k}}$:
\begin{equation}
\begin{aligned}
\mathcal{I}_{\boldsymbol{k}}&=-a_{1}\int\frac{dE}{2\pi}\int\frac{d\boldsymbol{k'}}{(2\pi)^{3}}\,\Big[\big(\hat{\sigma}_{0}+i\frac{\xi_{SO}}{k_{F}^{2}}\boldsymbol{n}\cdot\hat{\boldsymbol{\sigma}}\big)\hat{G}^{R}_{\boldsymbol{k'}E}(\hat{\sigma}_{0}-i\frac{\xi_{SO}}{k_{F}^{2}}\boldsymbol{n}\cdot\hat{\boldsymbol{\sigma}})\hat{g}_{\boldsymbol{k}}\hat{G}^{A}_{\boldsymbol{k}E}\\
&\qquad\qquad\qquad\qquad\qquad+\hat{G}^{R}_{\boldsymbol{k}E}\hat{g}_{\boldsymbol{k}}\big(\hat{\sigma}_{0}+i\frac{\xi_{SO}}{k_{F}^{2}}\boldsymbol{n}\cdot\hat{\boldsymbol{\sigma}}\big)\hat{G}^{A}_{\boldsymbol{k'}E}(\hat{\sigma}_{0}-i\frac{\xi_{SO}}{k_{F}^{2}}\boldsymbol{n}\cdot\hat{\boldsymbol{\sigma}})\Big],
\end{aligned}
\end{equation}
\noindent where 
\begin{equation}
\begin{aligned}
\int\frac{dE}{2\pi}\left[\hat{G}^{R}_{\boldsymbol{k'}E}\hat{g}_{\boldsymbol{k}}\hat{G}^{A}_{\boldsymbol{k}E}+\hat{G}^{R}_{\boldsymbol{k}E}\hat{g}_{\boldsymbol{k}}\hat{G}^{A}_{\boldsymbol{k'}E}\right]&=\pi\hat{g}_{\boldsymbol{k}}\delta_{T}+\pi\{\hat{g}_{\boldsymbol{k}},\hat{\boldsymbol{\sigma}}\cdot\boldsymbol{m}\}\delta_{J},
\end{aligned}
\end{equation}
\begin{equation}
\begin{aligned}
&\int\frac{dE}{2\pi} i\frac{\xi_{SO}}{k_{F}^{2}}\left[[\boldsymbol{n}\cdot\hat{\boldsymbol{\sigma}},\hat{G}^{R}_{\boldsymbol{k'}E}]\hat{g}_{\boldsymbol{k}}\hat{G}^{A}_{\boldsymbol{k}E}+\hat{G}^{R}_{\boldsymbol{k}E}\hat{g}_{\boldsymbol{k}}[\boldsymbol{n}\cdot\hat{\boldsymbol{\sigma}},\hat{G}^{A}_{\boldsymbol{k'}E}]\right]=\\
&=-\pi\frac{\xi_{SO}}{k_{F}^{2}}(\boldsymbol{n}\times\boldsymbol{m})\cdot\{\hat{g}_{\boldsymbol{k}},\hat{\boldsymbol{\sigma}}\}\delta_{J},
\end{aligned}
\end{equation}
\begin{equation}
\begin{aligned}
&\int\frac{dE}{2\pi} \frac{\xi^{2}_{SO}}{k_{F}^{4}}\left[\boldsymbol{n}\cdot\hat{\boldsymbol{\sigma}}\hat{G}^{R}_{\boldsymbol{k'}E}\boldsymbol{n}\cdot\hat{\boldsymbol{\sigma}}\hat{g}_{\boldsymbol{k}}\hat{G}^{A}_{\boldsymbol{k}E}+\hat{G}^{R}_{\boldsymbol{k}E}\hat{g}_{\boldsymbol{k}}\boldsymbol{n}\cdot\hat{\boldsymbol{\sigma}}\hat{G}^{A}_{\boldsymbol{k'}E}\boldsymbol{n}\cdot\hat{\boldsymbol{\sigma}}\right]=\\
&=\pi \frac{\xi^{2}_{SO}}{k_{F}^{4}}\boldsymbol{n}^{2}\hat{g}_{\boldsymbol{k}}\delta_{T}+\pi \frac{\xi^{2}_{SO}}{k_{F}^{4}}(2(\boldsymbol{n}\cdot\boldsymbol{m})\boldsymbol{n}-\boldsymbol{n}^{2}\boldsymbol{m})\cdot\{\hat{g}_{\boldsymbol{k}},\hat{\boldsymbol{\sigma}}\}\delta_{J},
\end{aligned}
\end{equation}
\noindent so we obtain:
\begin{equation}
\begin{aligned}
\mathcal{I}_{\boldsymbol{k}}&=-\pi a_{1}\int\frac{d\boldsymbol{k'}}{(2\pi)^{3}}\Big[\hat{g}_{\boldsymbol{k}}\delta_{T}+\{\hat{g}_{\boldsymbol{k}},\hat{\boldsymbol{\sigma}}\cdot\boldsymbol{m}\}\delta_{J}-\frac{\xi_{SO}}{k_{F}^{2}}(\boldsymbol{n}\times\boldsymbol{m})\cdot\{\hat{g}_{\boldsymbol{k}},\hat{\boldsymbol{\sigma}}\}\delta_{J}\\
&\qquad\qquad+\frac{\xi^{2}_{SO}}{k_{F}^{4}}\boldsymbol{n}^{2}\hat{g}_{\boldsymbol{k}}\delta_{T}+\frac{\xi^{2}_{SO}}{k_{F}^{4}}(2(\boldsymbol{n}\cdot\boldsymbol{m})\boldsymbol{n}-\boldsymbol{n}^{2}\boldsymbol{m})\cdot\{\hat{g}_{\boldsymbol{k}},\hat{\boldsymbol{\sigma}}\}\delta_{J}\Big].
\end{aligned}
\end{equation}
\noindent Finally, the collision integral is written as:
\begin{equation}
\begin{aligned}
coll&=\pi a_{1}\int\frac{d\boldsymbol{k'}}{(2\pi)^{3}}\Big[(\hat{g}_{\boldsymbol{k'}}-\hat{g}_{\boldsymbol{k}})\delta_{T}+\{\hat{g}_{\boldsymbol{k'}}-\hat{g}_{\boldsymbol{k}},\hat{\boldsymbol{\sigma}}\cdot\boldsymbol{m}\}\delta_{J}\\
&\qquad\qquad+i\frac{\xi_{SO}}{k_{F}^{2}}[\boldsymbol{n}\cdot\hat{\boldsymbol{\sigma}},\hat{g}_{\boldsymbol{k'}}]\delta_{T}+i\frac{\xi_{SO}}{k_{F}^{2}}\left(\boldsymbol{n}\cdot\hat{\boldsymbol{\sigma}}\hat{g}_{\boldsymbol{k'}}\boldsymbol{m}\cdot\hat{\boldsymbol{\sigma}}-\boldsymbol{m}\cdot\hat{\boldsymbol{\sigma}}\hat{g}_{\boldsymbol{k'}}\boldsymbol{n}\cdot\hat{\boldsymbol{\sigma}}  \right)\delta_{J}-\frac{\xi_{SO}}{k_{F}^{2}}\{(\boldsymbol{n}\times\boldsymbol{m})\cdot\hat{\boldsymbol{\sigma}},\hat{g}_{\boldsymbol{k'}}\}\delta_{J}\\
&\qquad\qquad+\frac{\xi^{2}_{SO}}{k_{F}^{4}}\boldsymbol{n}\cdot\hat{\boldsymbol{\sigma}}\hat{g}_{\boldsymbol{k'}}\boldsymbol{n}\cdot\hat{\boldsymbol{\sigma}}\delta_{T}
+\frac{\xi^{2}_{SO}}{k_{F}^{4}}\boldsymbol{n}\cdot\boldsymbol{m}\{\hat{g}_{\boldsymbol{k'}},\boldsymbol{n}\cdot\hat{\boldsymbol{\sigma}} \}\delta_{J}\\
&\qquad\qquad+i\frac{\xi^{2}_{SO}}{k_{F}^{4}}(\boldsymbol{n}\times\boldsymbol{m})\cdot\hat{\boldsymbol{\sigma}}\hat{g}_{\boldsymbol{k'}}\boldsymbol{n}\cdot\hat{\boldsymbol{\sigma}}\delta_{J}
-i\frac{\xi^{2}_{SO}}{k_{F}^{4}}\boldsymbol{n}\cdot\hat{\boldsymbol{\sigma}}\hat{g}_{\boldsymbol{k'}}(\boldsymbol{n}\times\boldsymbol{m})\cdot\hat{\boldsymbol{\sigma}}\delta_{J}\\
&\qquad\qquad-\frac{\xi^{2}_{SO}}{k_{F}^{4}}\boldsymbol{n}^{2}\hat{g}_{\boldsymbol{k}}\delta_{T}-\frac{\xi^{2}_{SO}}{k_{F}^{4}}(2(\boldsymbol{n}\cdot\boldsymbol{m})\boldsymbol{n}-\boldsymbol{n}^{2}\boldsymbol{m})\cdot\{\hat{g}_{\boldsymbol{k}},\hat{\boldsymbol{\sigma}}\}\delta_{J}\\
&\qquad\qquad+\frac{1}{2}\frac{\xi_{SO}}{k_{F}^{2}}\left\{(\boldsymbol{k'}-\boldsymbol{k})\times\hat{\boldsymbol{\sigma}},\nabla_{\boldsymbol{R}}\hat{g}_{\boldsymbol{k'}}\right\}\delta_{T}
+\frac{1}{2}\frac{\xi_{SO}}{k_{F}^{2}}\left\{(\boldsymbol{k'}-\boldsymbol{k})\times\hat{\boldsymbol{\sigma}},\{\nabla_{\boldsymbol{R}}\hat{g}_{\boldsymbol{k'}},\boldsymbol{m}\cdot\hat{\boldsymbol{\sigma}}\}\right\}\delta_{J}\Big]\\
&-\pi a_{2}\int\frac{d\boldsymbol{k'}}{(2\pi)^{3}}\Big[\{\hat{g}_{\boldsymbol{k'}},\boldsymbol{n}\cdot\hat{\boldsymbol{\sigma}}\}\delta_{T}+\left\{\boldsymbol{n}\cdot\hat{\boldsymbol{\sigma}},\{\hat{g}_{\boldsymbol{k'}},\hat{\boldsymbol{\sigma}}\cdot\boldsymbol{m}\}\right\}\delta_{J}\\
&\qquad\qquad-\beta\left(\hat{\boldsymbol{\sigma}}\cdot\boldsymbol{m}\hat{g}_{\boldsymbol{k'}}\boldsymbol{n}\cdot\hat{\boldsymbol{\sigma}}+\boldsymbol{n}\cdot\hat{\boldsymbol{\sigma}}\hat{g}_{\boldsymbol{k'}}\hat{\boldsymbol{\sigma}}\cdot\boldsymbol{m} \right) \delta_{T}-\beta\boldsymbol{n}\cdot\boldsymbol{m}\{\hat{g}_{\boldsymbol{k'}},\hat{\boldsymbol{\sigma}}\cdot\boldsymbol{m} \}\delta_{J}-\beta\boldsymbol{m}^{2}\{\hat{g}_{\boldsymbol{k'}},\boldsymbol{n}\cdot\hat{\boldsymbol{\sigma}} \}\delta_{J}\\
&\qquad\qquad-i\beta\left((\boldsymbol{n}\times\boldsymbol{m})\cdot\hat{\boldsymbol{\sigma}}\hat{g}_{\boldsymbol{k'}}\hat{\boldsymbol{\sigma}}\cdot\boldsymbol{m}-\hat{\boldsymbol{\sigma}}\cdot\boldsymbol{m} \hat{g}_{\boldsymbol{k'}} (\boldsymbol{n}\times\boldsymbol{m})\cdot\hat{\boldsymbol{\sigma}} \right)\delta_{J}\Big].
\end{aligned}
\label{eq:collision1}
\end{equation}
\noindent By neglecting higher order terms $\beta\delta_{J}\sim\beta^{2}$ in skew-scattering and introducing a more familiar distribution function $\hat{g}_{\boldsymbol{k}}=\hat{\sigma}_{0}-2\hat{h}_{\boldsymbol{k}}$, we get:
\begin{equation}
\begin{aligned}
coll&=-2\pi a_{1}\int\frac{d\boldsymbol{k'}}{(2\pi)^{3}}\Big[(\hat{h}_{\boldsymbol{k'}}-\hat{h}_{\boldsymbol{k}})\delta_{T}+\{\hat{h}_{\boldsymbol{k'}}-\hat{h}_{\boldsymbol{k}},\hat{\boldsymbol{\sigma}}\cdot\boldsymbol{m}\}\delta_{J}\\
&\qquad\qquad+i\frac{\xi_{SO}}{k_{F}^{2}}[\boldsymbol{n}\cdot\hat{\boldsymbol{\sigma}},\hat{h}_{\boldsymbol{k'}}]\delta_{T}+i\frac{\xi_{SO}}{k_{F}^{2}}(\boldsymbol{n}\cdot\hat{\boldsymbol{\sigma}}\hat{h}_{\boldsymbol{k'}}\boldsymbol{m}\cdot\hat{\boldsymbol{\sigma}}-\boldsymbol{m}\cdot\hat{\boldsymbol{\sigma}}\hat{h}_{\boldsymbol{k'}}\boldsymbol{n}\cdot\hat{\boldsymbol{\sigma}})\delta_{J}-\frac{\xi_{SO}}{k_{F}^{2}}\{(\boldsymbol{n}\times\boldsymbol{m})\cdot\hat{\boldsymbol{\sigma}},\hat{h}_{\boldsymbol{k'}}\}\delta_{J}\\
&\qquad\qquad+\frac{1}{2}\frac{\xi_{SO}}{k_{F}^{2}}\{(\boldsymbol{k'}-\boldsymbol{k})\times\hat{\boldsymbol{\sigma}},\nabla_{\boldsymbol{R}}\hat{h}_{\boldsymbol{k'}}\}\delta_{T}
+\frac{1}{2}\frac{\xi_{SO}}{k_{F}^{2}}\{(\boldsymbol{k'}-\boldsymbol{k})\times\hat{\boldsymbol{\sigma}},\{\nabla_{\boldsymbol{R}}\hat{h}_{\boldsymbol{k'}},\boldsymbol{m}\cdot\hat{\boldsymbol{\sigma}}\}\}\delta_{J}\\
&\qquad\qquad+\frac{\xi^{2}_{SO}}{k_{F}^{4}}\boldsymbol{n}\cdot\hat{\boldsymbol{\sigma}}\hat{h}_{\boldsymbol{k'}}\boldsymbol{n}\cdot\hat{\boldsymbol{\sigma}}\delta_{T}
+\frac{\xi^{2}_{SO}}{k_{F}^{4}}\boldsymbol{n}\cdot\boldsymbol{m}\{\hat{h}_{\boldsymbol{k'}},\boldsymbol{n}\cdot\hat{\boldsymbol{\sigma}} \}\delta_{J}\\
&\qquad\qquad+i\frac{\xi^{2}_{SO}}{k_{F}^{4}}(\boldsymbol{n}\times\boldsymbol{m})\cdot\hat{\boldsymbol{\sigma}}\hat{h}_{\boldsymbol{k'}}\boldsymbol{n}\cdot\hat{\boldsymbol{\sigma}}\delta_{J}
-i\frac{\xi^{2}_{SO}}{k_{F}^{4}}\boldsymbol{n}\cdot\hat{\boldsymbol{\sigma}}\hat{h}_{\boldsymbol{k'}}(\boldsymbol{n}\times\boldsymbol{m})\cdot\hat{\boldsymbol{\sigma}}\delta_{J}\\
&\qquad\qquad-\frac{\xi^{2}_{SO}}{k_{F}^{4}}\boldsymbol{n}^{2}\hat{h}_{\boldsymbol{k}}\delta_{T}-\frac{\xi^{2}_{SO}}{k_{F}^{4}}(2(\boldsymbol{n}\cdot\boldsymbol{m})\boldsymbol{n}-\boldsymbol{n}^{2}\boldsymbol{m})\cdot\{\hat{h}_{\boldsymbol{k}},\hat{\boldsymbol{\sigma}}\}\delta_{J}\Big]\\
&+2\pi a_{2}\int\frac{d\boldsymbol{k'}}{(2\pi)^{3}}\Big[\{\hat{h}_{\boldsymbol{k'}},\boldsymbol{n}\cdot\hat{\boldsymbol{\sigma}}\}\delta_{T}+\{\boldsymbol{n}\cdot\hat{\boldsymbol{\sigma}},\{\hat{h}_{\boldsymbol{k'}},\hat{\boldsymbol{\sigma}}\cdot\boldsymbol{m}\}\}\delta_{J}-\beta(\hat{\boldsymbol{\sigma}}\cdot\boldsymbol{m}\hat{h}_{\boldsymbol{k'}}\boldsymbol{n}\cdot\hat{\boldsymbol{\sigma}}+\boldsymbol{n}\cdot\hat{\boldsymbol{\sigma}}\hat{h}_{\boldsymbol{k'}}\hat{\boldsymbol{\sigma}}\cdot\boldsymbol{m} ) \delta_{T}\Big],
\end{aligned}
\label{eq:collision2}
\end{equation}
\noindent while the Keldysh equation~(\ref{eq:keld2}) is rewritten as:
\begin{equation}
-2\left(-i[\hat{h}_{\boldsymbol{k}},J\hat{\boldsymbol{\sigma}}\cdot\boldsymbol{m}]+\frac{\hbar^{2}}{m}(\boldsymbol{k}\cdot\nabla_{\boldsymbol{R}})\,\hat{h}_{\boldsymbol{k}} \right)=coll.
\label{eq:keld3}
\end{equation}

\section{Ferromagnetic solution without spin-orbit coupling}
\par Let us consider Eq. (\ref{eq:keld3}) without extrinsic spin-orbit coupling:
\begin{equation}
\begin{aligned}
-i[\hat{h}_{\boldsymbol{k}}, J\hat{\boldsymbol{\sigma}}\cdot\boldsymbol{m}]+\frac{\hbar^2}{m}(\boldsymbol{k} \cdot \nabla_{\boldsymbol{R}}) \hat{h}_{\boldsymbol{k}}= \pi a_{1}\int\frac{d\boldsymbol{k'}}{(2\pi)^{3}}\big[(\hat{h}_{\boldsymbol{k'}}-\hat{h}_{\boldsymbol{k}})\delta_{T}+\{\hat{h}_{\boldsymbol{k'}}-\hat{h}_{\boldsymbol{k}},\hat{\boldsymbol{\sigma}}\cdot\boldsymbol{m}\}\delta_{J}\big].
\end{aligned}
\end{equation}  
\noindent By introducing $\Omega=i\hbar/\tau_0$, $\h U = \hat{\boldsymbol{\sigma}} \cdot \boldsymbol{m}$, and: 
\begin{equation}
\begin{aligned}
\label{eq:k_fer}
\hat{K}=-\frac{\hbar^2}{m}(\boldsymbol{k}\cdot \nabla_{\boldsymbol{R}}) \hat{h}_{\boldsymbol{k}}+\pi a_{1}\int\frac{d\boldsymbol{k'}}{(2\pi)^{3}}\big[\hat{h}_{\boldsymbol{k'}}\delta_{T}+\{\hat{h}_{\boldsymbol{k'}},\hat{\boldsymbol{\sigma}}\cdot\boldsymbol{m}\}\delta_{J}\big],
\end{aligned}
\end{equation}
\noindent in the weak exchange coupling limit ($J\ll\varepsilon_{F}$) we have:
\begin{equation}
\begin{aligned}
\hat{h}_{\boldsymbol{k}} = \frac{i}{\Omega} \hat{K} + \frac{\beta}{2}\{\hat{h}_{\boldsymbol{k}}, \hat{U}\} + \frac{J}{\Omega}[\hat{U}, \hat{h}_{\boldsymbol{k}}].
\end{aligned}
\end{equation}
\noindent This equation is solved iteratively:
\begin{equation}
\begin{aligned}
\hat{h}_{\boldsymbol{k}} &= \frac{i}{\Omega}\left(1+2\frac{J^{2}}{\Omega^{2}}+8\frac{J^{4}}{\Omega^{4}}+32\frac{J^{6}}{\Omega^{6}}+... \right) \hat{K} + \frac{i}{\Omega}\frac{\beta^{2}}{2}\left(1+\beta^{2}+\beta^{4}+... \right)\hat{K} \\
&+ \frac{i}{\Omega}\frac{\beta}{2}\left(1+\beta^{2}+\beta^{4}+... \right)\{\hat{U},\hat{K}\}+\frac{i}{\Omega}\frac{J}{\Omega}\left(1+4\frac{J^{2}}{\Omega^{2}}+16\frac{J^{4}}{\Omega^{4}}+...  \right)[\hat{U},\hat{K}]\\
&+ \frac{i}{\Omega}\frac{\beta^{2}}{2}\left(1+\beta^{2}+\beta^{4}+... \right)\hat{U}\hat{K}\hat{U}-2\frac{i}{\Omega}\frac{J^{2}}{\Omega^{2}}\left(1+4\frac{J^{2}}{\Omega^{2}}+16\frac{J^{4}}{\Omega^{4}}+...  \right)\hat{U}\hat{K}\hat{U},
\end{aligned}
\end{equation}
\noindent or by using $1+x+x^{2}+x^{3}...=\frac{1}{1-x}$ for $x\ll 1$:
\begin{equation}
\begin{aligned}
\hat{h}_{\boldsymbol{k}} &=\frac{i}{\Omega}\left(\frac{\Omega^{2}-2J^{2}}{\Omega^{2}-4J^{2}}+\frac{\beta^{2}}{2(1-\beta^{2})} \right)\hat{K}+\frac{i}{\Omega}\frac{\beta}{2}\frac{1}{1-\beta^{2}}\{\hat{U},\hat{K}\}\\
&+\frac{iJ}{\Omega^{2}-4J^{2}}[\hat{U},\hat{K}]+\frac{i}{\Omega}\left(\frac{\beta^{2}}{2(1-\beta^{2})}-\frac{2J^{2}}{\Omega^{2}-4J^{2}}  \right)\hat{U}\hat{K}\hat{U}.
\end{aligned}
\end{equation}
\noindent Since $J^{2}/\Omega^{2}\ll1$ and $\beta^{2}\ll1$, this solution is well justified. By substituting $\Omega$, $\hat{K}$ and $\hat{U}$ and removing the delta-functions, we have:
\begin{equation}
\begin{aligned}
\frac{\hbar}{\tau_{0}}\hat{h}_{\boldsymbol{k}}=& -\frac{\hbar^{2}}{m}(\boldsymbol{k}\cdot\nabla_{\boldsymbol{R}})\hat{h}_{\boldsymbol{k}}+\frac{\tau_{0}^{2}}{m}\frac{2J^{2}}{1+\frac{4J^{2}\tau_{0}^{2}}{\hbar^{2}}}(\boldsymbol{k}\cdot\nabla_{\boldsymbol{R}})\left(\hat{h}_{\boldsymbol{k}}-\hat{\boldsymbol{\sigma}}\cdot\boldsymbol{m} \hat{h}_{\boldsymbol{k}}\hat{\boldsymbol{\sigma}}\cdot\boldsymbol{m} \right) \\
&-\frac{\beta^{2}}{2(1-\beta^{2})}\frac{\hbar^{2}}{m}(\boldsymbol{k}\cdot\nabla_{\boldsymbol{R}})\left(\hat{h}_{\boldsymbol{k}}+\hat{\boldsymbol{\sigma}}\cdot\boldsymbol{m} \hat{h}_{\boldsymbol{k}}\hat{\boldsymbol{\sigma}}\cdot\boldsymbol{m}  \right)-\frac{\beta}{2(1-\beta^{2})}\frac{\hbar^{2}}{m}(\boldsymbol{k}\cdot\nabla_{\boldsymbol{R}})\{\hat{h}_{\boldsymbol{k}},\hat{\boldsymbol{\sigma}}\cdot\boldsymbol{m} \}\\
&+\frac{\hbar}{\tau_{0}}\int\frac{d\check{\boldsymbol{k'}}}{4\pi}\,\hat{h}_{\boldsymbol{k'}} +\frac{\tau_{0}}{\hbar}\frac{2J^{2}}{1+\frac{4J^{2}\tau_{0}^{2}}{\hbar^{2}}}\left(\int \frac{d\check{\boldsymbol{k'}}}{4\pi}\,\hat{\boldsymbol{\sigma}}\cdot\boldsymbol{m} \hat{h}_{\boldsymbol{k'}}\hat{\boldsymbol{\sigma}}\cdot\boldsymbol{m}-\int \frac{d\check{\boldsymbol{k'}}}{4\pi}\,\hat{h}_{\boldsymbol{k'}}\right)\\
&-\frac{iJ}{1+\frac{4J^{2}\tau_{0}^{2}}{\hbar^{2}}} \int\frac{d\check{\boldsymbol{k'}}}{4\pi}\,[\hat{\boldsymbol{\sigma}}\cdot\boldsymbol{m},\hat{h}_{\boldsymbol{k'}}] +\frac{\tau_{0}\hbar}{m}\frac{iJ}{1+\frac{4J^{2}\tau_{0}^{2}}{\hbar^{2}}} (\boldsymbol{k}\cdot\nabla_{\boldsymbol{R}})[\hat{\boldsymbol{\sigma}}\cdot\boldsymbol{m},\hat{h}_{\boldsymbol{k}}], 
\label{eq:fer1}
\end{aligned}
\end{equation}
\noindent where $\check{\boldsymbol{k}}=\boldsymbol{k}/|\boldsymbol{k}|$. In the diffusive limit $v_{F}\tau\ll L$, where $L$ is the system size, we can partition the distribution function $\hat{h}_{\boldsymbol{k}}$ into the isotropic charge $\mu_{c}$ and spin $\boldsymbol{\mu}$ and anisotropic $\hat{\boldsymbol{j}}\cdot\check{\boldsymbol{k}}$ components, $\hat{h}_{\boldsymbol{k}}=\mu_{c}\hat{\sigma}_{0}+\boldsymbol{\mu}\cdot\hat{\boldsymbol{\sigma}}+\hat{\boldsymbol{j}}\cdot\check{\boldsymbol{k}}$. This form is nothing else but the generalized $p$-wave approximation for the distribution function. Upon integrating Eq. (\ref{eq:fer1}) multiplied by $\check{\boldsymbol{k}}$ over $d\check{\boldsymbol{k}}/4\pi$ and neglecting higher order terms $\sim \beta^{2}$, we obtain the following expression for $\hat{\boldsymbol{j}}$: 
\begin{equation}
\begin{aligned}
\label{eq:j_fer}
\frac{\hbar}{\tau_{0}}\hat{\boldsymbol{j}}=&-\frac{\hbar^{2}}{m}k\nabla\left(\mu_{c}\hat{\sigma}_{0}+\boldsymbol{\mu}\cdot\hat{\boldsymbol{\sigma}}+\beta\mu_{c}\hat{\boldsymbol{\sigma}}\cdot\boldsymbol{m}+\beta\boldsymbol{\mu}\cdot{\boldsymbol{m}}\hat{\sigma}_{0}\right)\\
&-\frac{\tau_{0}\hbar}{m}\frac{2J}{1+\frac{4J^{2}\tau_{0}^{2}}{\hbar^{2}}}k\nabla\,\hat{\boldsymbol{\sigma}}\cdot(\boldsymbol{m}\times\boldsymbol{\mu})-\frac{\tau_{0}^{2}}{m}\frac{4J^{2}}{1+\frac{4J^{2}\tau_{0}^{2}}{\hbar^{2}}}k\nabla\,\hat{\boldsymbol{\sigma}}\cdot(\boldsymbol{m}\times(\boldsymbol{m}\times\boldsymbol{\mu})).
\end{aligned}
\end{equation}
\par The charge and spin currents (its $j$th component in the spin space) can be defined as:
\begin{equation}
\tilde{\boldsymbol{j}}^{C}=\frac{1}{4}\int\frac{d\check{\boldsymbol{k}}}{4\pi}\mathrm{Tr}\,\{\hat{\boldsymbol{v}}_{\boldsymbol{k}},\hat{h}_{\boldsymbol{k}}\}=\frac{v_{F}}{6}\mathrm{Tr}\,\hat{\boldsymbol{j}},
\end{equation}
\noindent and
\begin{equation}
\tilde{\boldsymbol{J}}^{S}_{j}=\frac{1}{4}\int\frac{d\check{\boldsymbol{k}}}{4\pi}\mathrm{Tr}\,\big[\hat{\sigma}_{j}\{\hat{\boldsymbol{v}}_{\boldsymbol{k}},\hat{h}_{\boldsymbol{k}}\}\big]=\frac{v_{F}}{6}\mathrm{Tr}\,\big[\hat{\sigma}_{j}\,\hat{\boldsymbol{j}}\big],
\end{equation}
\noindent where the velocity operator is defined as $\hat{\boldsymbol{v}}_{\boldsymbol{k}}=\frac{\hbar}{m}\boldsymbol{k}\hat{\sigma}_{0}$, and $v_{F}=\frac{\hbar}{m} k_{F}$ is the Fermi velocity. Thus, neglecting higher order terms $\sim J^{3}$ gives:
\begin{equation}
\tilde{\boldsymbol{j}}^{C} = - D\nabla (\mu_c + \beta \boldsymbol{\mu} \cdot \boldsymbol{m}),
\label{eq:cc_fer}
\end{equation}
\noindent and
\begin{equation}
\frac{\tilde{\boldsymbol{J}}^{S}_{j}}{D}=  -\nabla( \mu_{j}+\beta\mu_c m_{j})-\frac{\tau_{0}}{\tau_{L}}\nabla(\boldsymbol{m}\times\boldsymbol{\mu})_{j}-\frac{\tau_{0}}{\tau_{\phi}}\nabla(\boldsymbol{m}\times(\boldsymbol{m}\times\boldsymbol{\mu}))_{j},
\label{eq:sc_fer}
\end{equation}
\noindent where $D=\tau_{0}v_{F}^{2}/3$ is the diffusion coefficient, $1/\tau_{L}=2J/\hbar$ is the Larmor precession time, and $1/\tau_{\phi}=4J^{2}\tau_{0}/\hbar^{2}$ is the spin dephasing time.
\par The corresponding equations for the charge and spin densities are obtained by integrating Eq.~(\ref{eq:fer1}) over $\check{\boldsymbol{k}}$ and neglecting terms $\sim J\nabla^{2}$ and $\sim\beta\nabla^{2}$:
\begin{equation}
-\frac{\hbar^{2}}{m}\frac{k}{3}\nabla \cdot \hat{\boldsymbol{j}}+\frac{\tau_{0}}{\hbar}\frac{4J^{2}}{1+\frac{4J^{2}\tau_{0}^{2}}{\hbar^{2}}}\hat{\boldsymbol{\sigma}}\cdot(\boldsymbol{m}\times(\boldsymbol{m}\times\boldsymbol{\mu}))+\frac{2J}{1+\frac{4J^{2}\tau_{0}^{2}}{\hbar^{2}}}\hat{\boldsymbol{\sigma}}\cdot(\boldsymbol{m}\times\boldsymbol{\mu})=0.
\end{equation}
\noindent Taking $\mathrm{Tr}\,[...]$ and $\mathrm{Tr}\,[\hat{\boldsymbol{\sigma}}...]$ and neglecting terms $\sim J^{3}$ leads to:
\begin{equation}
0=D\nabla^{2}(\mu_{c}+\beta\boldsymbol{\mu}\cdot\boldsymbol{m})=-\nabla\cdot\tilde{\boldsymbol{j}}^{C}
\end{equation}
\noindent and 
\begin{equation}
0=-\nabla\cdot\tilde{\boldsymbol{J}}^{S}+\frac{1}{\tau_{L}}(\boldsymbol{m}\times\boldsymbol{\mu})+\frac{1}{\tau_{\phi}}(\boldsymbol{m}\times(\boldsymbol{m}\times\boldsymbol{\mu}))
\end{equation}
\noindent for the charge and spin components, respectively. Finally, by recovering time-dependence from Eq. (\ref{eq:kelfull}) we obtain:
\begin{equation}
\partial_{T}\mu_{c}=-\nabla\cdot\tilde{\boldsymbol{j}}^{C}
\label{eq:ca_fer}
\end{equation}
\noindent and  
\begin{equation}
\partial_{T}\boldsymbol{\mu}=-\nabla\cdot\tilde{\boldsymbol{J}}^{S}+\frac{1}{\tau_{L}}(\boldsymbol{m}\times\boldsymbol{\mu})+\frac{1}{\tau_{\phi}}(\boldsymbol{m}\times(\boldsymbol{m}\times\boldsymbol{\mu})).
\label{eq:sa_fer}
\end{equation}
\noindent Thus, Eqs. (\ref{eq:cc_fer}), (\ref{eq:sc_fer}), (\ref{eq:ca_fer}) and (\ref{eq:sa_fer}) define a set of the drift-diffusion equations for ferromagnets in the absence of extrinsic spin-orbit coupling.

\section{Ferromagnetic solution with spin-orbit coupling}
\par To derive drift-diffusion equations including extrinsic spin-orbit coupling, we employ the same $p$-wave approximation for $\hat{h}_{\boldsymbol{k}}$. Then, we have:
\begin{gather}
-i[\hat{h}_{\boldsymbol{k}},J\hat{\boldsymbol{\sigma}}\cdot\boldsymbol{m}]=2J\hat{\boldsymbol{\sigma}}\cdot(\boldsymbol{\mu}\times\boldsymbol{m})-i[\hat{\boldsymbol{j}}\cdot\check{\boldsymbol{k}},J\hat{\boldsymbol{\sigma}}\cdot\boldsymbol{m}],\\
(\boldsymbol{k}\cdot\nabla_{\boldsymbol{R}})\hat{h}_{\boldsymbol{k}}=(\boldsymbol{k}\cdot\nabla_{\boldsymbol{R}})\mu_{c}\hat{\sigma}_{0}+(\boldsymbol{k}\cdot\nabla_{\boldsymbol{R}})\boldsymbol{\mu}\cdot\hat{\boldsymbol{\sigma}}+(\boldsymbol{k}\cdot\nabla_{\boldsymbol{R}})\hat{\boldsymbol{j}}\cdot\check{\boldsymbol{k}}
\end{gather}
\noindent  for the left-hand side of the Keldysh equation~(\ref{eq:keld3}), and:
\begin{gather}
\begin{aligned}
\{\nabla_{\boldsymbol{R}}\hat{h}_{\boldsymbol{k'}},(\boldsymbol{k'}-\boldsymbol{k})\times\hat{\boldsymbol{\sigma}}\}&=2\left(\nabla_{\boldsymbol{R}}\times(\boldsymbol{k'}-\boldsymbol{k}) \right)\cdot\hat{\boldsymbol{\sigma}}\mu_{c}-2(\boldsymbol{k'}-\boldsymbol{k})\cdot\left(\nabla_{\boldsymbol{R}}\times\boldsymbol{\mu}\right)\hat{\sigma}_{0}\\
&+\{\nabla_{\boldsymbol{R}}\,\hat{\boldsymbol{j}}\cdot\check{\boldsymbol{k'}},(\boldsymbol{k'}-\boldsymbol{k})\times\hat{\boldsymbol{\sigma}}\},
\end{aligned}\\
\begin{aligned}
\{(\boldsymbol{k'}-\boldsymbol{k})\times\hat{\boldsymbol{\sigma}},\{\nabla_{\boldsymbol{R}}\hat{h}_{\boldsymbol{k'}},\boldsymbol{m}\cdot\hat{\boldsymbol{\sigma}}\}\}&=4(\boldsymbol{k'}-\boldsymbol{k})\cdot(\boldsymbol{m}\times\nabla_{\boldsymbol{R}} \mu_{c})\hat{\sigma}_{0}+4\hat{\boldsymbol{\sigma}}\cdot(\nabla_{\boldsymbol{R}}\times(\boldsymbol{k'}-\boldsymbol{k}))\boldsymbol{\mu}\cdot\boldsymbol{m}\\
&+\{(\boldsymbol{k'}-\boldsymbol{k})\times\hat{\boldsymbol{\sigma}},\{\nabla_{\boldsymbol{R}}\,\hat{\boldsymbol{j}}\cdot\check{\boldsymbol{k'}},\boldsymbol{m}\cdot\hat{\boldsymbol{\sigma}}\}\}
\end{aligned}\\
\begin{aligned}
\boldsymbol{n}\cdot\hat{\boldsymbol{\sigma}}\hat{h}_{\boldsymbol{k'}}\boldsymbol{n}\cdot\hat{\boldsymbol{\sigma}}-\boldsymbol{n}^{2}\hat{h}_{\boldsymbol{k}}=2(\boldsymbol{n}\times(\boldsymbol{n}\times\boldsymbol{\mu}))\cdot\hat{\boldsymbol{\sigma}}+\boldsymbol{n}\cdot\hat{\boldsymbol{\sigma}}\hat{\boldsymbol{j}}\cdot\check{\boldsymbol{k'}}\boldsymbol{n}\cdot\hat{\boldsymbol{\sigma}}-\boldsymbol{n}^{2}\hat{\boldsymbol{j}}\cdot\check{\boldsymbol{k}}
\end{aligned}
\end{gather}
\noindent for the collision integral~(\ref{eq:collision2}). Upon integrating Eq. (\ref{eq:keld3}) over $d\check{\boldsymbol{k}}/4\pi$ and neglecting terms $\sim \xi_{SO}^{2}\beta$, we obtain in the limit $J\ll\varepsilon_{F}$: 
\begin{equation}
\begin{aligned}
\label{eq:gen1}
2J\hat{\boldsymbol{\sigma}}\cdot(\boldsymbol{\mu}\times\boldsymbol{m})+\frac{1}{3}\frac{\hbar^{2}k}{m}\,\nabla_{\boldsymbol{R}}\cdot\hat{\boldsymbol{j}}=&-\frac{8}{9}\frac{\hbar}{\tau_{0}}\frac{{k}^{2}}{k_{F}^{2}}\xi_{SO}^{2}\boldsymbol{\mu}\cdot\hat{\boldsymbol{\sigma}}\\
&+\frac{1}{6}\frac{\hbar}{\tau_{0}}\frac{\xi_{SO}}{k_{F}}\big(\nabla_{\boldsymbol{R}}\cdot(\hat{\boldsymbol{j}}\times\hat{\boldsymbol{\sigma}})-\nabla_{\boldsymbol{R}}\cdot(\hat{\boldsymbol{\sigma}}\times\hat{\boldsymbol{j}}) \big)\\
&+\frac{1}{6}\frac{\hbar}{\tau_{0}}\frac{\xi_{SO}}{k_{F}}\beta\big[\nabla_{\boldsymbol{R}}\cdot(\hat{\boldsymbol{j}}\times\hat{\boldsymbol{\sigma}})+\nabla_{\boldsymbol{R}}\cdot(\hat{\boldsymbol{\sigma}}\times\hat{\boldsymbol{j}}),\boldsymbol{m}\cdot\hat{\boldsymbol{\sigma}}\big]\\
&+\frac{2}{3}\frac{\hbar}{\tau_{0}}\frac{\xi_{SO}}{k_{F}}\beta\,\nabla_{\boldsymbol{R}}\cdot(\boldsymbol{m}\times\hat{\boldsymbol{j}}).
\end{aligned}
\end{equation}
\noindent One more equation is derived by averaging Eq.~(\ref{eq:keld3}) over $\check{\boldsymbol{k}}$ multiplied by $\check{\boldsymbol{k}}$ and neglecting terms $\sim \xi_{SO}^{2}\beta$:
\begin{equation}
\begin{aligned}
\label{eq:gen2}
-iJ\big[\,\hat{\boldsymbol{j}},\,\hat{\boldsymbol{\sigma}}\cdot\boldsymbol{m} \big]+\frac{\hbar^{2}k}{m}\nabla_{\boldsymbol{R}}(\mu_{c}\hat{\sigma}_{0}+\boldsymbol{\mu}\cdot\hat{\boldsymbol{\sigma}})=&-\frac{\hbar}{\tau_{0}}\Big(1+\frac{2}{3}\frac{k^{2}}{k_{F}^{2}}\xi_{SO}^{2}\Big)\hat{\boldsymbol{j}}+\frac{1}{2}\frac{\hbar}{\tau_{0}}\beta\big\{\hat{\boldsymbol{j}},\hat{\boldsymbol{\sigma}}\cdot\boldsymbol{m} \big\}\\
&-\frac{i}{3}\frac{\hbar}{\tau_{0}}\frac{k}{k_{F}}\xi_{SO}\big(\hat{\boldsymbol{j}}\times\hat{\boldsymbol{\sigma}}+\hat{\boldsymbol{\sigma}}\times\hat{\boldsymbol{j}}  \big)\\
&+\frac{i}{3}\frac{\hbar}{\tau_{0}}\frac{k}{k_{F}}\xi_{SO}\beta\big(\boldsymbol{m}\cdot\hat{\boldsymbol{\sigma}}\,\hat{\boldsymbol{j}}\times\hat{\boldsymbol{\sigma}}+\hat{\boldsymbol{\sigma}}\times\hat{\boldsymbol{j}}\,\boldsymbol{m}\cdot\hat{\boldsymbol{\sigma}}\big)\\
&+\frac{1}{3}\frac{\hbar}{\tau_{0}}\frac{k}{k_{F}}\xi_{SO}\beta\big(\hat{\boldsymbol{\sigma}}\cdot\hat{\boldsymbol{j}}+\hat{\boldsymbol{j}}\cdot\hat{\boldsymbol{\sigma}})\boldsymbol{m}
-\frac{1}{3}\frac{\hbar}{\tau_{0}}\frac{k}{k_{F}}\xi_{SO}\beta\big\{\hat{\boldsymbol{\sigma}},\hat{\boldsymbol{j}}\cdot\boldsymbol{m} \big\}\\
&+\frac{\hbar}{\tau_{0}}\frac{k}{k_{F}}\frac{\xi_{SO}}{k_{F}}\big(\nabla_{\boldsymbol{R}}\times\hat{\boldsymbol{\sigma}}\,(\mu_{c}-\beta \boldsymbol{\mu}\cdot\boldsymbol{m})+\nabla_{\boldsymbol{R}}\times(\boldsymbol{\mu}-\beta\mu_{c}\boldsymbol{m})\,\hat{\sigma}_{0} \big)\\
&+\frac{1}{6}\frac{m}{\pi\hbar}\frac{v_{i}}{\tau_{0}}\xi_{SO}k\big(\hat{\boldsymbol{\sigma}}\times\hat{\boldsymbol{j}}-\hat{\boldsymbol{j}}\times\hat{\boldsymbol{\sigma}}\big)\\
&+\frac{1}{3}\frac{m}{\pi\hbar}\frac{v_{i}}{\tau_{0}}\xi_{SO}\beta k\big(\boldsymbol{m}\cdot\hat{\boldsymbol{\sigma}}\,\hat{\boldsymbol{j}}\times\hat{\boldsymbol{\sigma}}-\hat{\boldsymbol{\sigma}}\times\hat{\boldsymbol{j}}\,\boldsymbol{m}\cdot\hat{\boldsymbol{\sigma}}\big)\\
&+\frac{1}{6}\frac{m}{\pi\hbar}\frac{v_{i}}{\tau_{0}}\xi_{SO}\beta k\big(\boldsymbol{m}\cdot\hat{\boldsymbol{\sigma}}\,\hat{\boldsymbol{\sigma}}\times \hat{\boldsymbol{j}}-\hat{\boldsymbol{j}}\times\hat{\boldsymbol{\sigma}}\,\boldsymbol{m}\cdot\hat{\boldsymbol{\sigma}}\big)\\
&-\frac{2}{3}\frac{m}{\pi\hbar}\frac{v_{i}}{\tau_{0}}\xi_{SO}\beta k\,\boldsymbol{m}\times \hat{\boldsymbol{j}}.
\end{aligned}
\end{equation}
\noindent The equations above define a set of the generalized drift-diffusion equations, which can now be solved approximately while keeping leading orders in $\xi_{SO}$ and $\beta$. Then, starting from a ferromagnetic solution given by Eq.~(\ref{eq:j_fer}) the anisotropic component of the density matrix is obtained by solving Eq.~(\ref{eq:gen2}):
\begin{equation}
\begin{aligned}
\hat{\boldsymbol{j}}&=-\tau_{0}v_{F}\nabla\hat{\mu}_{0}+\left(\frac{\xi_{SO}}{k_{F}}+\frac{\tau_{0}v_{i}k_{F}^{2}}{3\pi\hbar}\xi_{SO} \right)\nabla\times\boldsymbol{\mu}\hat{\sigma}_{0}-\frac{\tau_{0}v_{i}k_{F}^{2}}{3\pi\hbar}\xi_{SO}\beta (\nabla\times\boldsymbol{m})\mu_{c}\hat{\sigma}_{0}\\
&-\left(\frac{2}{3}\xi_{SO}\beta\tau_{0}v_{F}-\frac{\tau_{0}v_{i}k_{F}^{2}}{3\pi\hbar}\xi_{SO}\frac{\tau_{0}}{\tau_{L}} \right)\nabla\times(\boldsymbol{m}\times\boldsymbol{\mu})\hat{\sigma}_{0}+\left(\frac{\xi_{SO}}{k_{F}}+ \frac{\tau_{0}v_{i}k_{F}^{2}}{3\pi\hbar}\xi_{SO}\right)\nabla\times\hat{\boldsymbol{\sigma}}\mu_{c}\\
&-\frac{\xi_{SO}}{k_{F}}\beta\nabla\times(\boldsymbol{m}\times(\hat{\boldsymbol{\sigma}}\times\boldsymbol{\mu}))-\frac{\tau_{0}v_{i}k_{F}^{2}}{3\pi\hbar}\xi_{SO}\beta\nabla\times((\boldsymbol{m}\times\hat{\boldsymbol{\sigma}})\times\boldsymbol{\mu})-\frac{\tau_{0}v_{i}k_{F}^{2}}{3\pi\hbar}\xi_{SO}\beta(\nabla\times\hat{\boldsymbol{\sigma}})\boldsymbol{\mu}\cdot\boldsymbol{m}\\
&+\left(\frac{\xi_{SO}}{k_{F}}+\frac{\tau_{0}v_{i}k_{F}^{2}}{3\pi\hbar} \right)\frac{\tau_{0}}{\tau_{L}}\nabla\times(\hat{\boldsymbol{\sigma}}\times\boldsymbol{m})\mu_{c} - \frac{2}{3}\tau_{0}v_{F}\xi_{SO}\nabla\times(\hat{\boldsymbol{\sigma}}\times(\boldsymbol{\mu}-\beta\mu_{c}\boldsymbol{m}))\\
&-\frac{2}{3}\tau_{0}v_{F}\xi_{SO}\frac{\tau_{0}}{\tau_{L}}\nabla\times\left((\hat{\boldsymbol{\sigma}}\times\boldsymbol{m})\times\boldsymbol{\mu}+\hat{\boldsymbol{\sigma}}\times(\boldsymbol{m}\times\boldsymbol{\mu})  \right),
\end{aligned}
\end{equation}
\noindent where 
\begin{equation}
\hat{\mu}_{0}=(\mu_{c}+\beta\boldsymbol{\mu}\cdot\boldsymbol{m})\hat{\sigma}_{0}+(\boldsymbol{\mu}+\beta\mu_{c}\boldsymbol{m})\cdot\hat{\boldsymbol{\sigma}}+\frac{\tau_{0}}{\tau_{L}}(\boldsymbol{m}\times\boldsymbol{\mu})\cdot\hat{\boldsymbol{\sigma}}+\frac{\tau_{0}}{\tau_{\phi}}(\boldsymbol{m}\times(\boldsymbol{m}\times\boldsymbol{\mu}))\cdot\hat{\boldsymbol{\sigma}}.
\end{equation}
\noindent Here, the first term of $\hat{\boldsymbol{j}}$ comes from the ferromagnetic solution by moving the right-hand side of Eq.~(\ref{eq:gen2}) into Eq.~(\ref{eq:k_fer}). Plugging this solution in Eq.~(\ref{eq:gen1}) leads to:
\begin{equation}
\begin{aligned}
\label{eq:j_full}
\frac{2J}{\hbar}(\boldsymbol{\mu}\times\boldsymbol{m})\cdot\hat{\boldsymbol{\sigma}}&+\frac{8}{9}\frac{\xi_{SO}^{2}}{\tau_{0}}\boldsymbol{\mu}\cdot\hat{\boldsymbol{\sigma}}=\\
&=D\nabla\cdot\Big[ \nabla\hat{\mu}_{0}+\alpha_{sj}\big[\hat{\boldsymbol{\sigma}}\times\nabla(2\mu_{c}-\beta\boldsymbol{\mu}\cdot\boldsymbol{m})-\nabla\times(2\boldsymbol{\mu}-\beta\mu_{c}\boldsymbol{m})\big]\\
&+\alpha_{sk}\big[\hat{\boldsymbol{\sigma}}\times\nabla(\mu_{c}-\beta\boldsymbol{\mu}\cdot\boldsymbol{m})-\nabla\times(\boldsymbol{\mu}-\beta\mu_{c}\boldsymbol{m})\big]+\alpha_{sw}\nabla\times(\hat{\boldsymbol{\sigma}}\times\boldsymbol{\mu})\\
&-\nabla\times\big[(\alpha_{sj}\frac{\tau_{0}}{\tau_{L}}+\alpha_{sk}\frac{\tau_{0}}{\tau_{L}}+\alpha_{sw}\beta)(\hat{\boldsymbol{\sigma}}\times\boldsymbol{m})\mu_{c} + (\alpha_{sj}\frac{\tau_{0}}{\tau_{L}}+\alpha_{sk}\frac{\tau_{0}}{\tau_{L}}-\alpha_{sw}\beta)(\boldsymbol{m}\times\boldsymbol{\mu}) \big]\\
&+\nabla\times\big[(\alpha_{sj}\beta+\alpha_{sk}\beta+\alpha_{sw}\frac{\tau_{0}}{\tau_{L}})\hat{\boldsymbol{\sigma}}\times(\boldsymbol{m}\times\boldsymbol{\mu})  - (\alpha_{sj}\beta+\alpha_{sk}\beta-\alpha_{sw}\frac{\tau_{0}}{\tau_{L}})(\hat{\boldsymbol{\sigma}}\times\boldsymbol{m})\times\boldsymbol{\mu}  \big]\Big],
\end{aligned}
\end{equation}
\noindent where $\alpha_{sw}=\frac{2\xi_{SO}}{3}$, $\alpha_{sj}=\frac{\xi_{SO}}{l_{F}k_{F}}$ and $\alpha_{sk}=\frac{v_{i}mk_{F}}{3\pi\hbar^{2}}\xi_{SO}$ are the spin swapping, side-jump and skew-scattering coefficients, respectively, and $l_{F}=\tau_{0}v_{F}$ is the mean-free path. As seen, Eq.~(\ref{eq:j_full}) can be regarded as a generalized continuity equation for the density matrix $\mu_{c}\hat{\sigma}_{0}+\boldsymbol{\mu}\cdot\hat{\boldsymbol{\sigma}}$, and its right-hand side is nothing else but the divergence of the full current $\boldsymbol{j}^{C}\hat{\sigma}_{0}+\boldsymbol{J}^{S}\cdot\hat{\boldsymbol{\sigma}}$, where the dot product is over spin components. Thus, the corresponding expressions for the charge and spin currents (its $j$th spin component) can be readily written as:
\begin{equation}
\begin{aligned}
\label{eq:finalcharge}
\boldsymbol{j}^{C}/D&=\tilde{\boldsymbol{j}}^{C}/D+\alpha_{sj}\nabla\times(2\boldsymbol{\mu}-\beta\mu_{c}\boldsymbol{m})+\alpha_{sk}\nabla\times(\boldsymbol{\mu}-\beta\mu_{c}\boldsymbol{m})+(\alpha_{sj}\frac{\tau_{0}}{\tau_{L}}+\alpha_{sk}\frac{\tau_{0}}{\tau_{L}}-\alpha_{sw}\beta)\nabla\times(\boldsymbol{m}\times\boldsymbol{\mu})
\end{aligned}
\end{equation}
\noindent and
\begin{equation}
\begin{aligned}
\label{eq:finalspin}
\boldsymbol{J}_{j}^{S}/D&=\tilde{\boldsymbol{J}}_{j}^{S}/D+\alpha_{sj}\nabla\times\boldsymbol{e}_{j}(2\mu_{c}-\beta\boldsymbol{\mu}\cdot\boldsymbol{m})+\alpha_{sk}\nabla\times\boldsymbol{e}_{j}(\mu_{c}-\beta\boldsymbol{\mu}\cdot\boldsymbol{m})-\alpha_{sw}\nabla\times(\boldsymbol{e}_{j}\times\boldsymbol{\mu})\\
&+\nabla\times (\alpha_{sj}\frac{\tau_{0}}{\tau_{L}}+\alpha_{sk}\frac{\tau_{0}}{\tau_{L}}+\alpha_{sw}\beta)(\boldsymbol{e}_{j}\times\boldsymbol{m})\mu_{c} - \nabla\times(\alpha_{sj}\beta+\alpha_{sk}\beta+\alpha_{sw}\frac{\tau_{0}}{\tau_{L}})(\boldsymbol{e}_{j}\times(\boldsymbol{m}\times\boldsymbol{\mu})) \\
& + \nabla\times(\alpha_{sj}\beta+\alpha_{sk}\beta-\alpha_{sw}\frac{\tau_{0}}{\tau_{L}})((\boldsymbol{e}_{j}\times\boldsymbol{m})\times\boldsymbol{\mu}),
\end{aligned}
\end{equation}
\noindent or
\begin{equation}
\begin{aligned}
J_{ij}^{S}/D&=\tilde{J}_{ij}^{S}/D-\alpha_{sj}\epsilon_{ijk}\nabla_{k}(2\mu_{c}-\beta\mu_{n}m_{n})-\alpha_{sk}\epsilon_{ijk}\nabla_{k}(\mu_{c}-\beta\mu_{n}m_{n})-\alpha_{sw}(\delta_{ij}\nabla_{k}\mu_{k}-\nabla_{j}\mu_{i})\\
&+(\alpha_{sj}\frac{\tau_{0}}{\tau_{L}}+\alpha_{sk}\frac{\tau_{0}}{\tau_{L}}+\alpha_{sw}\beta)(\delta_{ij}\nabla_{k}m_{k}-m_{i}\nabla_{j})\mu_{c} \\
& - (\alpha_{sj}\beta+\alpha_{sk}\beta+\alpha_{sw}\frac{\tau_{0}}{\tau_{L}})\epsilon_{ikn}\nabla_{k}(m_{n}\mu_{j}-\mu_{n}m_{j}) \\
& +(\alpha_{sj}\beta+\alpha_{sk}\beta-\alpha_{sw}\frac{\tau_{0}}{\tau_{L}})(\epsilon_{ikn}\nabla_{k}m_{n}\mu_{j}+\epsilon_{ijk}\nabla_{k}m_{n}\mu_{n}),
\end{aligned}
\end{equation}
\noindent where $\epsilon_{ijk}$ is the Levi-Civita symbol, and summation over repeated indexes is implied. Here, the first and second subscripts correspond to the spatial and spin components, respectively. Finally, by recovering time dependence in Eq.~(\ref{eq:j_full}) we obtain the remaining equations for the charge and spin densities:
\begin{equation}
\label{eq:finalchden}
\partial_{T}\mu_{c}=D\nabla^{2}(\mu_{c}+\beta\boldsymbol{\mu}\cdot\boldsymbol{m})=-\nabla\cdot\boldsymbol{j}^{C}
\end{equation}
\noindent and
\begin{equation}
\label{eq:finalspden}
\partial_{T}\boldsymbol{\mu}=-\nabla\cdot\boldsymbol{J}^{S}+\frac{1}{\tau_{L}}(\boldsymbol{m}\times\boldsymbol{\mu})+\frac{1}{\tau_{\phi}}(\boldsymbol{m}\times(\boldsymbol{m}\times\boldsymbol{\mu}))-\frac{1}{\tau_{sf}}\boldsymbol{\mu},
\end{equation}
\noindent where $1/\tau_{sf}=8\xi_{SO}^{2}/9\tau_{0}$ is the spin-flip relaxation time. The set of Eqs.~(\ref{eq:finalcharge}), (\ref{eq:finalspin}), (\ref{eq:finalchden}) and (\ref{eq:finalspden}) is the central result of this work.

\section{Spin swapping symmetry}
\par In this section we verify the symmetry of the spin swapping term and compare our results with the previous ones derived for normal metals. 
\par In their original work Dyakonov and Lifshits give the following definition of the spin current $q_{ij}$ due to scattering off the spin-orbit coupling potential:\cite{perel} 
\begin{equation}
\label{lifshits}
q_{ij}=q_{ij}^{(0)}-\alpha_{sh}\epsilon_{ijk}q^{(0)}_{k}+\alpha_{sw}(q_{ji}^{(0)}-\delta_{ij}q_{kk}^{(0)}),
\end{equation}
\noindent where $q^{(0)}_{k}$ and $q_{ij}^{(0)}$ stand for the primary charge and spin currents in the absence of spin-orbit coupling, respectively, and $\alpha_{sh}$ represents the overall spin Hall effect. Thus, it is argued that the spin swapping effect always appears in the form given above.
\par Let us consider our solution in the case of normal metals ($\beta=0$ and $J=0$):
\begin{equation}
\begin{aligned}
\boldsymbol{J}_{j}^{S}&=-D\nabla\mu_{j}+D\alpha_{sh}\nabla\times\boldsymbol{e}_{j}\mu_{c}-D\alpha_{sw}\nabla\times(\boldsymbol{e}_{j}\times\boldsymbol{\mu})
\end{aligned}
\end{equation}
\noindent or
\begin{equation}
J_{ij}^{S}=-D\nabla_{i}\mu_{j}-D\alpha_{sh}\epsilon_{ijk}\nabla_{k}\mu_{c}+D\alpha_{sw}(\nabla_{j}\mu_{i}-\delta_{ij}\nabla_{k}\mu_{k}).
\end{equation}
\noindent Taking into account that $q_{ij}^{(0)}\approx-D\nabla_{i}\mu_{j}$ and $q_{k}^{(0)}\approx-D\nabla_{k}\mu_{c}$, it is seen that Eq.~(\ref{eq:finalspin}) displays the correct symmetry up to a sign coming from the definition of the spin-orbit coupling potential, Eq.~(\ref{eq:imppot}). This form is also in agreement with some previously published results.\cite{brataas, raimondi}
\par Finally, it is worth comparing our equations with those that fail to include spin swapping in the form given by Eq.~(\ref{lifshits}). For example, in Ref.~[\onlinecite{manchon1}] the spin swapping term appeared with the following symmetry:
\begin{equation}
\begin{aligned}
e^{2}\boldsymbol{J}_{j}^{S}/\sigma_{N}&=-\nabla\mu_{j}/2+\alpha_{sj}\boldsymbol{e}_{j}\times\nabla\mu_{c}-\alpha_{sw}\boldsymbol{e}_{j}\times(\nabla\times\boldsymbol{\mu})/2\\
&=-\nabla\mu_{j}/2+\alpha_{sj}\boldsymbol{e}_{j}\times\nabla\mu_{c}-\alpha_{sw}(\nabla\mu_{j}-\nabla_{j}\boldsymbol{\mu})/2,
\end{aligned}
\end{equation}
\noindent where $\sigma_{N}$ is the bulk conductivity. It is clear that the symmetry of spin swapping is wrong: e.g. $q_{xx}$ should contain a term $\sim\alpha_{sw}(-q^{(0)}_{yy}-q^{(0)}_{zz})$, which is absent in the expression above. There is the same symmetry problem in Eq.~(2) of Ref.~[\onlinecite{brataas2}], where the spin swapping term reads as $-\alpha_{sw}\hat{\boldsymbol{\sigma}}\times\nabla\times\boldsymbol{\mu}$ (or $-\alpha_{sw}\boldsymbol{e}_{j}\times\nabla\times\boldsymbol{\mu}$ for the spin current component).
